\documentclass[twocolumn]{aastex701}

\begin{document}

\title{A Detailed Model Atmosphere Analysis of Hot White Dwarfs in DESI DR1}

\author[orcid=0000-0001-6098-2235,sname='Kilic']{Mukremin Kilic} 
\affiliation{Homer L. Dodge Department of Physics and Astronomy, University of Oklahoma, 440 W. Brooks St., Norman, OK, 73019 USA}
\email[show]{kilic@ou.edu}

\author[orcid=0000-0003-2368-345X,sname='Bergeron']{Pierre Bergeron}
\affiliation{D\'epartement de Physique, Universit\'e de Montr\'eal, C.P. 6128, Succ. Centre-Ville, Montr\'eal, QC H3C 3J7, Canada}
\email[show]{bergeron@astro.umontreal.ca}

\author[0000-0001-7143-0890]{Adam Moss} 
\affiliation{Department of Astronomy, University of Florida, Bryant Space Science Center, Stadium Road, Gainesville, FL 32611, USA}
\email{Adam.G.Moss-1@ou.edu}

\author[orcid=0000-0002-9632-1436,sname='Blouin']{Simon Blouin}
\affiliation{Department of Physics and Astronomy, University of Victoria, Victoria BC V8W 2Y2, Canada}
\email{sblouin@uvic.ca}

\author[0000-0002-0948-4801]{Matthew J. Green}
\affiliation{Homer L. Dodge Department of Physics and Astronomy, University of Oklahoma, 440 W. Brooks St., Norman, OK, 73019 USA}
\email{matthew.j.green-2@ou.edu}

\author[orcid=0009-0009-9105-7865,sname='Jewett']{Gracyn Jewett}
\affiliation{Homer L. Dodge Department of Physics and Astronomy, University of Oklahoma, 440 W. Brooks St., Norman, OK, 73019 USA}
\email{gjewett@ou.edu}

\author[0000-0002-6153-9304,sname='Barrientos']{Manuel Barrientos} 
\affiliation{Homer L. Dodge Department of Physics and Astronomy, University of Oklahoma, 440 W. Brooks St., Norman, OK, 73019 USA}
\email{mbarrientos@ou.edu }

\author[0009-0000-7416-5228]{Alexander L. Albright}
\affiliation{Homer L. Dodge Department of Physics and Astronomy, University of Oklahoma, 440 W. Brooks St., Norman, OK, 73019 USA}
\email{Alexander.L.Albright-1@ou.edu}

\author[0000-0002-4462-2341,sname='Brown']{Warren R.\ Brown}
\affiliation{Center for Astrophysics, Harvard \& Smithsonian, 60 Garden Street, Cambridge, MA 02138, USA}
\email{wbrown@cfa.harvard.edu}

\begin{abstract}

We present a detailed model atmosphere analysis of hot white dwarfs in the Dark Energy Spectroscopic Instrument (DESI) Data Release 1. Our sample includes 19,321 unique targets with $G_{\rm BP}-G_{\rm RP}\leq0$. We use the DESI spectra along with Gaia parallaxes and SDSS, Pan-STARRS, and SkyMapper photometry to perform spectroscopic and photometric fits. We find a significant discrepancy between the photometric and spectroscopic masses for DA white dwarfs (a systematic offset of 0.05-$0.06~M_\odot$), indicating problems with the broad hydrogen line profiles in DESI spectroscopy data. Our photometric fits are consistent with a peak at the canonical mass of $0.6~M_\odot$. A remarkable feature of the mass distribution is the prevalence of magnetic white dwarfs among the ultramassive DA population and that of warm DQs in the non-DA distribution. We identify 70 DQs in the DESI hot white dwarf sample, including 9 DAQs with carbon and hydrogen atmospheres. We constrain the ratio of non-DA to DA white dwarfs as a function of temperature, and discuss the implications for the spectral evolution of white dwarfs in the temperature range $10^5-10^4$ K. We also discuss unusual objects in the sample, including metal-rich white dwarfs and extremely low mass white dwarfs. This analysis provides the first look at the large sample of Gaia-selected white dwarf candidates that will be observed with multiplexed spectroscopic surveys like DESI, SDSS-V, 4MOST, and WEAVE over the next several years.

\end{abstract}

\keywords{\uat{White dwarf stars}{1799} --- \uat{DA stars}{348} --- \uat{DB stars}{358} --- \uat{DO stars}{397} --- \uat{DQ stars}{1849} --- \uat{DZ stars}{1848}}

\section{Introduction}

Gaia Data Release 2 \citep{gaia18} revolutionized the white dwarf research field
by finally revealing the hidden population of faint white dwarfs in the solar neighborhood \citep{tremblay24}.
Thanks to Gaia Data Release 3 \citep{gaia20}, we know now nearly $4\times10^5$ high-confidence white dwarf candidates \citep{gentile21}. 

Another revolution is in progress;
several wide-field multiplexed spectroscopic surveys, i.e. the Sloan Digital Sky Survey \citep[SDSS-V,][]{kollmeier19}, Dark Energy Spectroscopic Instrument \citep[DESI,][]{desi25}, 4-metre Multi-Object Spectroscopic Telescope \citep[4MOST,][]{dejong19}, and the William Herschel Telescope Enhanced Area Velocity Explorer \citep[WEAVE,][]{jin22}, are taking advantage of recent developments in robotic fiber positioners to obtain spectroscopy of millions of objects, including the Gaia DR3 white dwarf sample. This is the beginning of a new era, in which the number of spectroscopically confirmed white dwarfs will increase by an order of magnitude. Such a large sample will not only provide better characterization of the white dwarf sample in the solar neighborhood, but also reveal rare classes of white dwarfs that would not be found otherwise.

Over the past two decades, SDSS provided the largest samples of spectroscopically confirmed white dwarfs. For example,
\citet{eisenstein06} identified 9316 white dwarfs in SDSS Data Release 4, while \citet{kleinman13} found 20407 white dwarf spectra in Data Release 7 \citep[also see][]{kepler15,kepler19,kepler21}. These studies concentrated mainly on normal $\sim0.6~M_\odot$ white dwarfs and on the spectral properties of H-atmosphere DA and He-atmosphere DB white dwarfs, and were later followed up by detailed model atmosphere analysis of various spectral types \citep[e.g.,][]{tremblay11,genest19,koester19,coutu19,bedard20}. 

Among the recently initiated wide-field multiplexed spectroscopic surveys, SDSS-V and DESI are the first surveys to release their data.
\citet{manser24} presented the identification of 2706 spectroscopically confirmed white dwarfs in DESI Early Data Release, including 1958 DA, 141 DB,
55 DQ, and 152 metal-rich white dwarfs. This study was based on survey validation observations that include $\sim$$4\times10^5$ stellar spectra. For
comparison, the recent DESI Data Release 1 (DR1) includes $\sim$$4\times10^6$ stellar spectra \citep{koposov25}. Here we take advantage of the publicly
available DESI DR1 and provide a detailed model atmosphere analysis of 19,321 unique targets, including a tailored analysis for each spectral type. 
Our sample selection is discussed in Section \ref{sample}, whereas the model atmosphere analysis and the results for various spectral types are presented in Section \ref{secmodel}. Section \ref{secres}
presents the results for the overall sample properties, including mass and temperature distributions and the constraints on spectral evolution.
We conclude in Section \ref{seccon}. 

\section{DESI Data, Sample Selection, and Spectral Classification}
\label{sample}

DESI is mounted on the 4m Mayall telescope at the Kitt Peak National Observatory. \citet{desi25} provide the details of survey operations and
Data Release 1. Briefly, DESI uses three spectrographs to cover the wavelength ranges 3600-5930, 5600-8000, and 7470-9800 \AA. The spectra are
extracted at a linearized pixel scale of 0.8 \AA/pixel. The flux calibration is based on spectra of stars observed in each field. DESI DR1 is based on 13
months of observations that cover the observing dates 2021 May through 2022 June, and include both the survey validation data and the main DESI survey.

We cross matched DESI DR1 with the Gaia DR3 white dwarf catalog of \citet{gentile21}, and found 47,795 objects in common. \citet{caron23} noted
that even though progress can only be made through a tailored analysis of individual objects, they felt that they reached the limit of human
capacity to individually analyze each object given their sample size of nearly 3000 white dwarfs. Given this previous observation by \citet{caron23}
and the significantly larger number of objects with spectra in DESI DR1, we decided to limit our study to only those objects with
$G_{\rm BP} - G_{\rm RP}\leq0$ to keep the sample size manageable. This color choice is not arbitrary, as it corresponds to
$T_{\rm eff}\approx10,000$ K, above which H and He lines are visible, making a detailed model atmosphere analysis possible based on the presence
or absence of H and He lines in the DESI data.

\begin{figure}
\includegraphics[width=3.3in]{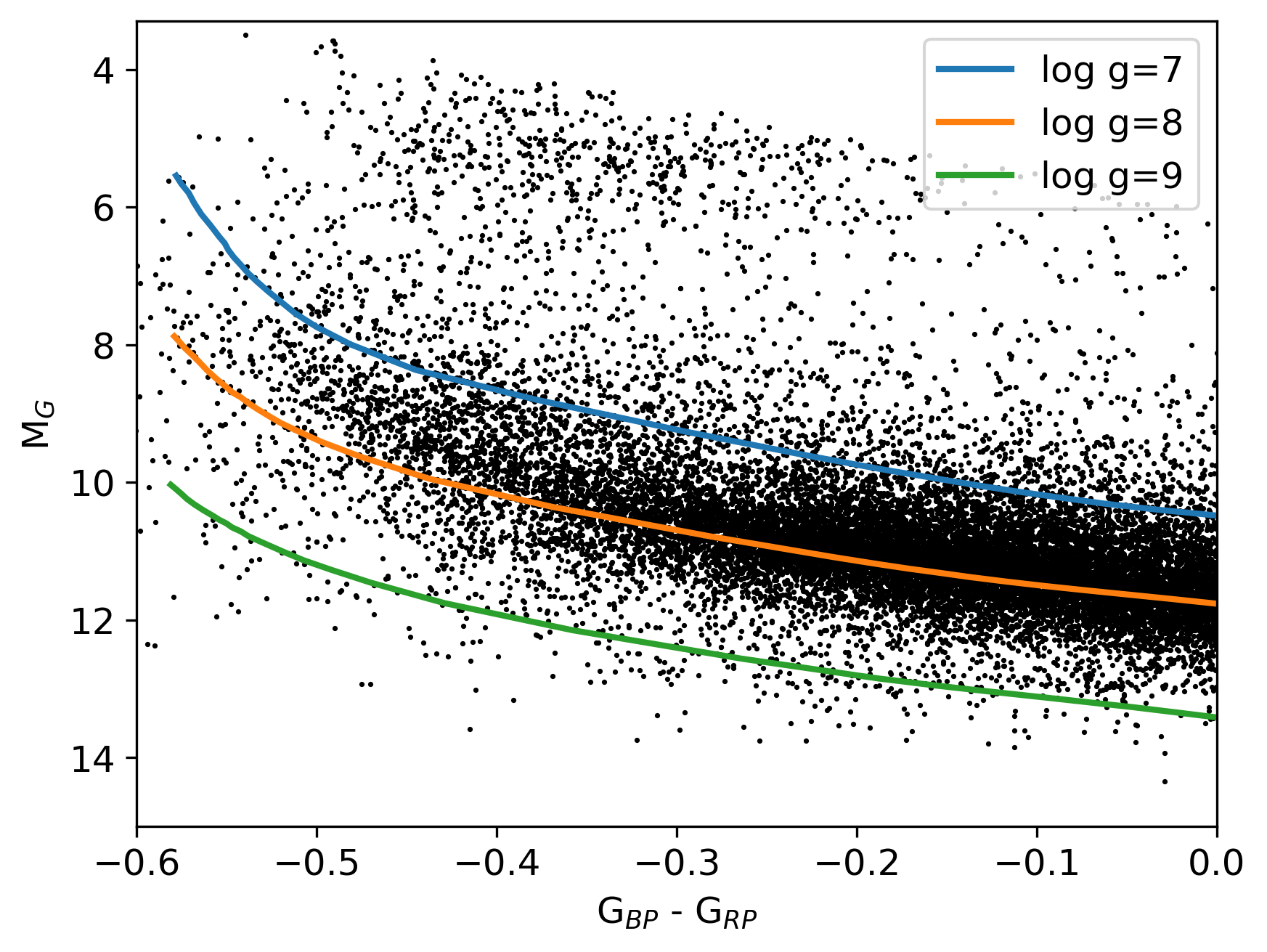}
\caption{Our selected DESI DR1 white dwarf sample in the H-R diagram. Evolutionary tracks for pure H atmosphere white dwarfs with $\log{g}=$ 7, 8, and 9 \citep{holberg06,bedard20} are shown for comparison.}
\label{fighr} 
\end{figure}

Figure \ref{fighr} shows the H-R diagram for our selected sample of 19,411 white dwarf candidates with $G_{\rm BP} - G_{\rm RP}\leq0$. The colored
lines show the evolutionary tracks for pure H atmosphere white dwarfs with $\log{g}=$ 7, 8, and 9. Our sample of objects clearly span a broad range of
surface gravities (masses). In addition, since DESI is a magnitude limited survey, overluminous objects like subdwarfs (with $M_{\rm G}\sim5$) and
low-mass white dwarfs are over-represented. An inspection of the DESI data for these 19,411 targets shows that 19,321 have usable spectra that enable spectral classification and a detailed model atmosphere analysis. Many objects have multiple spectra in DESI; we find 23,411 spectra for these targets, but we always use the spectrum with the best signal-to-noise (S/N) ratio for our analysis.

\begin{figure}
\center
\includegraphics[width=3in]{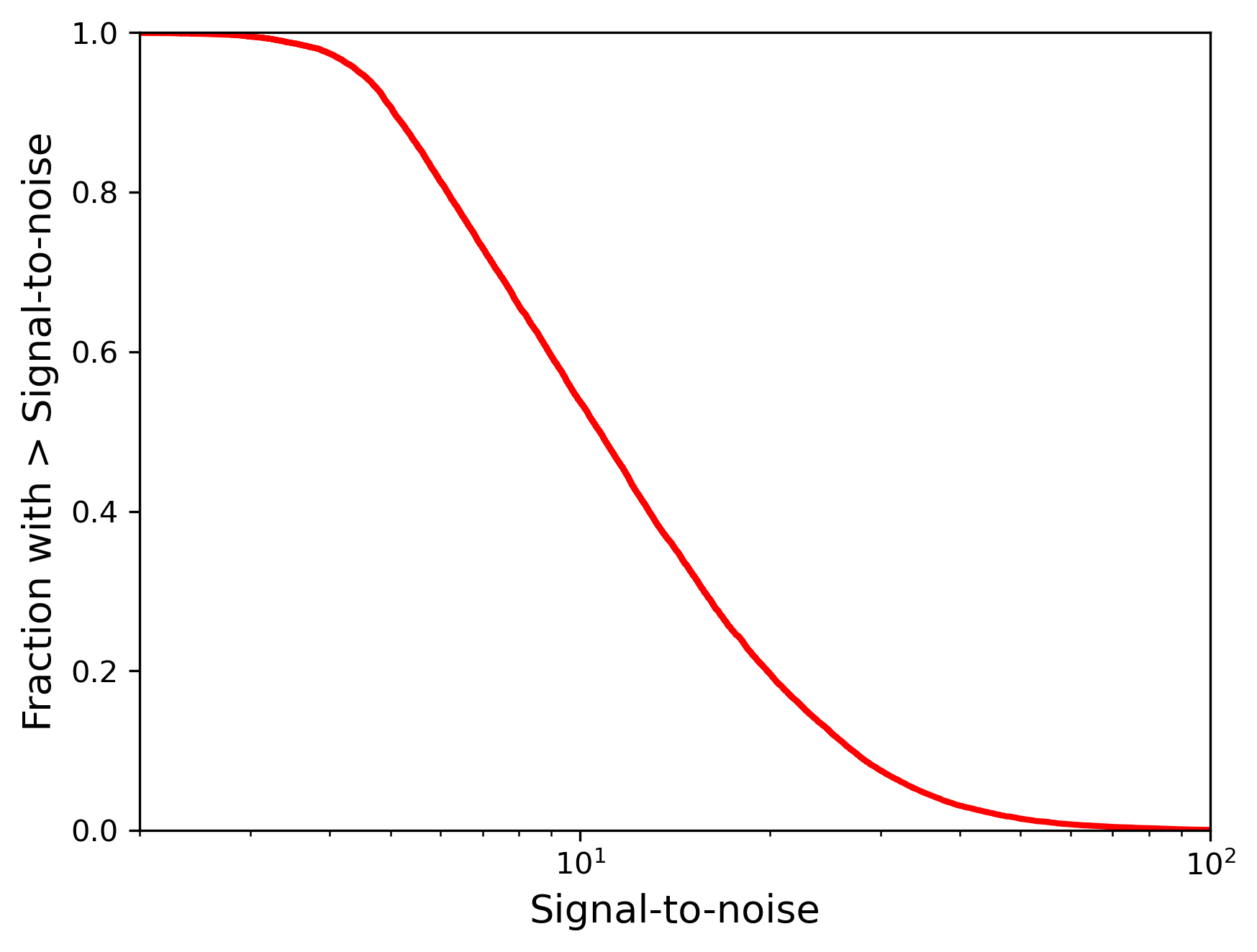}
\caption{Cumulative distribution of the signal-to-noise ratios of the DESI DR1 spectra of our sample of targets.}
\label{figsn} 
\end{figure}

Figure \ref{figsn} shows the cumulative S/N ratio distribution of the DESI spectra for our sample. 
The average S/N is calculated in two regions, 4450 to 4750 \AA\ for DA stars and 4175 to 4300 \AA\ for all other spectral types.
The DESI spectra have S/N ranging from 2 to 180, with a median at S/N = 11. About 91, 54, and 20\% of the targets have S/N $\geq5$, 10, and 20, respectively.

For spectral classification, we first used the photometric method (see below) to find the best-fitting H and He atmosphere models, and compared the predicted model spectra against the observed DESI spectra. We visually inspected all of these model fits to identify subdwarfs, DA, DB, DC, DO, DQ, and DZ white dwarfs plus unusual objects like He-DA, magnetic white dwarfs, and CVs. We then performed a tailored analysis of each spectral type as described below, and re-inspected all of the model fits to verify and also identify additional subtypes like DAO, DBA, DBAZ, DZA etc. Finally, we visually re-inspected all of the model fits again to verify the final spectral classifications. 

\begin{figure}
\center
\includegraphics[width=3in]{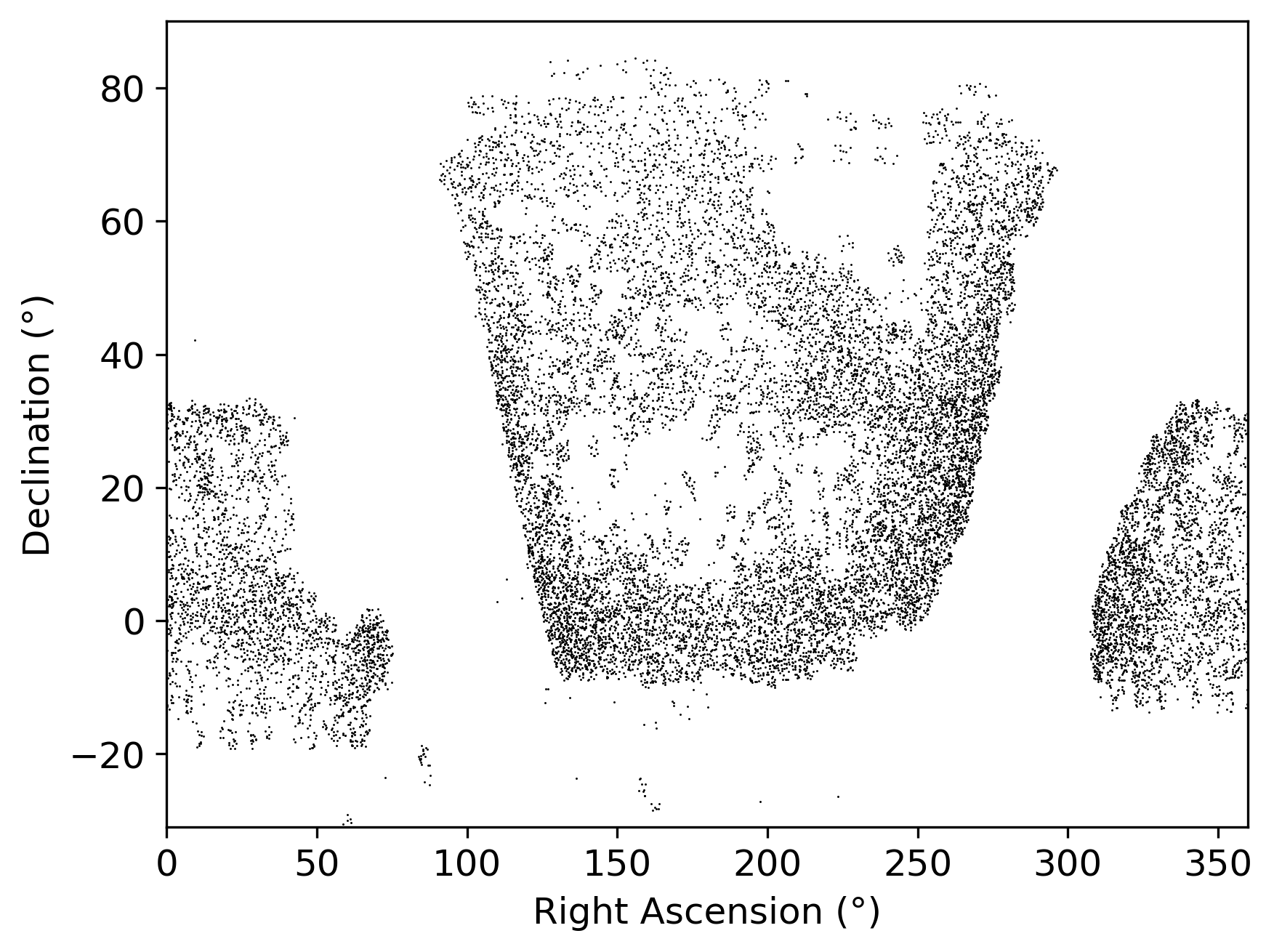}
\caption{Sky coordinates of our DESI DR1 targets.}
\label{figfoot} 
\end{figure}

Figure \ref{figfoot} shows the sky coordinates of our selected white dwarf sample, which mirror the DESI DR1 footprint. Unsurprisingly, DESI's
target selection is limited to northern objects with $\delta\geq-30^\circ$. In addition, given the emphasis on cosmology, DESI DR1 avoided the Galactic plane. This highlights a major issue for the white dwarf research field, as even with the ongoing and upcoming highly multiplexed spectroscopic surveys,
we are unlikely to obtain follow-up spectroscopy of a significant population of white dwarfs near the Galactic plane. 
There are other gaps visible within the current DESI footprint. However, those gaps should be filled in within the 5 year timeline of the DESI survey. 

\section{Model Atmosphere Analysis}
\label{secmodel}

\subsection{Theoretical  Models}

For DA white dwarfs, we rely on the pure H composition LTE models described in \citet{tremblay09} with improved Stark broadening profiles. Our model grid covers $T_{\rm eff}$ ranging from 8000 K to 120,000 K and $\log{g}$ from 5.0 to 9.5. We employ the 3D atmosphere model corrections from \citet{tremblay13,tremblay15} to the resulting temperature and surface gravity estimates from our spectroscopic fits.

For hot DAs, the parameters remain uncertain above a certain temperature due to our use of the LTE models. \citet{liebert05} fitted the Balmer lines of the LTE models with the NLTE model spectra to derive corrections (see their Figure 3). They found that the LTE models overestimate both $T_{\rm eff}$ and $\log{g}$, though the impact on $\log{g}$ is relatively small ($\leq0.03$ dex). The differences between the LTE and NLTE models become significant for $T_{\rm eff}\geq40,000$ K. For example, at $T_{\rm eff}=80,000$ K, the LTE models overestimate the temperature by about 5000 K. 

For the hot helium-rich stars in the sample (DAO, DO, DAB, and subdwarfs), we use a grid of LTE models with $T_{\rm eff}$ ranging from 20,000 K to 120,000 K, $\log{g}$ from 5.0 to 9.5, with pure H and pure He compositions, and also with mixed compositions of $\log {\rm H/He} = -5$ to +5 in steps of 1 dex. 
Note that all abundances given in this paper are by number.
Given the low $\log{g}$ values for subdwarfs, the mass-radius relation for white dwarfs is not applicable, and thus photometric solutions are not provided for subdwarfs.
Furthermore, many of the fits for hot DA/DAO/DO stars show the well known high-Balmer problem where the observed lines are much deeper than predicted \citep[see Figure 11 in][for examples]{gianninas10}. This discrepancy is due to the presence of metals \citep[CNO,][]{werner96} in the atmosphere, as shown for instance in Figure 5 of \citet{gianninas10}. Our models are metal-free. Hence, a similar problem exists for DAO and DO stars as well.

For the DB/DBA stars in the sample, we use the LTE model atmospheres described in \citet{bergeron11},
with improvements to the van der Waals broadening discussed in \citet{genest19}.
Our model grid covers the $T_{\rm eff}$ range from 11,000 K to 50,000 K, $\log{g}$ from 7 to 9, and it includes atmosphere models with pure He composition and mixed atmospheres with $\log{\rm H/He} = -6$ to $+1.5$ and 0.5 dex resolution. The 3D corrected parameters for DB and DBA white dwarfs  still
show discrepancies with the photometrically derived parameters \citep{cuka21}. This is likely because the treatment of microphysics, including Stark and neutral
line broadening at cooler temperatures, needs to be revisited \citep[e.g.,][]{cuka21}. Hence, we refrain from applying the 3D corrections  to the spectroscopic parameters of DB/DBA white dwarfs.

For warm DQs, due to the absence of helium features in their spectra, there is no way to tell how much He is present in the atmosphere \citep{koester19,kilic24}.
We can at best put an upper limit on the atmospheric He abundance, but the abundance could be as low as zero. In addition, the
helium abundance has minimal impact on the physical parameters of warm DQs \citep{kilic25b}. Hence, we adopt He-free models in this work.
We rely on an extended version of the C+H (no helium) model atmosphere grid presented in \citet{kilic25b}. Our model grid covers the $T_{\rm eff}$ range from 11,000 to 26,000 K, $\log{g}$ from 8.0 to 9.5, and $\log {\rm C/H}$ from $-1.5$ to +3.0. 
 
Finally, for metal-rich DBZ/DZ white dwarfs, we rely on an extended version of the model grids presented in \citet{blouin18a,blouin18b}. This model grid covers $T_{\rm eff}$ ranging from 4000 to 19,500 K, $\log{g}$ from 7 to 9, $\log {\rm H/He}$ from $-6$ to $0$ as well as H/He = 0 (no hydrogen), and $\log {\rm Ca/He}$ from $-12$ to $-6$.
The abundances for all other elements are scaled to Ca to match the abundance ratios of CI chondrites.

Evolutionary sequences are required in our analysis to convert the radius into mass and $\log{g}$ for the photometric method, and $\log{g}$ into mass for the spectroscopic method. For DA white dwarfs, we rely on the evolutionary models from \citet{bedard20} with CO cores, $q({\rm He})\equiv \log M_{\rm He}/M_{\star}=10^{-2}$ and $q({\rm H})=10^{-4}$, which are representative of H-atmosphere white dwarfs. For objects with $T_{\rm eff}\lesssim 30,000$~K and $M_{\rm CO}< 0.35~M_\odot$, we instead rely on the \citet{althaus13} evolutionary models involving He-core white dwarfs. For non-DA white dwarfs, we rely on the \citet{bedard20} evolutionary models with $q({\rm He})=10^{-2}$ and $q({\rm H})=10^{-10}$, which are representative of He-atmosphere white dwarfs.

\subsection{DA White Dwarfs}

There are two independent methods to constrain the physical parameters of DA white dwarfs, one based
on photometry + distance and the other using spectroscopy. Prior to Gaia, most studies used
the spectroscopic method \citep{bergeron92}, where 1D atmosphere models are compared against the normalized Balmer line profiles to constrain $T_{\rm eff}$ and $\log{g}$. Those results require 3D model corrections, which are now well established \citep{tremblay13,tremblay15}. 

Since Gaia Data Release 2 \citep{gaia18},
most studies took advantage of the photometric method \citep{bergeron19}, which uses broadband photometry from various surveys along with the Gaia parallaxes to constrain the effective temperature
and the solid angle, which constrain the radius directly thanks to the distance measurement.
The radius constraints can then be used to infer the white dwarf mass through evolutionary models.
Both the photometric and spectroscopic methods have their advantages and disadvantages. For example,
optical photometry only samples the Rayleigh-Jeans tail of the spectral energy distribution for hot
white dwarfs. Hence, the photometric method becomes unreliable for hot white dwarfs, whereas the spectroscopic method is unreliable for cooler ($T_{\rm eff}\lesssim7000$~K) DA white dwarfs with diminishingly weaker lines. 

We use both methods to obtain independent constraints on the physical parameters of the DA white dwarfs in our sample. For the photometric fits, we use all available photometry from the SDSS, Pan-STARRS, and SkyMapper surveys along with the Gaia DR3 parallaxes to constrain the effective temperature and the solid angle $\pi (R/D)^2$, where $R$ is the radius of the star and $D$ is its  distance. Since the distance is known from Gaia parallaxes, we constrain the radius of the star directly, and therefore the mass and $\log{g}$ using the evolutionary models. We exclude Gaia photometry from our fits, unless no other photometric data is available. Gaia $G_{\rm BP}, G$ and $G_{\rm RP}$ measurements are not independent \citep{huang24}, as they are derived from the same observations. Furthermore, \citet{bergeron19} found a much better agreement between the spectroscopic and photometric masses based on $ugrizy$ photometry.

\begin{figure*}
\center
\includegraphics[width=3in, clip=true, trim=0.4in 0.8in 0.1in 1.1in]{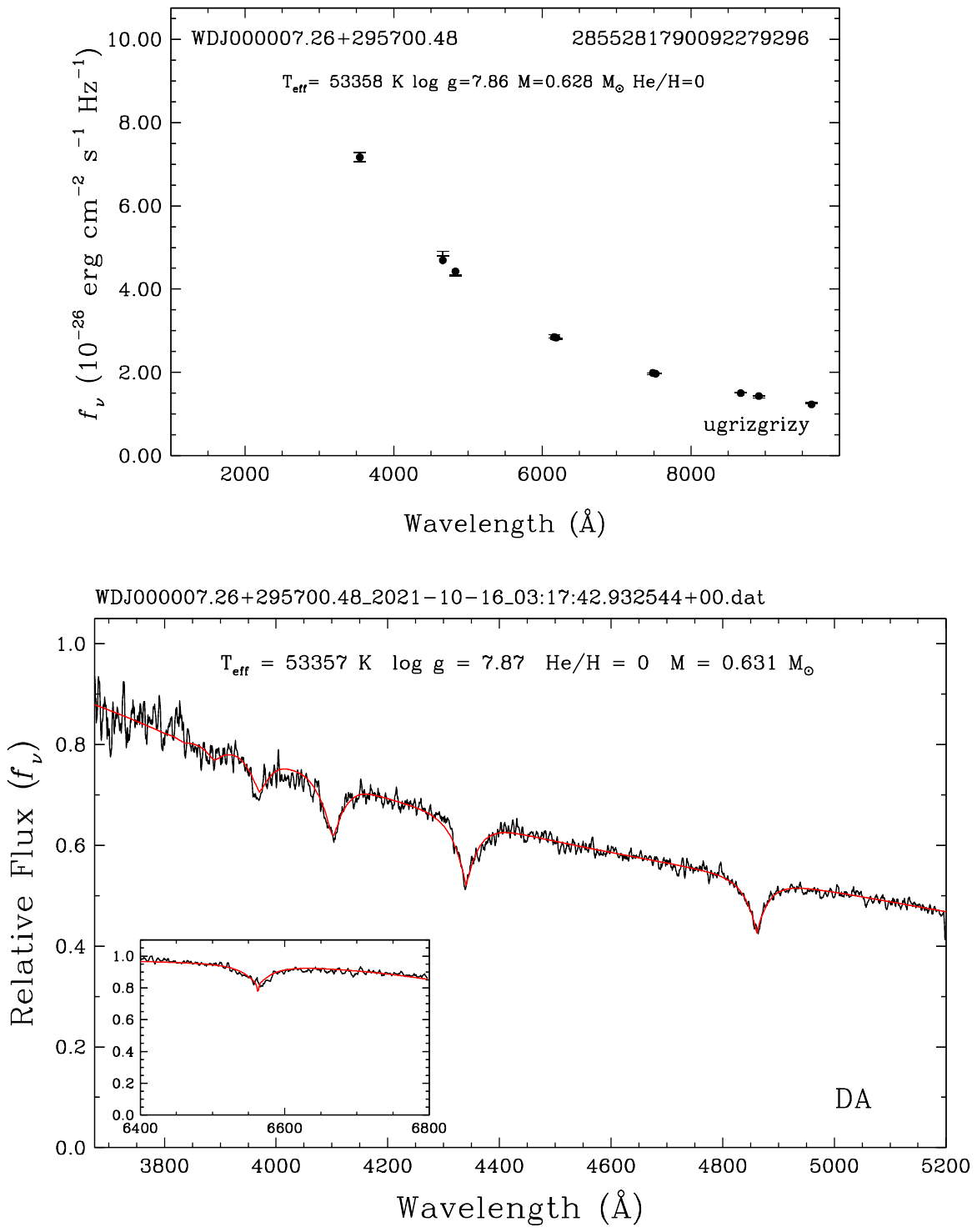}
\includegraphics[width=3in, clip=true, trim=0.4in 0.8in 0.1in 1.1in]{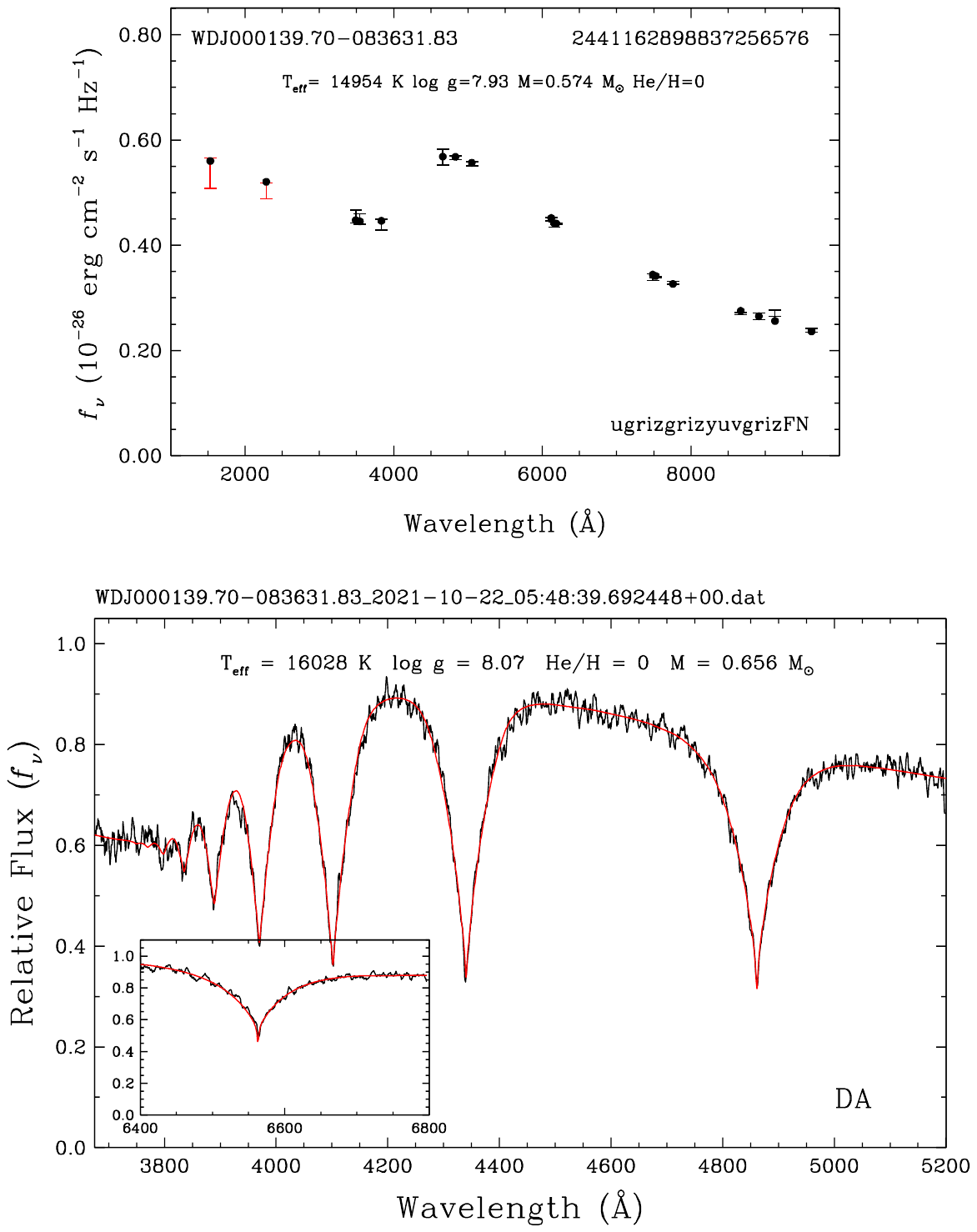}
\caption{Model fits to a hot (left) and a warm (right) DA white dwarf. The top panels show the best-fitting pure H (filled dots) atmosphere models to the photometry (error bars). These panels also include the white dwarf name, Gaia Source ID, the file name of the spectrum (including the DESI observation date), and the photometry used in the fitting. The bottom panels show the fits to the DESI spectra. Since the photometric method becomes unreliable for hot white dwarfs, we force the spectroscopic temperature on their photometric fits. Here and in the following figures, we smooth the DESI spectra by 5 points (with 0.8 \AA\ per point) for display purposes only.}
\label{fitda} 
\end{figure*}

\begin{deluxetable*}{lrcccccccc}
\tablecolumns{10} \tablewidth{0pt}
\tablefontsize{\tiny}
\tablecaption{Physical Parameters of the Hot White Dwarfs in DESI DR1.\label{tabpar}}
\tablehead{\colhead{Object} & \colhead{Gaia SourceID} & \colhead{Type} & \colhead{Comp} & \colhead{$T_{\rm eff,phot}$} & \colhead{$\log{g, phot}$} & \colhead{Mass,phot} & \colhead{$T_{\rm eff,spec}$} & \colhead{$\log{g,spec}$} & \colhead{Mass,spec}\\
 & & & & (K) & (cm s$^{-2}$) & ($M_\odot$) & (K) & (cm s$^{-2}$) & ($M_\odot$)}
\startdata
WDJ000007.26+295700.48   &  2855281790092279296 &  DA &  H &  53358 $\pm$ \nodata &  7.862 $\pm$ 0.007 &  0.628 $\pm$ 0.004 &  53357 &  7.870 &  0.631 \\
WDJ000022.88$-$000635.71 &  2449953455645886464 &  DA &  H &  21255 $\pm$ 355 &  7.435 $\pm$ 0.119 &  0.377 $\pm$ 0.037 &  23600 &  7.509 &  0.410 \\
WDJ000034.07$-$010820.03 &  2449631676695853440 &  DA &  H &  12600 $\pm$ 127 &  7.998 $\pm$ 0.029 &  0.606 $\pm$ 0.024 &  12950 &  8.084 &  0.658 \\
WDJ000047.68$-$125135.38 &  2421123333052628864 &  DB &  He &  11885 $\pm$ 220 &  7.617 $\pm$ 0.312 &  0.391 $\pm$ 0.173 &  \nodata & \nodata & \nodata \\
WDJ000059.17$-$123745.25 &  2421163327788222976 &  DB &  He &  13418 $\pm$ 643 &  7.795 $\pm$ 0.235 &  0.476 $\pm$ 0.166 &  13188 &  7.288 &  0.281 \\
WDJ000100.41$-$042742.85 &  2447719110579100032 &  DA &  H &  14847 $\pm$ 314 &  7.334 $\pm$ 0.085 &  0.365 $\pm$ 0.034 &  16823 &  7.725 &  0.474 \\
WDJ000100.48+310124.76   &  2873413595827680512 &  DA &  H &  16781 $\pm$ 430 &  8.222 $\pm$ 0.074 &  0.751 $\pm$ 0.064 &  18039 &  8.268 &  0.782 \\
WDJ000104.06+000355.88   &  2546033488965366784 &  DA &  H &  13221 $\pm$ 254 &  8.081 $\pm$ 0.074 &  0.657 $\pm$ 0.062 &  12653 &  7.885 &  0.543 \\
WDJ000109.21+294904.89   &  2861268729688392448 &  DA &  H &  20187 $\pm$ 682 &  7.338 $\pm$ 0.280 &  0.391 $\pm$ 0.072 &  19215 &  7.794 &  0.514 \\
WDJ000115.77+285647.36   &  2855153009792862592 &  DA &  H &  14015 $\pm$ 506 &  8.211 $\pm$ 0.153 &  0.739 $\pm$ 0.131 &  13271 &  8.173 &  0.714 \\
\enddata
\tablecomments{This table is available in its entirety in machine-readable format in the online journal. A portion is shown here for guidance regarding its form and content. For WDJ000007.26+295700.48 and other hot white dwarfs with missing $T_{\rm eff, phot}$ errors, we forced the spectroscopic temperature.}
\end{deluxetable*}

We include GALEX photometry, if available, in the figures, but not in the fits themselves. We also omit problematic photometry, where contamination from nearby objects is clearly visible in our model fits. This is especially a problem for SkyMapper data, which has worse seeing than the SDSS and Pan-STARRS.
The omitted photometry is shown in red in the photometric fits.
We correct for reddening based on the mean $A_V$ for each object from \citet{gentile21} with $A_V/E(B-V) = 3.1$ and assuming a \citet{fitzpatrick99} reddening law. For the GALEX bands, we use the extinction coefficients from \citet{wall19}. We convert the observed magnitudes into average fluxes, and compare with the synthetic fluxes calculated from pure H model atmospheres. We minimize the $\chi^2$ difference between the observed and model fluxes over all band passes using the nonlinear least-squares method of Levenberg-Marquardt \citep{press86} to obtain the best fitting photometric parameters and their uncertainties. 

Balmer lines are strongest at $T_{\rm eff}\sim14,000$ K in DA white dwarfs. Hence, spectroscopic model fits usually have a degeneracy between a hot and a cold solution above and below this temperature. To overcome this problem, we adopt the photometric $T_{\rm eff}$ as an initial estimate of the spectroscopic $T_{\rm eff}$ in our model fits to the DESI spectra. Even though this method works well for most stars, it sometimes fails for unresolved double degenerate binaries where contamination from a cool companion impacts the photometric solution. In those cases, we force a hot or cool initial temperature seed based on visual inspection of the model fits. 

The spectroscopic method uses the normalized Balmer line profiles for parameter estimation \citep{bergeron92,bergeron19}. Here, we use a different approach for DESI spectra,
where we fit the full spectrum (and a polynomial to account for flux calibration issues) instead of the normalized line profiles. We compared the parameters using both methods and the results agree with no systematic trends, giving us confidence that this method (fitting the full DESI spectra) could then be applied to other spectral types as well. Radial velocity is included as a free parameter in the model fits.

Figure \ref{fitda} shows our photometric (top) and spectroscopic (bottom) fits to two DA white dwarfs. 
We provide the model fits for all stars in our sample on Zenodo, which can be accessed via the DOI \href{https://doi.org/10.5281/zenodo.18332706}{10.5281/zenodo.18332706}.
The top left panel shows the SDSS $ugriz$ and Pan-STARRS $grizy$ photometry (error bars) along with the
predicted fluxes from the best-fitting pure H (filled dots) model for the hot white dwarf WDJ000007.26+295700.48. The labels in the same panel give the white dwarf name, Gaia Source ID, and the photometry used in the fitting. The bottom panel shows the spectroscopic fit, which provides an excellent match to the DESI spectrum of this object. Since the photometric method becomes unreliable for hot white dwarfs, we force the spectroscopic temperature in the photometric fits if $T_{\rm phot}$ is above 50,000 K. Hence, for hot white dwarfs, our photometric solutions are dependent on the spectroscopic temperature, but this still results in an independent photometric mass estimate. 

Table \ref{tabpar} presents the physical parameters for all white dwarfs in our sample, including the results from both the photometric and spectroscopic fits. For certain spectral types, only one set of solutions is presented. For example, only the spectroscopic parameters are given for subdwarfs, since the photometric parameters rely on the mass-radius relation for white dwarfs. On the other hand,  for metal-rich DBZ/DBAZ and DZ white dwarfs, only the photometric solutions are given, since we use a hybrid technique that uses both photometry and spectroscopy for those stars (see section \ref{secdz}). 

Table  \ref{tabpar}  does not include
any errors for the spectroscopic parameters. As discussed in \citet{liebert05},  the internal errors of the spectroscopic fitting technique are arbitrarily small for high
S/N spectra. The actual error budget is dominated by the external uncertainties, mainly due to the flux calibration problems. \citet{liebert05} used multiple observations of 126 DA white dwarfs to estimate the external uncertainties in their survey, and found average errors of 1.2\% in $T_{\rm eff}$ and  0.038 dex in $\log{g}$. Performing a similar analysis of the 353 DA white dwarfs in our sample with multiple S/N $>20$ DESI spectra, we find average errors of 4.7\% in $T_{\rm eff}$, 0.067 dex in $\log{g}$, and $0.036~M_\odot$ in mass. Hence, these errors can be used as the uncertainties for the spectroscopic parameters given in Table  \ref{tabpar}. Given the lower S/N of the DESI spectra compared to the \citet{liebert05} sample, it is not surprising that these errors are larger.

\begin{figure}
\center
\includegraphics[width=2.5in, clip=true, trim=1.5in 0.8in 1.7in 1.1in]{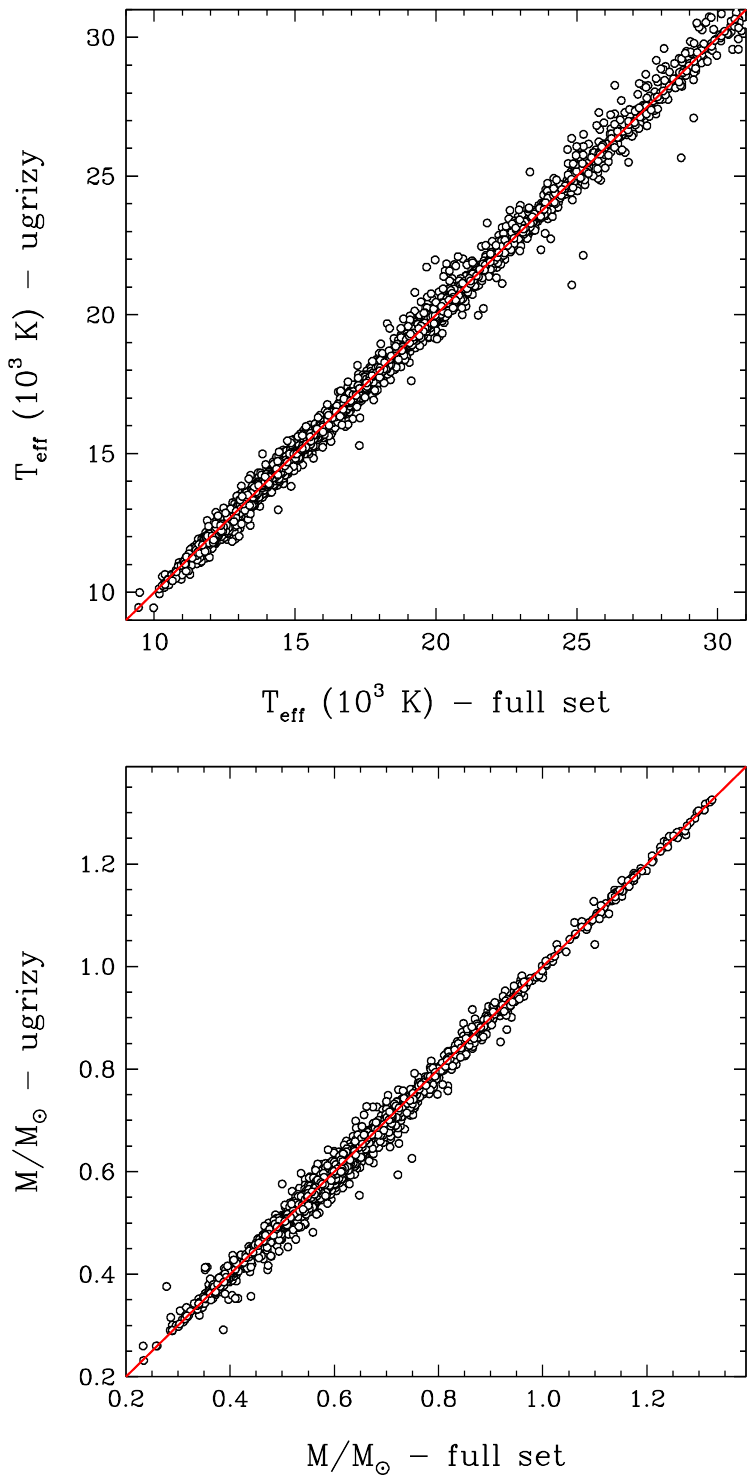}
\caption{Physical parameters of DA white dwarfs estimated based on the SDSS $u$ and Pan-STARRS $grizy$ photometry versus using the full set of SDSS $ugriz$,\\ Pan-STARRS $grizy$, and SkyMapper $uvgriz$ photometry. Here the sample is restricted to objects with distance accuracy better than 10\%.}
\label{figphotugrizy} 
\end{figure}

\begin{figure*}
\center
\includegraphics[width=3in, clip=true, trim=0.4in 0.8in 0.1in 1.1in]{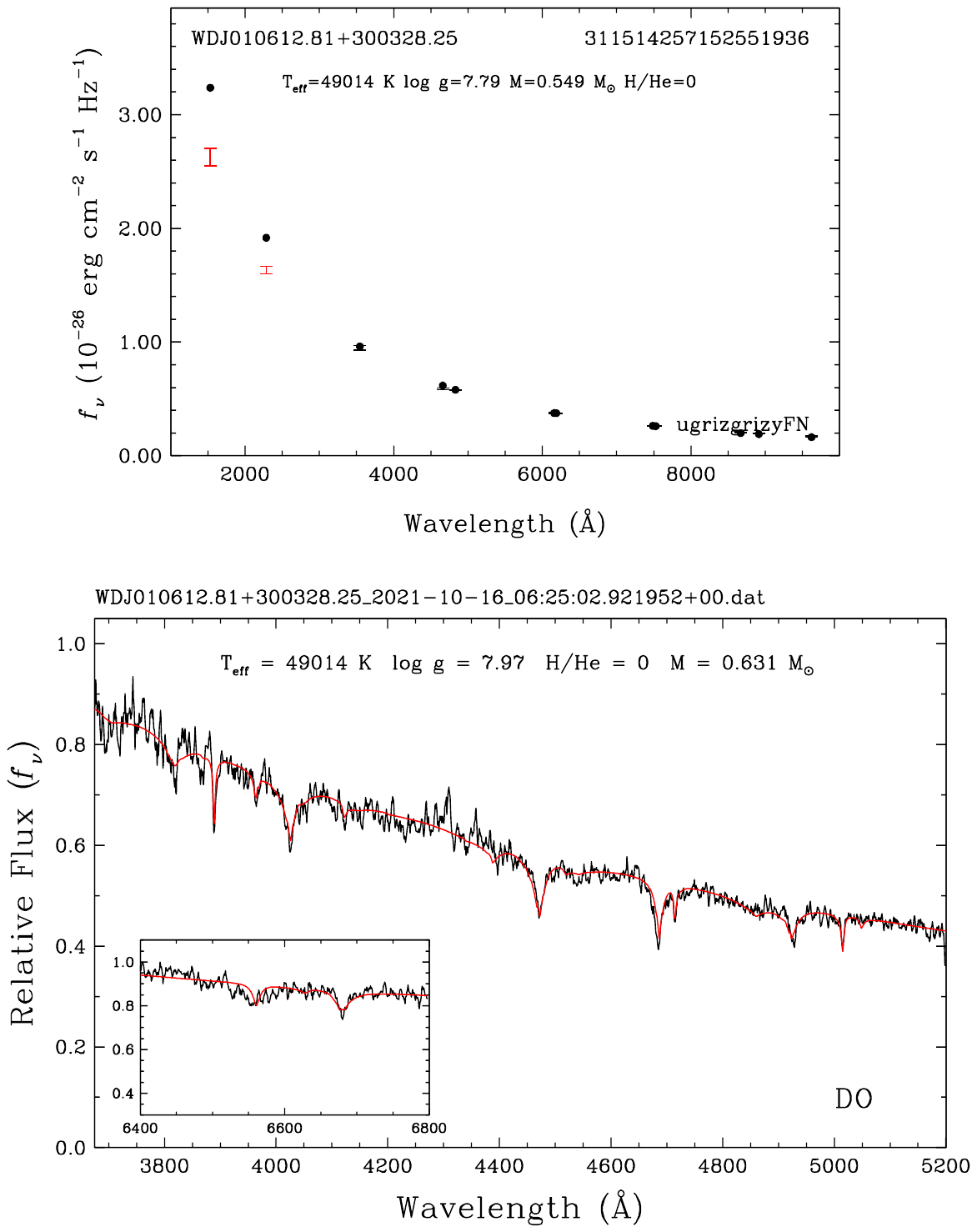}
\includegraphics[width=3in, clip=true, trim=0.4in 0.8in 0.1in 1.1in]{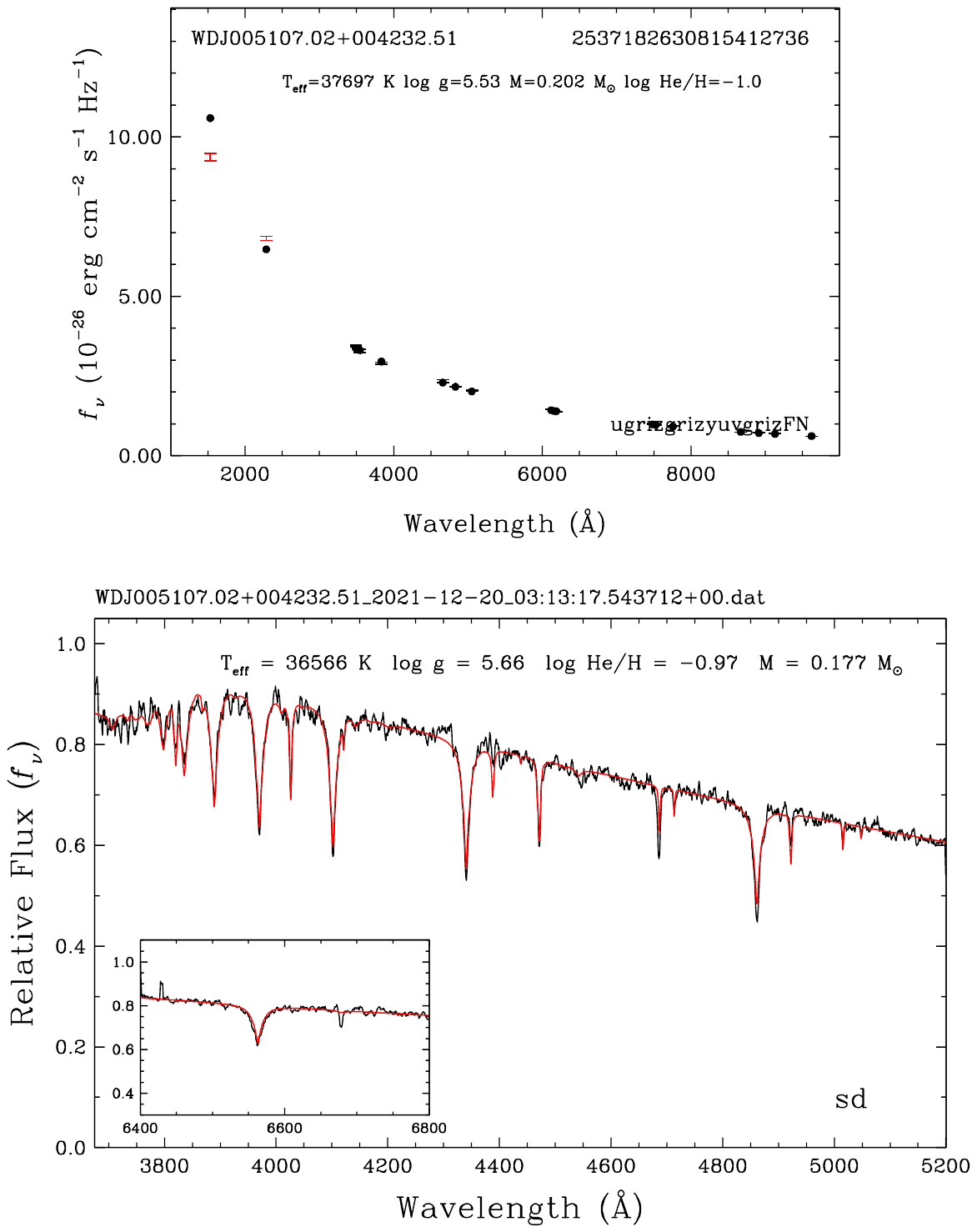}
\caption{Model fits to a DO white dwarf (left) and a subdwarf (right). The bottom panels show
the spectroscopic fit, where we constrain the H/He ratio, which is then used in the photometric fits
shown in the top panels.}
\label{fitdosd} 
\end{figure*}

The right panels in Figure \ref{fitda} shows the model fits to the warm DA WDJ000139.70-083631.83, where the photometric and spectroscopic solutions are totally independent. This star has SDSS $ugriz$, Pan-STARRS $grizy$, SkyMapper $uvgriz$, and GALEX FUV/NUV photometry available. Both the photometric and spectroscopic fits provide an excellent match to the data. Note that the spectroscopic temperature and mass are higher than the photometric temperature estimates for this star. The problem is not unique to this object, and we discuss it further below. As mentioned previously, some of our hot DA fits show Balmer lines that are deeper than predicted. This is the famous Balmer-line problem \citep{napiwotzki92,bergeron94a,gianninas10} due to the presence of metals, which are unaccounted for in our models.

In previous papers, we relied on SDSS $u$ and Pan-STARRS $grizy$ photometry for the photometric fits \citep[e.g.,][]{bergeron19,kilic20,kilic25a}. In order to see if our use
of the full photometric set introduces a bias, in Figure \ref{figphotugrizy} we compare the physical parameters of a subsample of the DA white dwarfs in
our sample based on the SDSS $u$ and Pan-STARRS $grizy$ versus using the full set of the SDSS, Pan-STARRS, and SkyMapper photometry. We see no significant differences between the temperature and mass estimates. Hence, we use all available SDSS, Pan-STARRS, and SkyMapper photometry in our model fits.

Some of the white dwarfs in DESI DR1 only have a red spectrum available, covering the H$\alpha$ region.
Without the higher order Balmer lines in the blue, a spectroscopic fit is not meaningful for those sources. In such cases, we display the spectroscopic fit using the photometric parameters.

\subsection{Hot non-DA White Dwarfs and Subdwarfs}

The Gaia color-magnitude diagram shown in Figure \ref{fighr} reveals a large number of subdwarf candidates (with $M_G\sim5$). Based on visual inspection of the DESI spectra and our model fits, we identify hot white dwarfs that are not pure DAs, including DAO, DAB, and DO white dwarfs, and 815 subdwarfs. We did not bother to go further in the spectral classification of subdwarfs into subtypes like sdO or sdB, since our main goal in this paper is to study the white dwarf population in DESI. 

We use the spectroscopic method to determine $T_{\rm eff}$, $\log{g}$, and the He/H ratio. 
We loop over the pure He and mixed He/H compositions and pick the best $\chi^2$ solution.
Then we adopt the spectroscopic He/H value (rounded off to 1 dex) in our photometric fits, which
provide an independent photometric mass estimate. Above 40,000 K, we force $T_{\rm phot} = T_{\rm spec}$ for the photometric fits. 

Figure \ref{fitdosd} shows the fits to a DO white dwarf (left panels) and a subdwarf (right panels) for comparison. The spectrum of WDJ010612.81+300328.25 is reproduced exceptionally well by a pure He atmosphere white dwarf model with $T_{\rm eff}=49,014$ K and $\log{g}=7.97$ (bottom left panel). Similarly, our models provide an excellent match to the DESI spectrum of WDJ005107.02+004232.51, which is a subdwarf with $T_{\rm eff}=36,566$ K, $\log{g}=5.66$, and $\log {\rm He/H} = -0.97$. 

\begin{figure}
\hspace{-0.2in}
\includegraphics[width=3.5in, clip=true, trim=1in 0.8in 1in 0.4in]{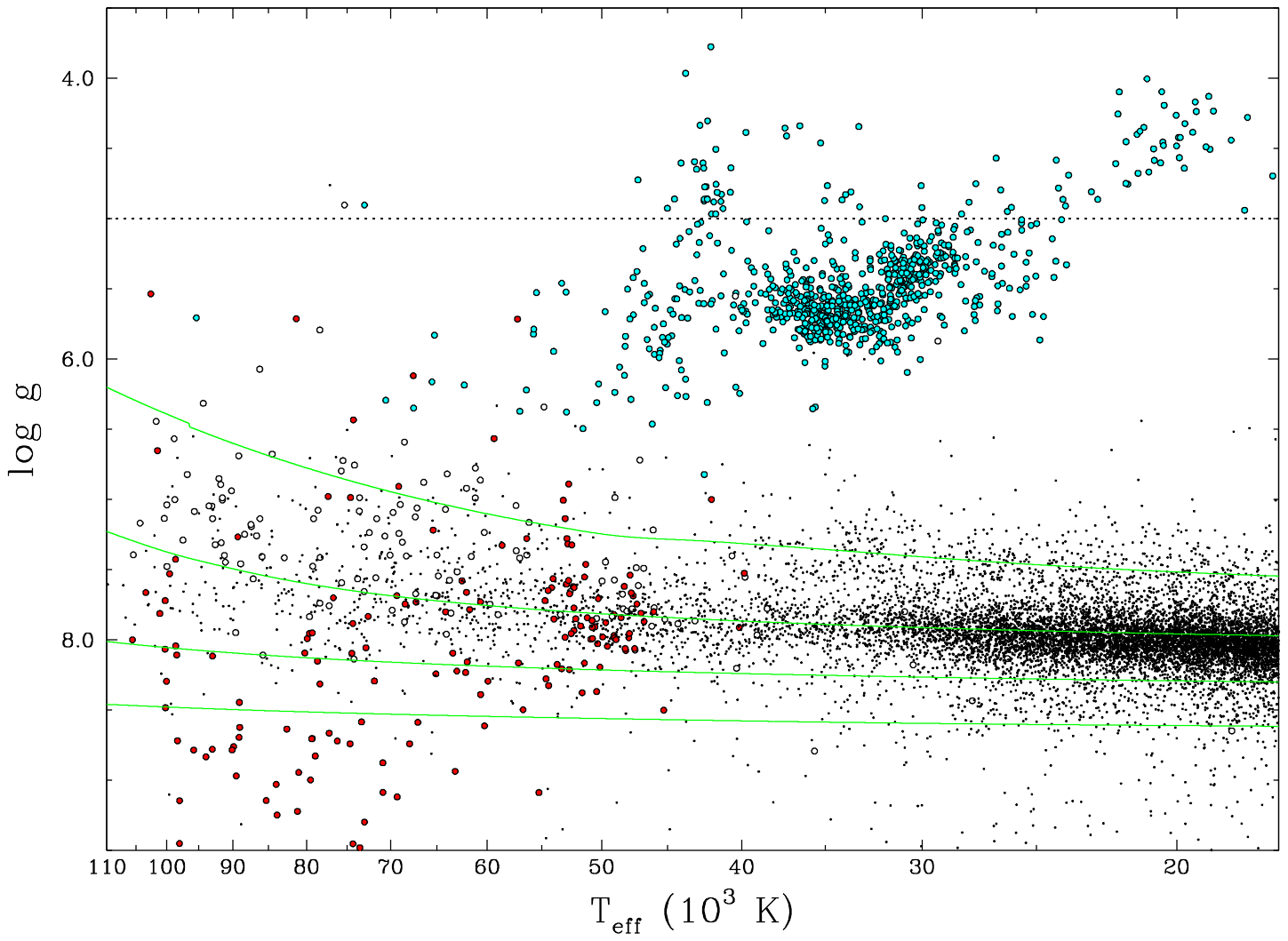}
\caption{Spectroscopic parameters of the DO (red circles), DAO (white circles, including a few cool DAB stars), DA (small dots), and subdwarfs (cyan circles) in our sample. Green lines show the evolutionary tracks for pure H atmosphere white dwarfs with $M=0.4$, 0.6, 0.8, and $1.2~M_{\odot}$ (from top to bottom), respectively. The parameters for the objects above the dotted line ($\log{g}<5$) are extrapolated.}
\label{figtg} 
\end{figure}

Figure \ref{figtg} shows the spectroscopic parameters of the hot DA, DAO, DAB, DO, and subdwarfs in our sample. We had difficulty in classifying some objects as white dwarfs vs subdwarfs when their parameters start to overlap in this diagram, and higher quality spectra would be helpful to definitely confirm the spectral types and parameters of those objects. In addition, the parameters for the objects above the dotted line
($\log{g}<5$) are extrapolated, and therefore should be used with caution. Below 20,000 K, many of the subdwarfs (including sdB and sdA) and pre-ELM white dwarfs show only H lines. We used a pure H atmosphere DA model grid to fit their spectra (see Section \ref{secelm} for further discussion of the ELM sample). 

Despite these caveats and the use of the LTE models, our results are very similar to those of \citet[][see their Figure 13]{bedard20}.
Cooler DA white dwarfs roughly fall between the 0.6-$0.8~M_\odot$ model tracks shown here, but at high temperatures, DAO white dwarfs have lower-than-average and DO white dwarfs have higher-than-average surface gravities. \citet{bedard20} found average masses of 0.508 and $0.540~M_\odot$ for their hydrogen-rich sample above and below 60,000 K, respectively. On the other hand, they found average masses of 0.695 and $0.575~M_\odot$ for helium-rich white dwarfs above and below 50,000 K, respectively. Similar problems in the mass distributions of hot white dwarfs have been reported previously \citep{kepler19,genest19}. 

\begin{figure*}
\center
\includegraphics[width=3in, clip=true, trim=0.4in 0.8in 0.1in 1.1in]{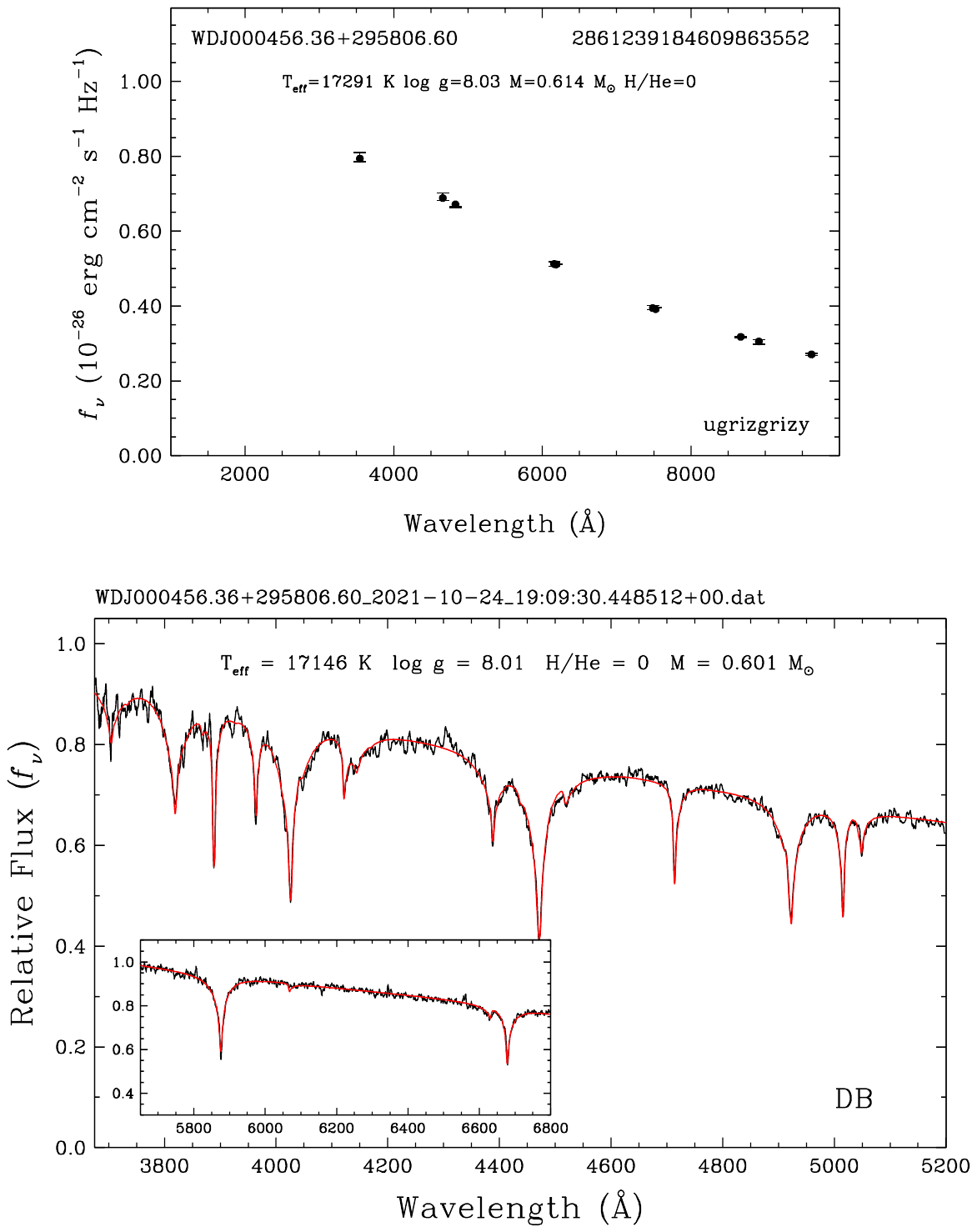}
\includegraphics[width=3in, clip=true, trim=0.4in 0.8in 0.1in 1.1in]{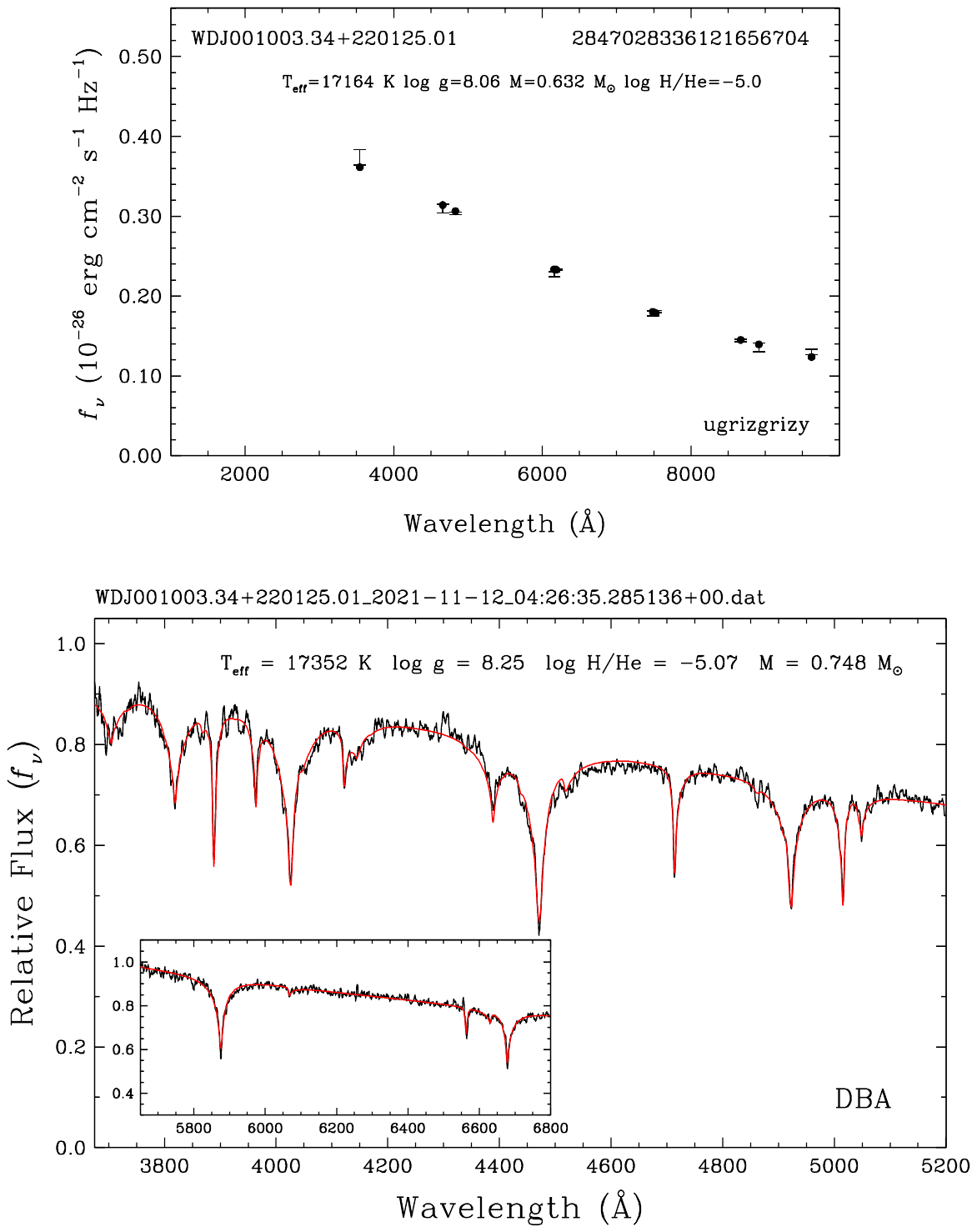}
\caption{Example model fits to DB (left) and DBA (right) white dwarfs.}
\label{fitdb} 
\end{figure*}

\citet{bedard20} show that the photometric method, where the spectroscopic temperature and the H/He abundance are forced and only the radius is considered a free parameter, mostly resolves the discrepancies between the hotter and cooler samples. The reason for forcing the spectroscopic values of temperature and the composition is because the optical photometry is weakly sensitive to temperature for these hot white dwarfs. \citet{bedard20} discuss potential reasons for the observed discrepancies in
the spectroscopic fits. The most likely culprits are the spectral line profiles. For example, the high-Balmer problem likely impacts a large number of systems \citep{gianninas10}, but is not easily observed in noisier data in the SDSS and DESI. Hence, our use of metal-free atmospheres likely leads to the observed mass offsets in DA/DAO white dwarfs. For the DO white dwarfs, problems with the Stark broadening profiles for \ion{He}{2} or the potential impacts from ultra-hot circumstellar winds \citep{reindl19} could lead to over-estimated masses \citep{bedard20}. Our sample includes 46 DOZ white dwarfs; for which the use of metal-rich atmosphere models would be helpful for better constraints on their parameters. 

\subsection{DB and DBA White Dwarfs}

Our sample includes 2244 DB white dwarfs, including 346 DBA, 83 DBAZ, 40 DBZ, and 10 magnetic DBs. We use the H lines (if visible) to constrain the H/He ratio, and force this composition (to nearest 0.5 dex value) in the photometric fits. Otherwise, the photometric temperatures and masses are determined independently. Except, for a number of cool DBs with $T_{\rm eff}\sim11,000$ K and relatively weak He lines, spectroscopic fits are not meaningful; we forced the pure He atmosphere photometric solution for those targets.

Figure \ref{fitdb} shows our model fits to two DB white dwarfs. On the left panels, we have a DB white dwarf where no H$\alpha$ is visible. The best-fitting pure He atmosphere model with $T_{\rm eff}=17,146$ K and $M=0.601~M_\odot$ provides an excellent match to the DESI spectrum of this target. The photometric fit under the assumption of a pure helium atmosphere indicates very similar parameters with $T_{\rm eff}=17,291$ K and $M=0.614~M_\odot$. 

The right panels show the fits for a DBA white dwarf, where the observed H$\alpha$ line constrain the composition to $\log {\rm H/He} = -5$. Forcing this composition in the photometric fit results in a very similar temperature estimate, but a significantly different surface gravity and mass. There are known issues with the spectroscopic and photometric masses for DB white dwarfs \citep{genest19}, and we leave this topic for further discussion below.

\begin{figure}
\center
\includegraphics[width=3in, clip=true, trim=0.4in 0.8in 0.1in 1in]{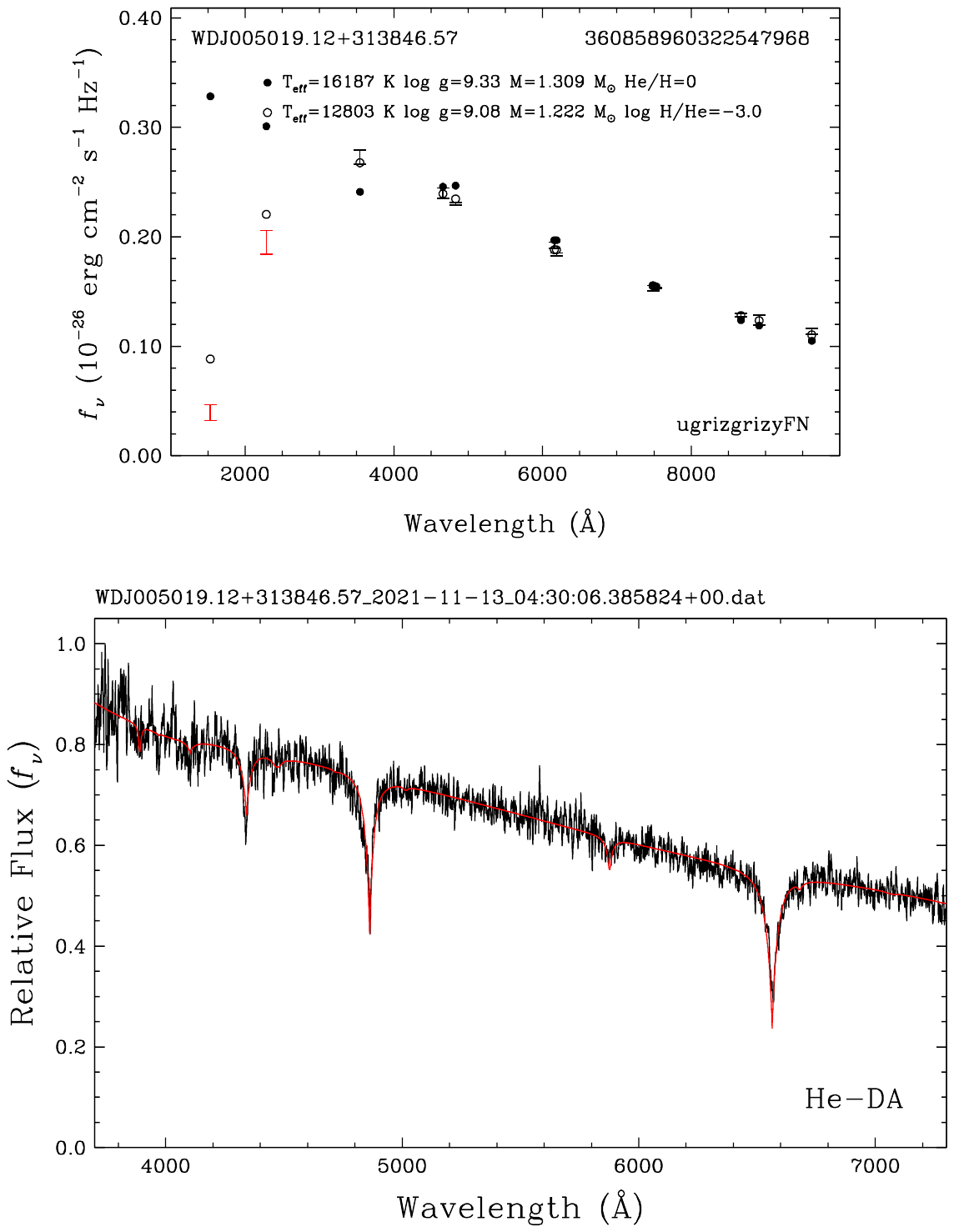}
\caption{Example model fits to a He-DA white dwarf.}
\label{fitheda} 
\end{figure}

We identify 9 He-DA white dwarfs where the spectra are dominated by Balmer lines, but the pure H atmosphere solution predicts H lines that are much stronger than observed, and the spectral energy distribution is better reproduced by He-dominated atmosphere models. Figure \ref{fitheda} shows our model fits to one of these targets, WDJ005019.12+313846.57. The top panel shows both the pure H and mixed H/He solutions, and the bottom panel shows the predicted spectroscopic fit with the H/He abundance that best represents the spectrum. That value is not fitted but visually determined. 
Here the photometric temperature assuming a pure H atmosphere (16,187 K) is too high, and clearly ruled out by GALEX UV photometry and the DESI spectrum. On the other
hand, the mixed H/He solution with $T_{\rm eff}=12,803$ K, $M=1.222~M_\odot$, and $\log$ H/He = $-3.0$
provides an excellent match to both the photometry and the observed Balmer line profiles in this unusual white dwarf. 

\subsection{DQ White Dwarfs}
\label{warmdq}

\begin{figure*}
\center
\includegraphics[width=3in, clip=true, trim=0.4in 0.8in 0.1in 1.1in]{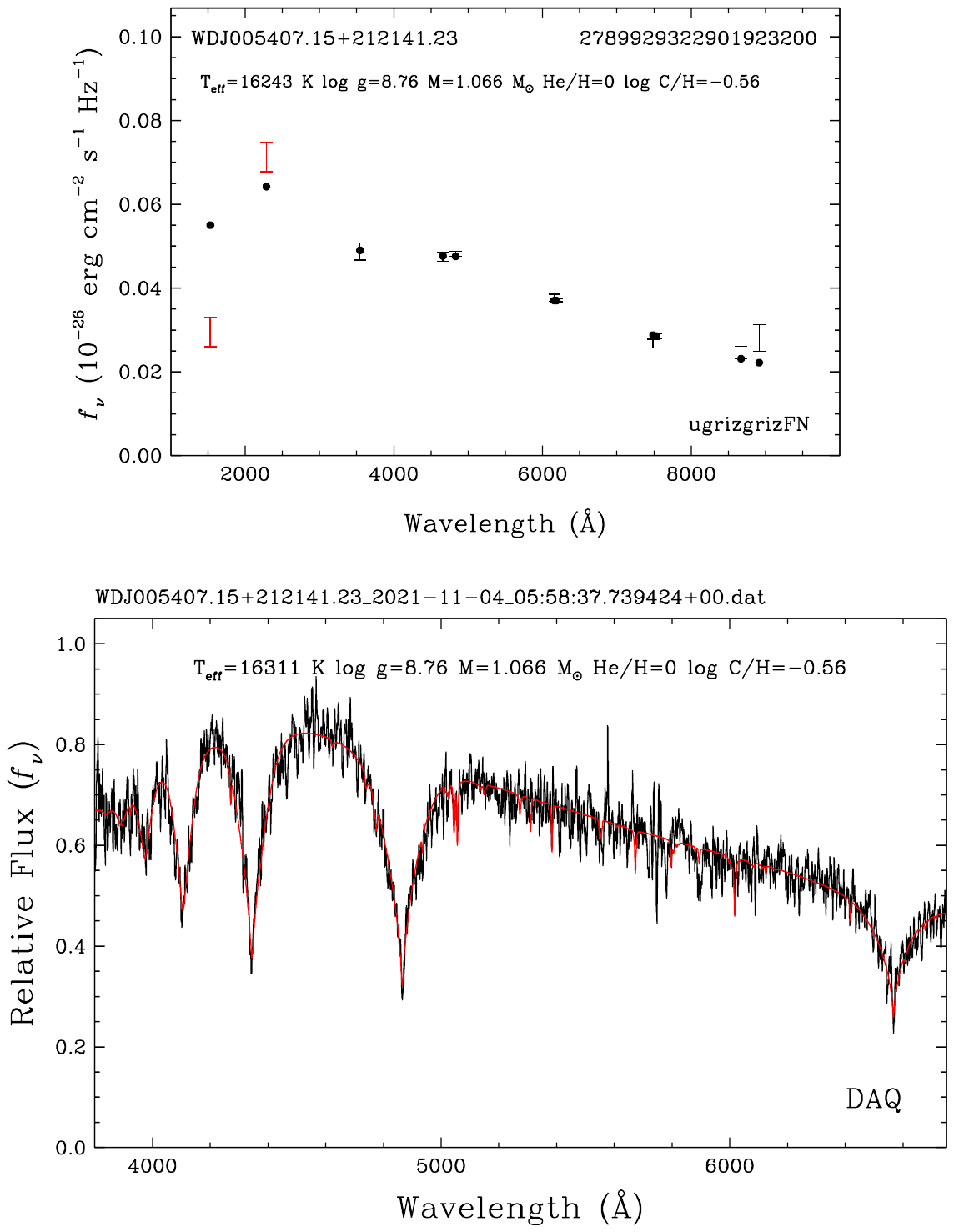}
\includegraphics[width=3in, clip=true, trim=0.4in 0.8in 0.1in 1.1in]{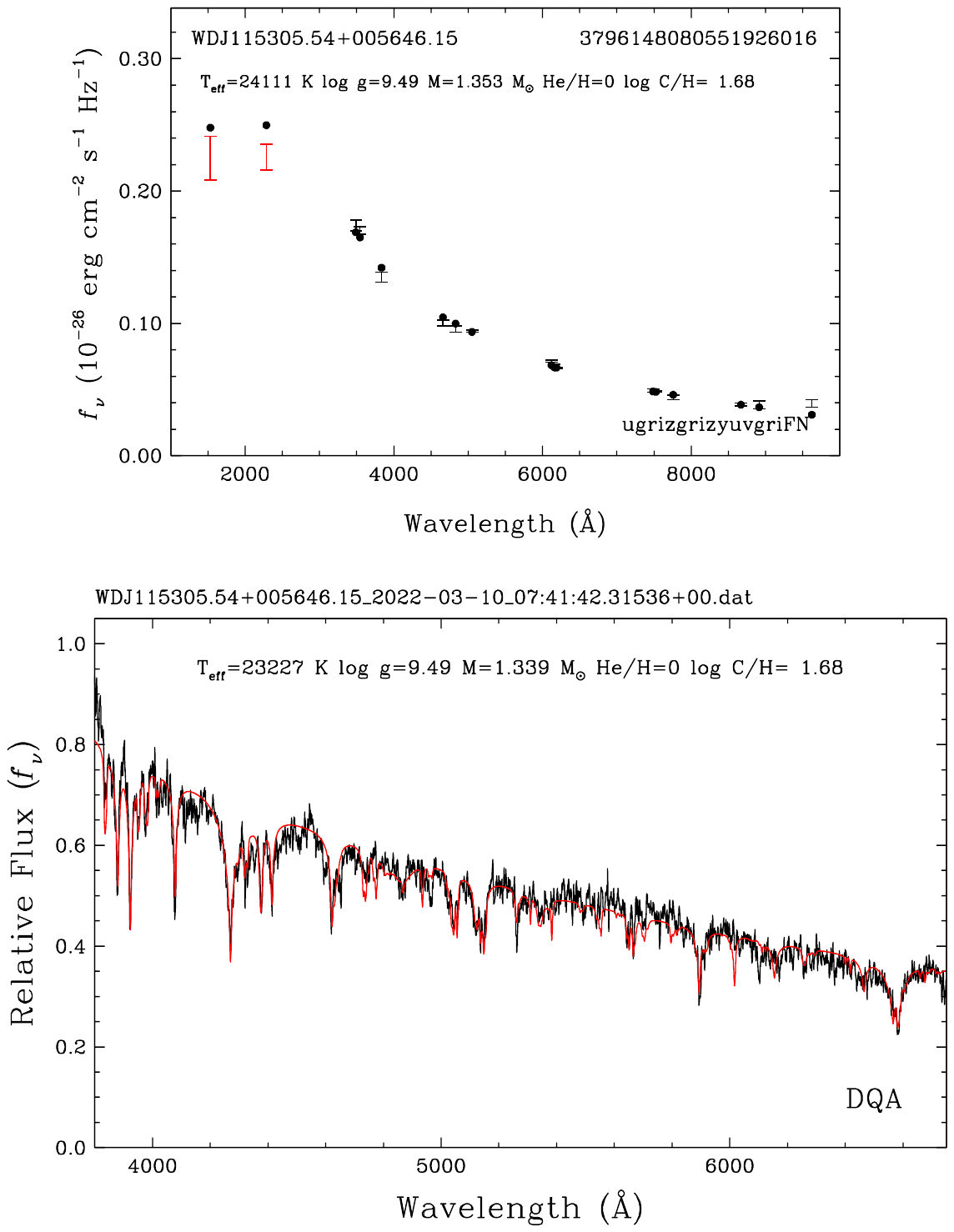}
\caption{Model atmosphere fits to a DAQ (left) and a DQA (right) white dwarf. The top and bottom panels show the photometric and spectroscopic fits, respectively.}
\label{fitdq} 
\end{figure*}

We identify 70 DQs in the DESI hot white dwarf sample, including nine DAQ white dwarfs with spectra dominated by Balmer lines and weaker C features \citep{liebert83,hollands20,kilic24,jewett24}. Six of these DAQs are new discoveries. In addition, two of the white dwarfs in our sample show molecular carbon features that are best-explained by models with $T_{\rm eff}=9824 \pm 109$ K, $M=0.588 \pm 0.048$, and $\log$ (C/He) =  $-3.87$ (for WDJ085239.66+042804.48) and  $10443 \pm 279$ K, $0.774 \pm 0.046~M_\odot$, and  $\log$ (C/He) =  $-3.37$ (for WDJ175631.55+372827.41).
These two stars belong to the canonical, cool DQ population. 

\begin{figure}
\center
\includegraphics[width=3in]{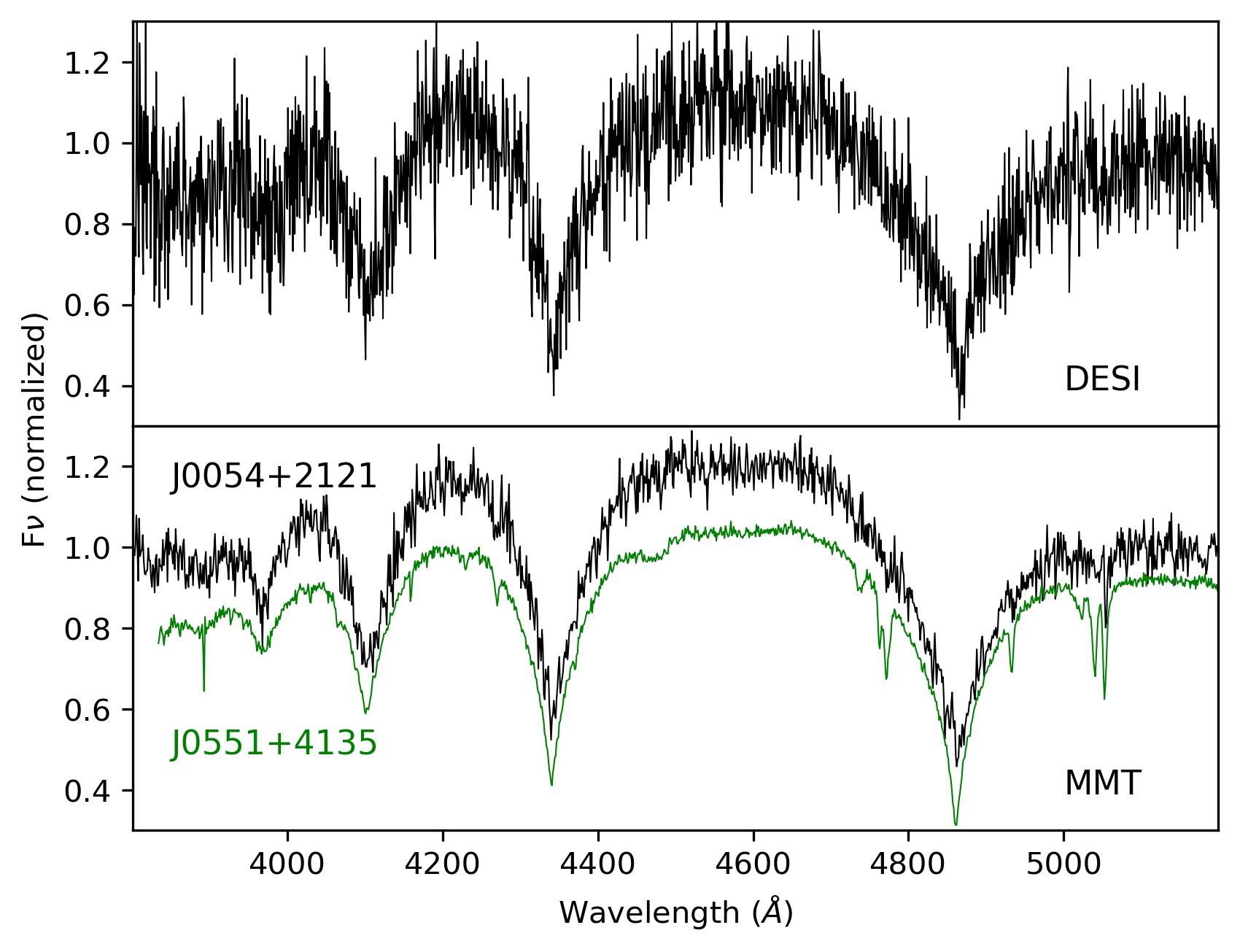}
\caption{DESI (top) and MMT (bottom) spectra of the newly identified DAQ white dwarf WDJ005407.15+212141.23, compared to the MMT spectrum of the previously known DAQ J0551+4135 shown in green.}
\label{figmmt} 
\end{figure}

Figure \ref{fitdq} shows example fits for DAQ (left) and DQA (right) white dwarfs. We first fit the DESI spectrum with $T_{\rm spec}$ and $\log$ C/H as free parameters and $\log{g}$ determined from photometry. 
We force the photometric $\log{g}$ because we do not believe that the DQ models are accurate enough to use the spectroscopic $\log{g}$ as a free parameter.
We then fit the photometry with $T_{\rm phot}$ and $\log{g}$ (i.e., the radius) as free parameters and $\log$ C/H determined from spectroscopy. We repeat the spectroscopic and photometric fits until a consistent solution is found. In the end, this iterative process provides internally consistent values of $\log{g}$ and $\log$ C/H for both solutions, and excellent fits for both spectroscopy and photometry, but different temperatures $T_{\rm spec}$ and $T_{\rm phot}$. The difference between these two is a measure of the model accuracy. 
Note that this procedure is different from what we used in all previous analyses where we forced $T_{\rm spec}=T_{\rm phot}$. The problem with this procedure is that for hotter stars we get a runaway process, where $T_{\rm phot}$ keeps increasing as the C/H ratio gets updated from the spectroscopic fit. This is why we decided to treat the photometric and spectroscopic temperatures separately.  

\begin{figure*}
\center
\includegraphics[width=3in, clip=true, trim=0.4in 0.8in 0.1in 1.1in]{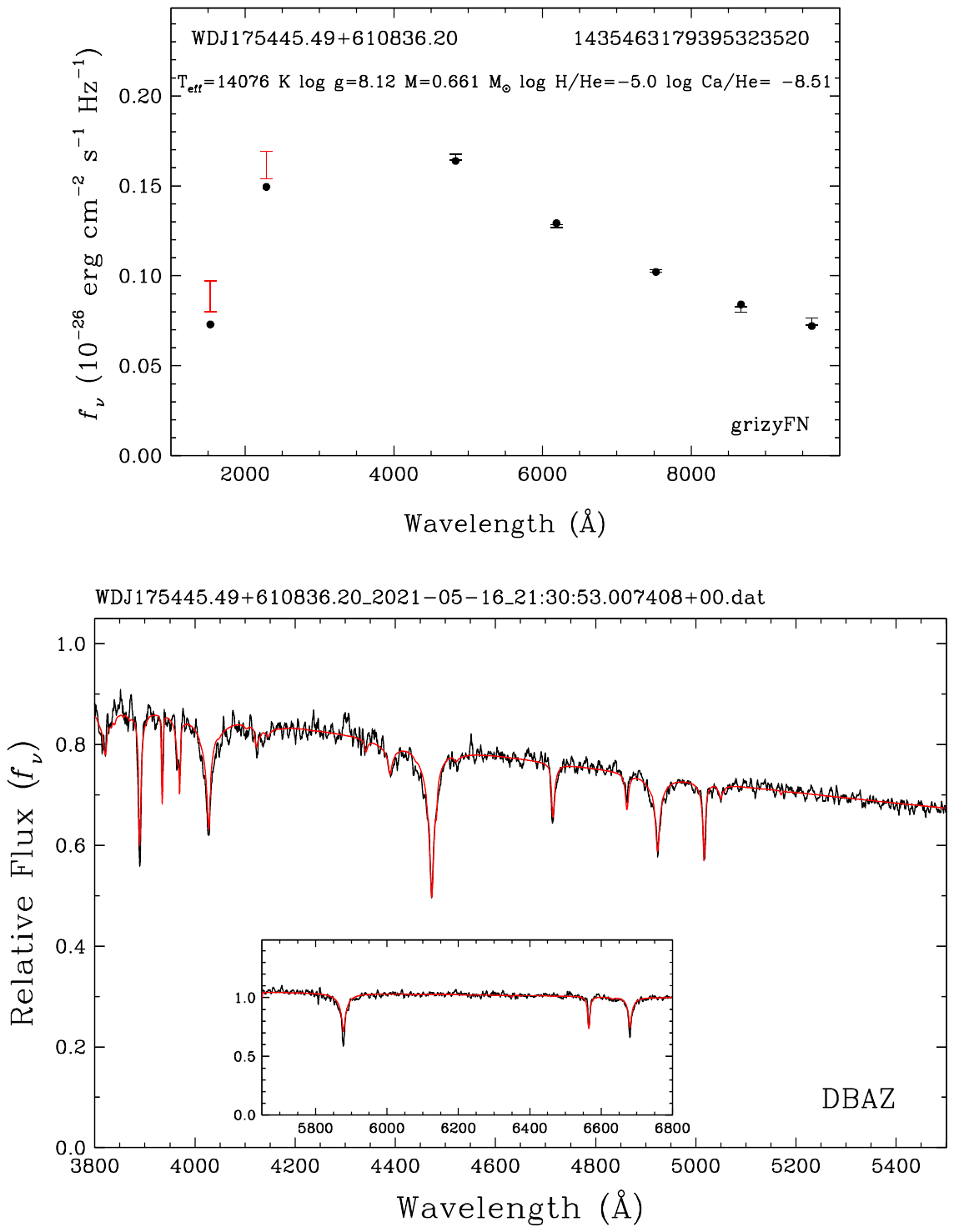}
\includegraphics[width=3in, clip=true, trim=0.4in 0.8in 0.1in 1.1in]{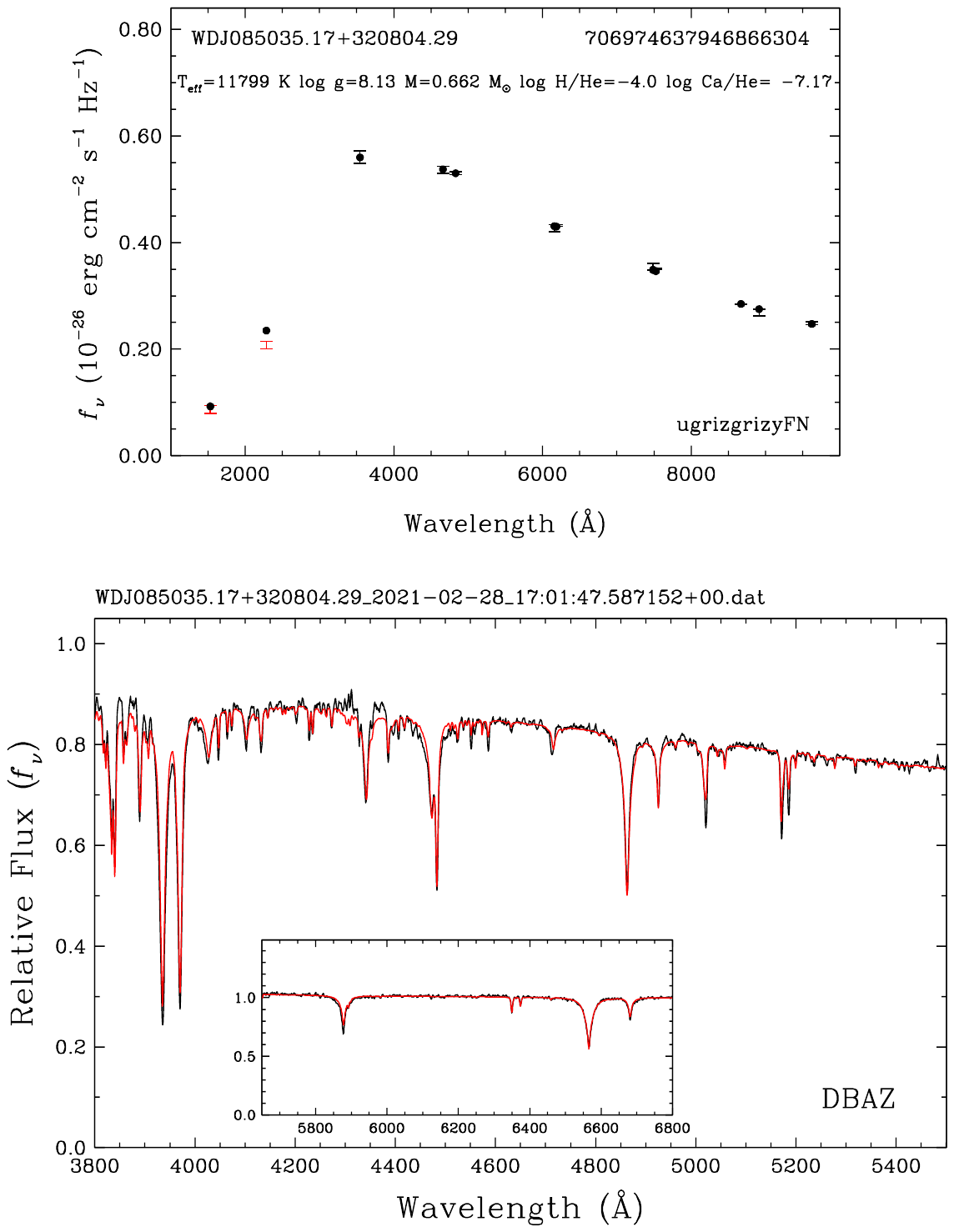}
\caption{Example model fits to two metal-rich white dwarfs. The spectroscopic parameters are not included in the bottom panels because all parameters, except the metal abundance, are determined photometrically.}
\label{fitdbaz} 
\end{figure*}

For the DAQ white dwarf shown on the left, spectroscopic and photometric temperatures are in excellent agreement, whereas for the DQA shown on the right, the temperatures differ slightly ($\approx4$\%). We find a relatively good agreement between $T_{\rm spec}$ and $T_{\rm phot}$ for the DAQ and DQA stars in our sample, where H is clearly visible and helps constrain the C/H ratio in the atmosphere (see Section \ref{secdq}). 

WDJ005407.15+212141.23, the DAQ shown in the left panels of Figure \ref{fitdq}, turns out to be one of the most H-rich DAQs known with
$\log$ C/H = $-0.56$. Figure \ref{figmmt} shows the original (unsmoothed) DESI spectrum of this target in the top panel.
In fact, the atomic C features are so weak that they are not visible in the raw DESI spectrum. We initially classified this object as a potential DAQ based on the smoothed version of the DESI spectrum shown in Figure \ref{fitdq}. To confirm its spectral type, we obtained a higher quality spectrum at the 6.5m MMT equipped with the Blue Channel Spectrograph on UT 2025 Oct 24 with a spectral resolution of 4.7 A. The bottom panel in Figure \ref{figmmt} shows this spectrum (black) compared to the spectrum of the DAQ white dwarf J0551+4135 \citep{kilic24}. The C feature near 5050 \AA\ is clearly detected in the MMT spectrum of WDJ005407.15+212141.23, confirming it as a DAQ white dwarf. In addition, there are several other weaker C lines near 4933 and 4772 \AA\ that are also present. WDJ005407.15+212141.23's C/H ratio is almost identical to J0551+4135, but it is $>3000$ K hotter, which significantly impacts the strengths of its C features. Hence, it is possible that there could be other DAQ white dwarfs that we missed in the DESI sample given the relatively noisy spectra.

\begin{figure*}
\includegraphics[width=2.4in, clip=true, trim=0.4in 0.8in 0.1in 1.1in]{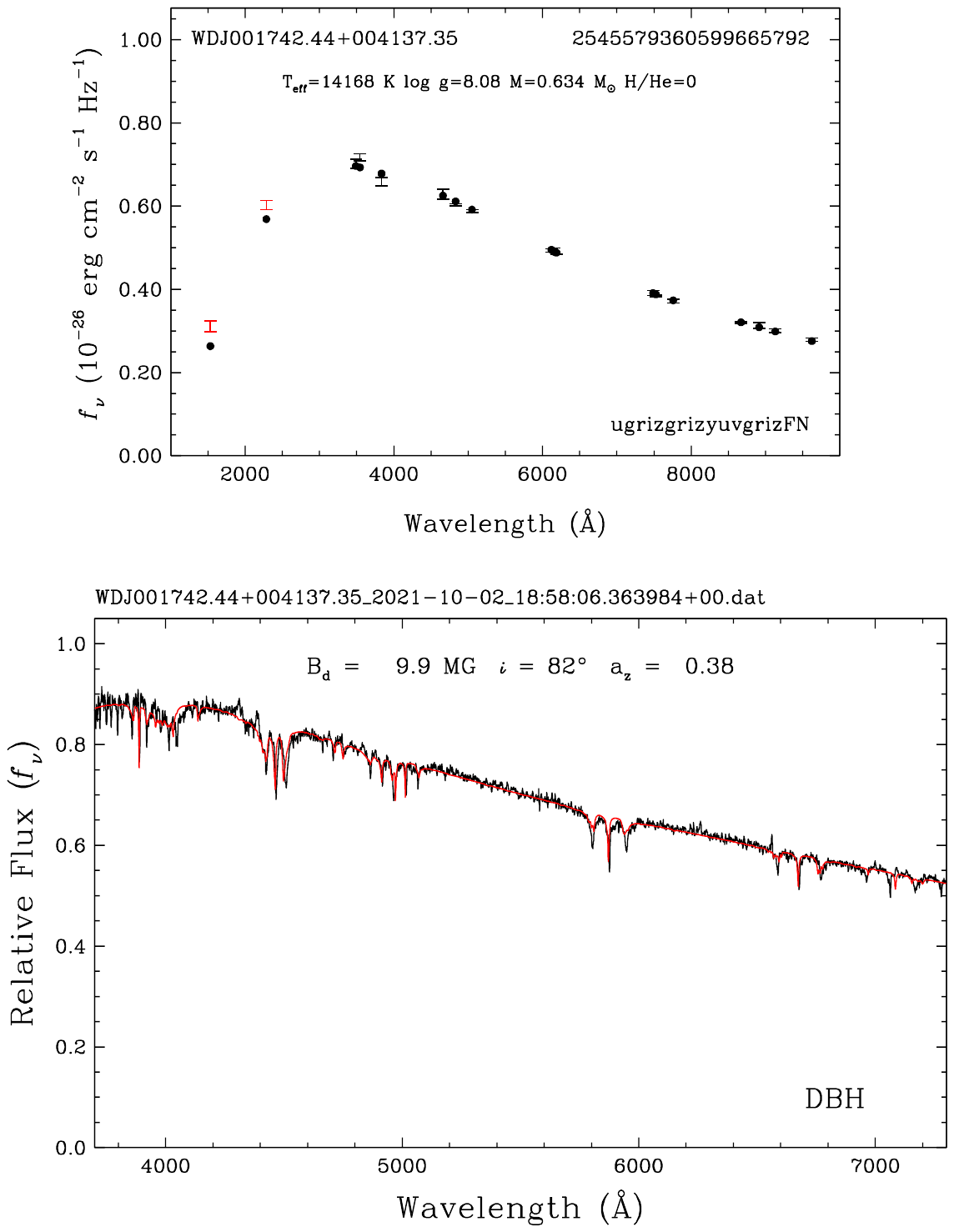}
\includegraphics[width=2.4in, clip=true, trim=0.4in 0.8in 0.1in 1.1in]{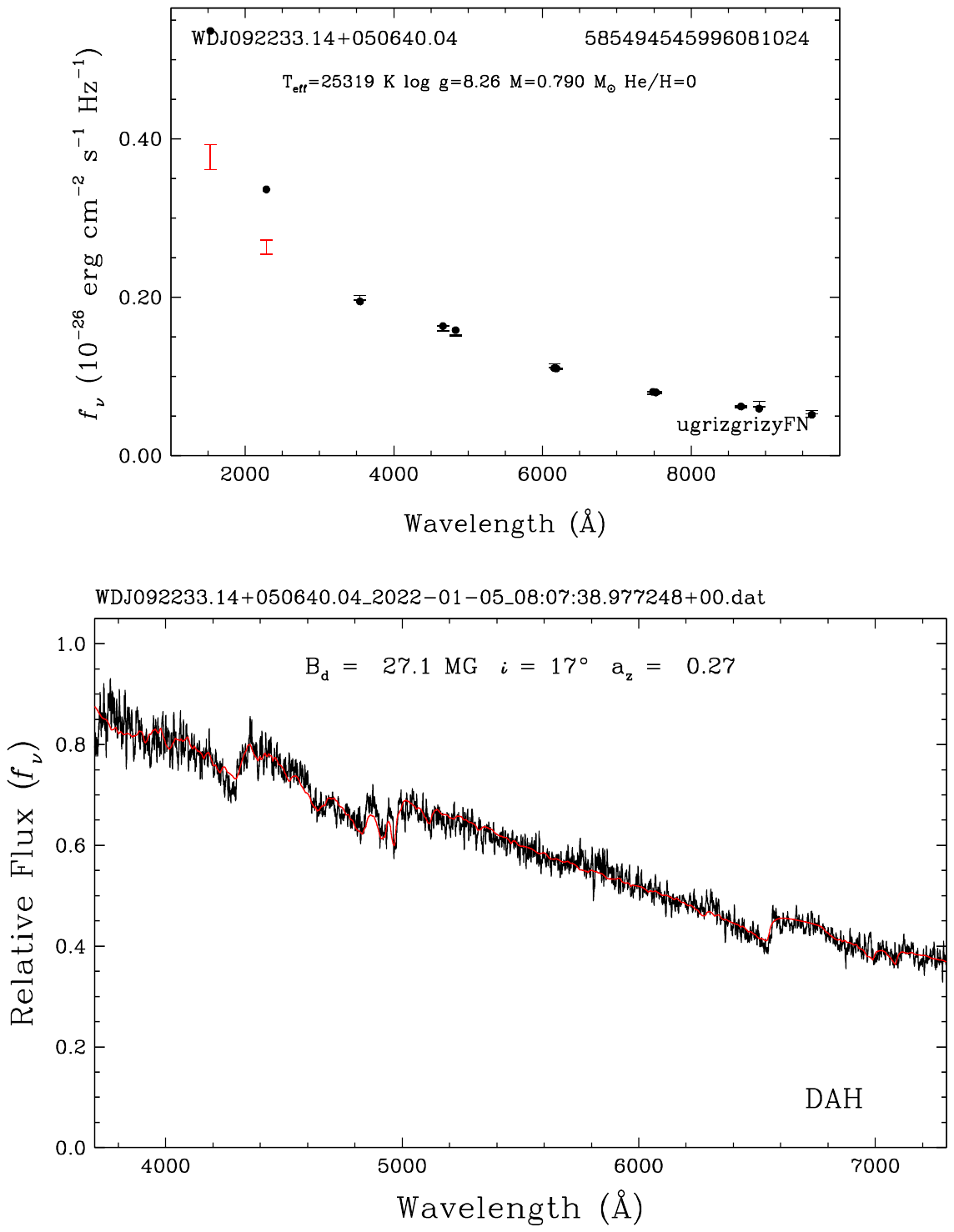}
\includegraphics[width=2.4in, clip=true, trim=0.4in 0.8in 0.1in 1.1in]{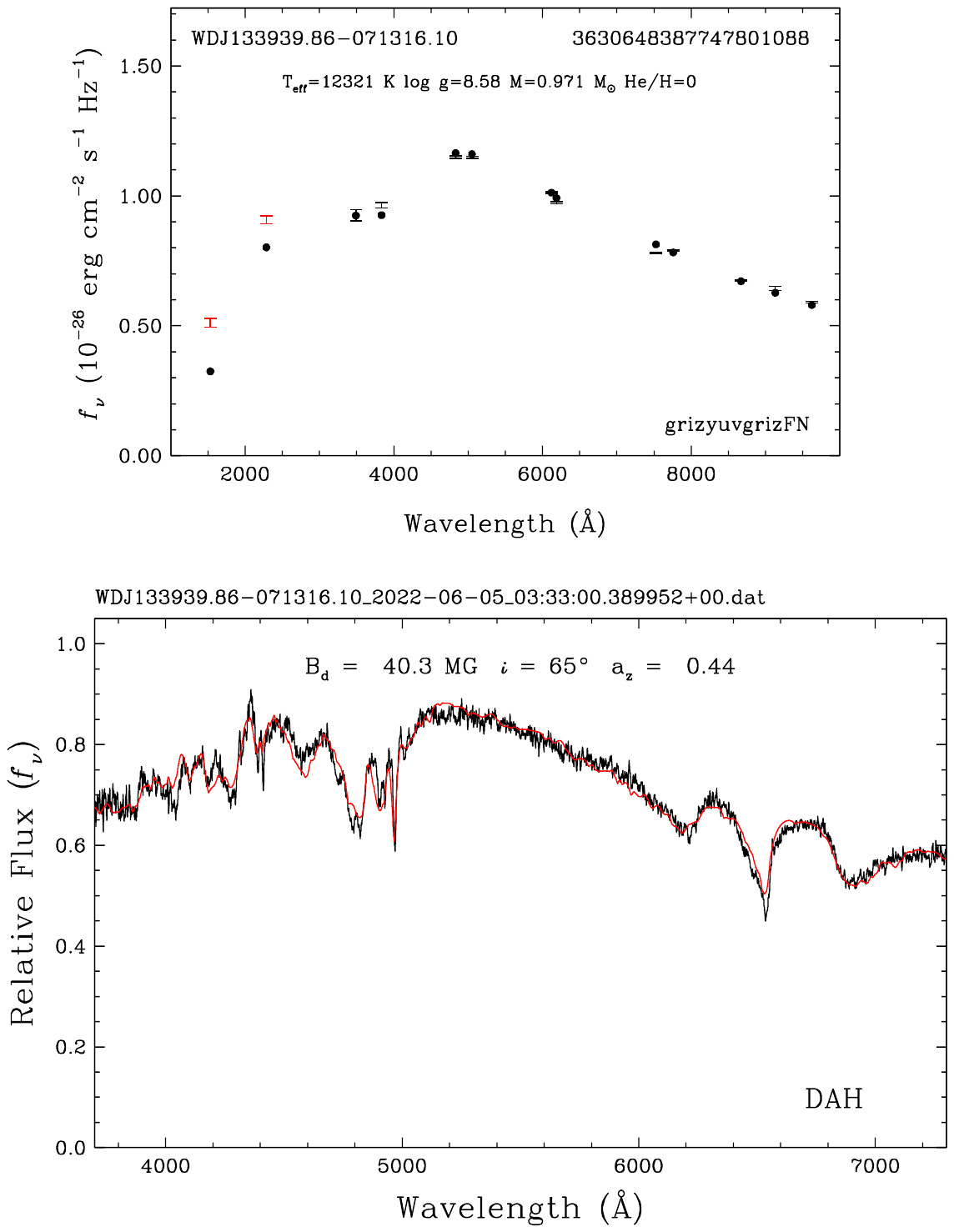}
\includegraphics[width=2.4in, clip=true, trim=0.4in 0.8in 0.1in 1.1in]{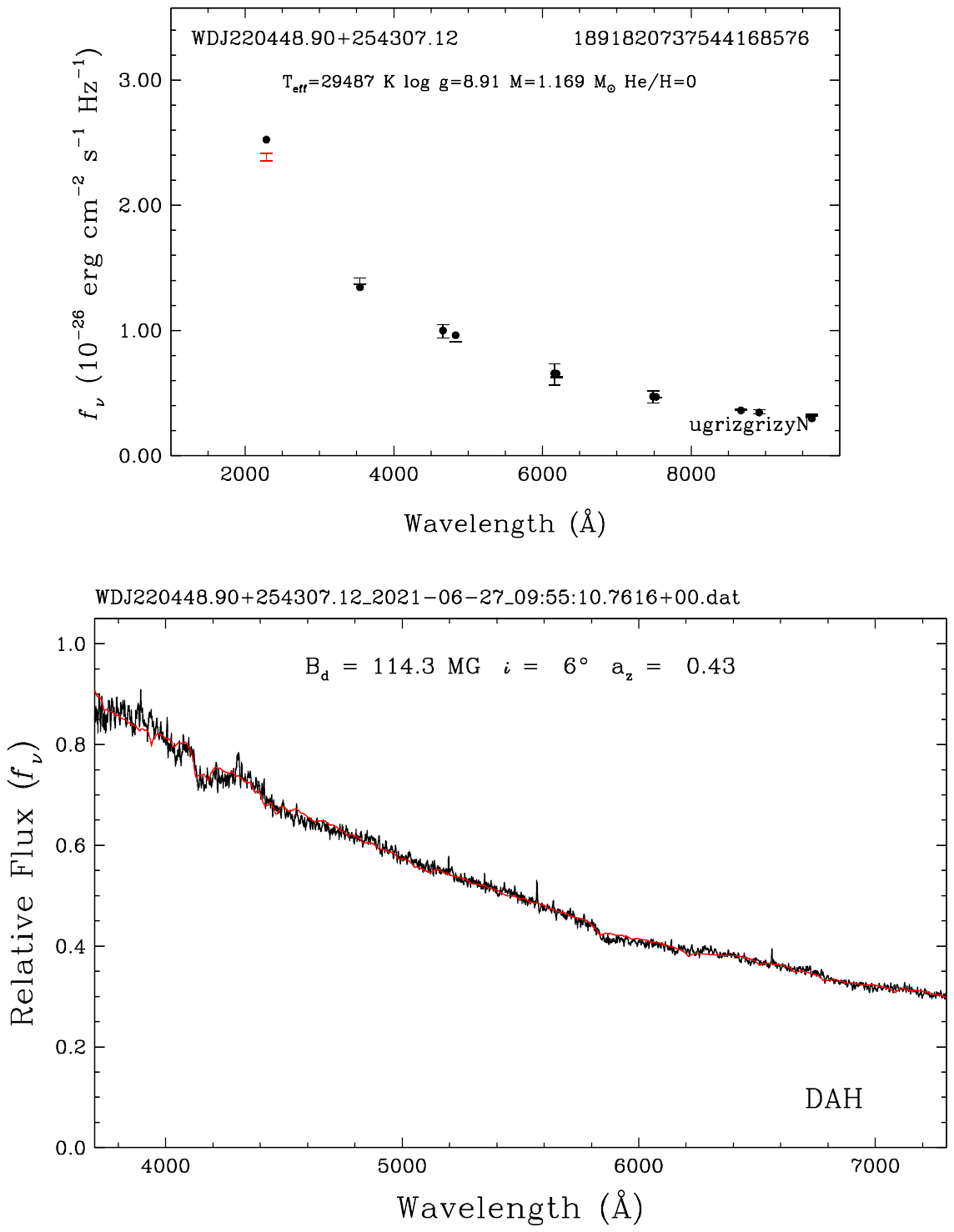}
\includegraphics[width=2.4in, clip=true, trim=0.4in 0.8in 0.1in 1.1in]{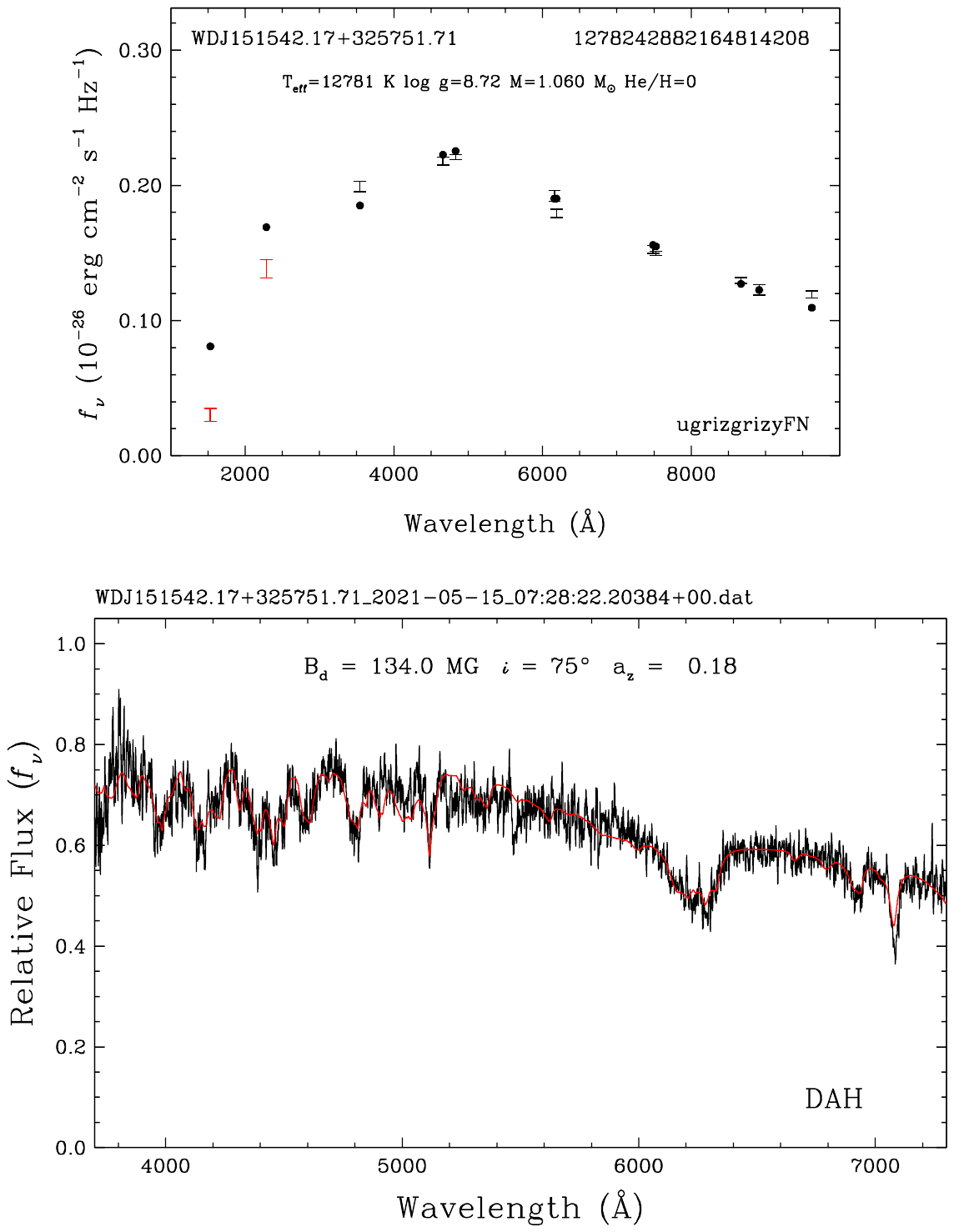}
\includegraphics[width=2.4in, clip=true, trim=0.4in 0.8in 0.1in 1.1in]{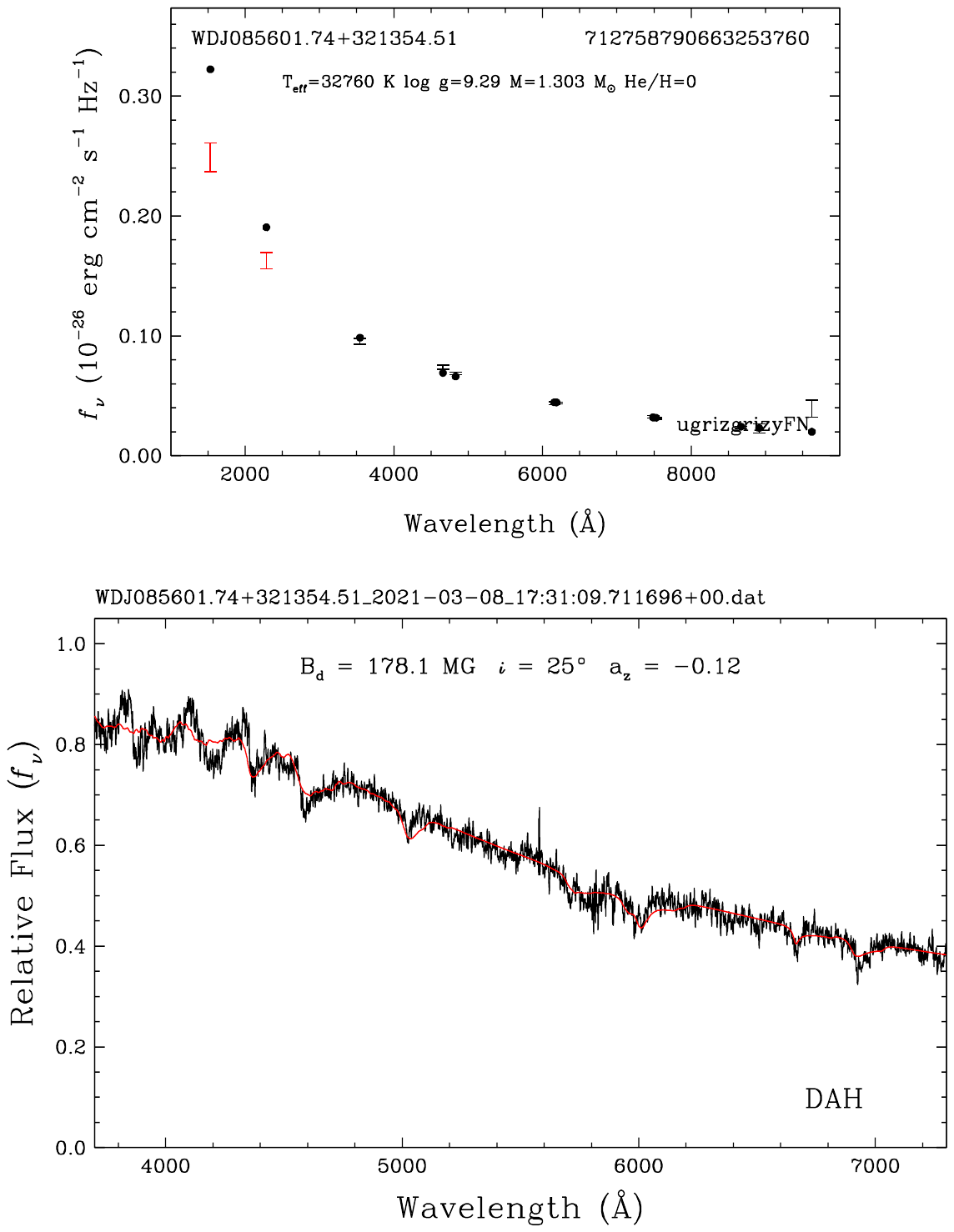}
\caption{Model fits to magnetic DB (top left) and DA white dwarfs with field strengths ranging from 10 to 178 MG.}
\label{fitdah} 
\end{figure*}

\subsection{Metal-Rich White Dwarfs}
\label{secdz}

For metal-rich white dwarfs in our sample, we use a hybrid method to determine their parameters. We rely on the photometric technique to determine the temperature and surface gravity, and fit the DESI spectrum to constrain the Ca/He ratio, forcing $T_{\rm phot}$ and log $g_{\rm phot}$ in the spectroscopic fit. The abundance ratios of the other heavy elements are assumed to match the CI chondrites. The resulting abundance is then used to improve the photometric fit at that abundance in an iterative fashion. We repeat this entire procedure for all H/He abundances in our models grids and adopt the best overall fit, which is mostly driven by H$\alpha$.

Figure \ref{fitdbaz} shows our model fits to two metal-rich white dwarfs. WDJ175445.49+610836.20 (left panels) is a relatively hot DBAZ white dwarf with $T_{\rm eff}=14,076$ K, $M=0.661~M_\odot$, and $\log$ H/He = $-5$. It shows relatively weak Ca H and K lines, which are best fit by a model with $\log$ Ca/He = $-8.51$. The right panels show the model fits to a cooler DBAZ with stronger Ca lines. A model with $T_{\rm eff}=11,799$ K, $M=0.662~M_\odot$, $\log$ H/He = $-4$, and $\log$ Ca/He = $-7.17$ provides an excellent match to both photometry and DESI spectrum of WDJ085035.17+320804.29. The chondritic 
elemental ratios seem to provide a remarkable match to the observed metal lines in both stars shown here.

Note that we classify the majority of the metal- and hydrogen-rich DBs as DBAZ white dwarfs. Depending on the strength of the hydrogen, helium, and the metal lines, various spectral types are possible: DABZ, DAZB, DZAB, DBAZ, DBZA etc. However, these classifications require equivalent width measurements for various lines and they can get complicated in noisy spectra. The presence of He lines in these stars demonstrate that they are clearly He-dominated. If there were no metal lines present, we would classify these objects as DBA white dwarfs, even if the H lines are stronger, e.g. WDJ085035.17+320804.29 shown in the right panels in Figure \ref{fitdbaz}. H is the most abundant element after He in these objects. Hence, we simply classify them as DBAZ white dwarfs to stress the fact that they have He-dominated atmospheres with trace amounts of H and lower traces of metals. 
 
\subsection{Magnetic White Dwarfs}

We identify 298 magnetic white dwarfs in our sample, including 10 confirmed or likely magnetic DBs and 7 magnetic DQs. 
We classify the confirmed magnetic DA and DB white dwarfs as DAH and DBH, respectively. We classify two targets as DAH:, where the detection of Zeeman splitting is questionable. In addition, we classify 66 objects as DAH?, 3 as DBH?, and 2 as DQH?, where the objects are clearly magnetic, but the atmospheric composition is uncertain. Since we do not have magnetic DQ models, we simply use non-magnetic models to fit their spectra (see \S \ref{warmdq}). 

For the magnetic DA and DB white dwarfs, we use the photometric technique to constrain the effective temperature and solid angle, thus the radius and mass. For each object, we construct a grid of specific intensities at the surface by solving the radiative transfer equation for various field strengths. The line displacements and oscillator strengths of the Zeeman components of hydrogen and helium lines are kindly provided by S. Jordan \citep[see also][]{hardy23a,hardy23b,moss24}. The total line opacity is calculated as the sum of the individual Stark-broadened Zeeman components, and is normalized to that resulting from the zero-field solution. The emergent Eddington flux $H_\nu$ is then calculated on the fly during the fitting process by numerically integrating the specific intensity $I_\nu$ over the visible surface of the disc given values of the dipole field strength $B_d$, dipole offset $a_z$, and viewing angle $i$ ($0^\circ$ for a pole-on configuration). We use the {\tt PIKAIA} genetic algorithm \citep{pikaia} to find the best-fitting magnetic model.

With increasing field strength, the absorption features are shifted and blended so much that it may be difficult to identify each line, or they may create broad absorption features, and sometimes featureless DC-like spectra. Such magnetic white dwarfs with featureless spectra can still be identified above $T_{\rm eff}=11,000$ K, since H and He lines should be visible for non-magnetic white dwarfs at those temperatures. 
Even with these caveats, our model fits using {\tt PIKAIA} provide excellent constraints on the field strength and geometry for a large number of magnetic white dwarfs in our sample. Figure \ref{fitdah} shows fits to six magnetic white dwarfs with fields ranging from $B_d=10$ to 178 MG. The fits here are remarkable; these models provide excellent fits to even subtle features in some of these spectra. The top left panels show the model fits to a magnetic DB, where Zeeman split He lines including $\lambda\lambda$4471, 5876, and 6678 \AA\ are clearly visible. Our magnetic DB model with a field strength of $B_d=9.9$ MG provides an excellent fit to this spectrum. We also highlight WDJ085601.74+321354.51 (bottom right panels) here. This object was originally classified as a DQ by \citet{manser24} due to the broad absorption features. However, our magnetic DA model fits clearly show that this is a strongly magnetic DA white dwarf where the observed features are explained by a field strength of 178 MG. 

\begin{figure}
\includegraphics[width=3.2in,clip=true, trim=0.3in 3in 0.2in 3.1in]{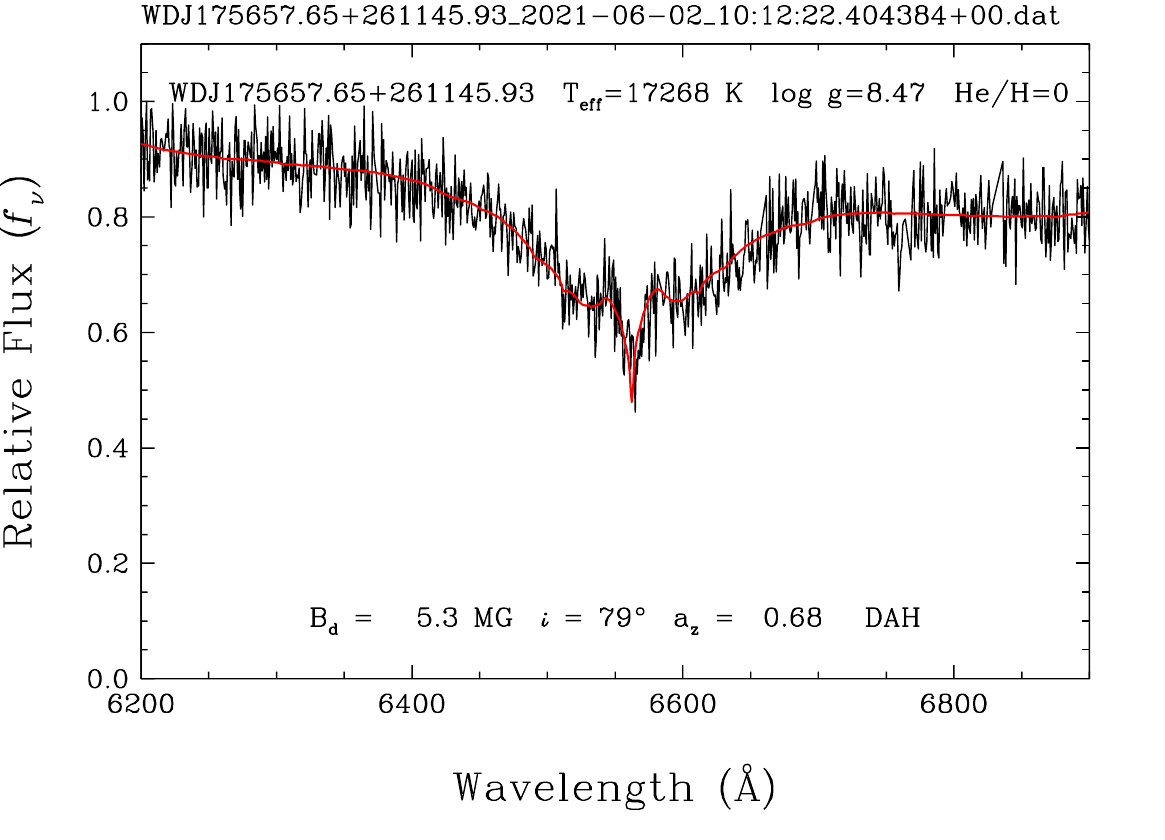}
\caption{Magnetic model fits to a DA white dwarf with a relatively low-field strength of $B=5.3$ MG.}
\label{fitdahlow} 
\end{figure}

\begin{figure}
\includegraphics[width=3.2in]{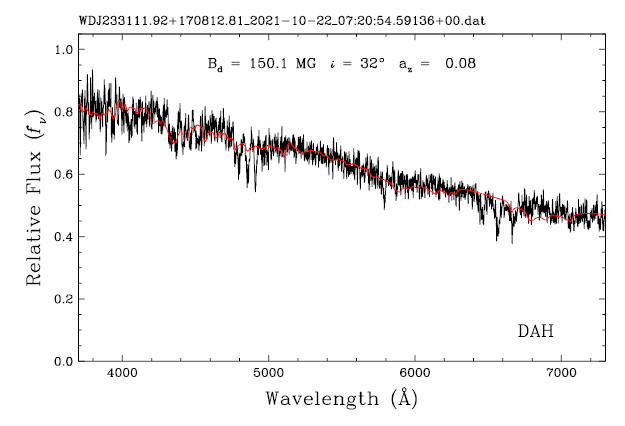}
\includegraphics[width=3.3in, clip=true, trim=0.3in 3in 0.2in 3in]{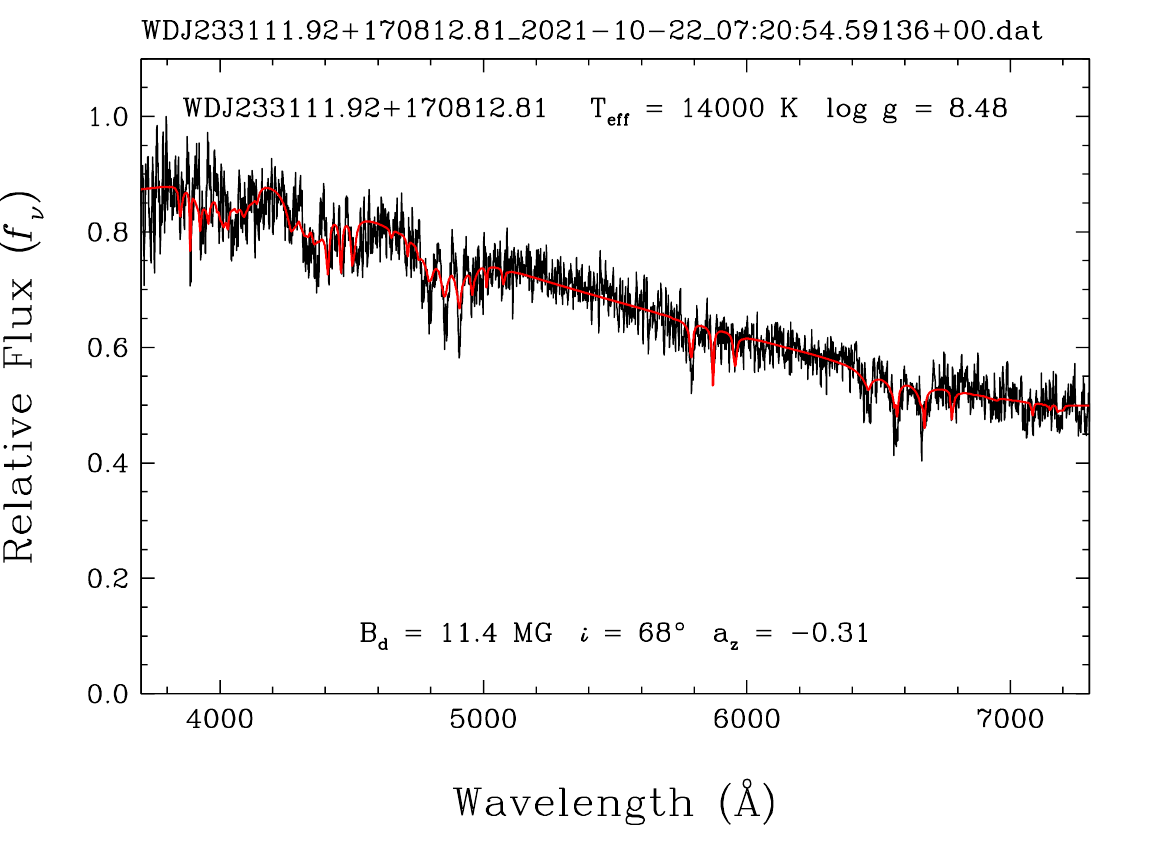}
\caption{Model fits to a magnetic white dwarf with relatively sharp lines. The top panel shows our model fits assuming a homogeneous pure H  atmosphere,
whereas the bottom panel shows the fits using a patchy atmosphere with H caps and a He belt.}
\label{figpatch} 
\end{figure}

Some of the magnetic white dwarfs in our sample require a more refined analysis. This is especially true for low-field objects where Zeeman splitting is only visible in the line cores. Our usual method of fitting the entire spectrum does not work for these objects, since the $\chi^2$ is dominated by the broad absorption features and the continuum. To fit the narrowly split absorption features, we had to increase the wavelength resolution of the models. 
Figure \ref{fitdahlow} shows an example fit to a magnetic DA with a relatively low field strength of $B_d = 5.3$ MG. Here we only show the H$\alpha$ region to highlight the modest Zeeman splitting observed. To accurately constrain the field strength in this system, we had to perform a tailored analysis of only the H$\alpha$ region with a higher wavelength resolution.  A more refined analysis for the entire sample of magnetic white dwarfs in our sample is beyond the scope of this paper and it will be presented elsewhere. 

Some of the magnetic spectra show sharper lines than predicted. Figure \ref{figpatch} shows our model fits to one such system, WDJ233111.92+170812.81. The top panel shows our model fits under the assumption of a homogeneous, pure H atmosphere. This model fit utterly fails to reproduce the observed lines. However, this problem could be solved by using patchy atmospheres \citep{moss24,moss25}. The bottom panel shows our model fits using a patchy atmosphere with H caps and an equatorial He belt. We define $\theta_c$ as the extent of the polar cap from the pole down to the equator, where $\theta_c=90^\circ$ means the entire hemisphere is H. This patchy model with a field strength of 11.4 MG and H-caps extending $37^\circ$ from the poles provides a remarkable fit to the sharp Zeeman split components, and also confirms that the features near 5876 \AA\ are indeed from He. 

The split H$\beta$ components still appear underestimated in the patchy atmosphere model compared with the data. This is likely because our patchy atmosphere model using H caps and He belts is a proxy for the actual surface abundance and field distributions, which are likely more complicated than our simple assumption of caps and belts.  We also note that if the field strength was significantly higher, we would likely not be able to identify He lines. Hence, there could be other patchy atmosphere white dwarfs in our sample with even higher field strengths that we may be missing. There are other instances of objects where our magnetic models totally fail, a situation that \citet{kilic25b} also encountered in their GALEX selected white dwarf sample. 

\begin{figure}
\includegraphics[width=3.2in]{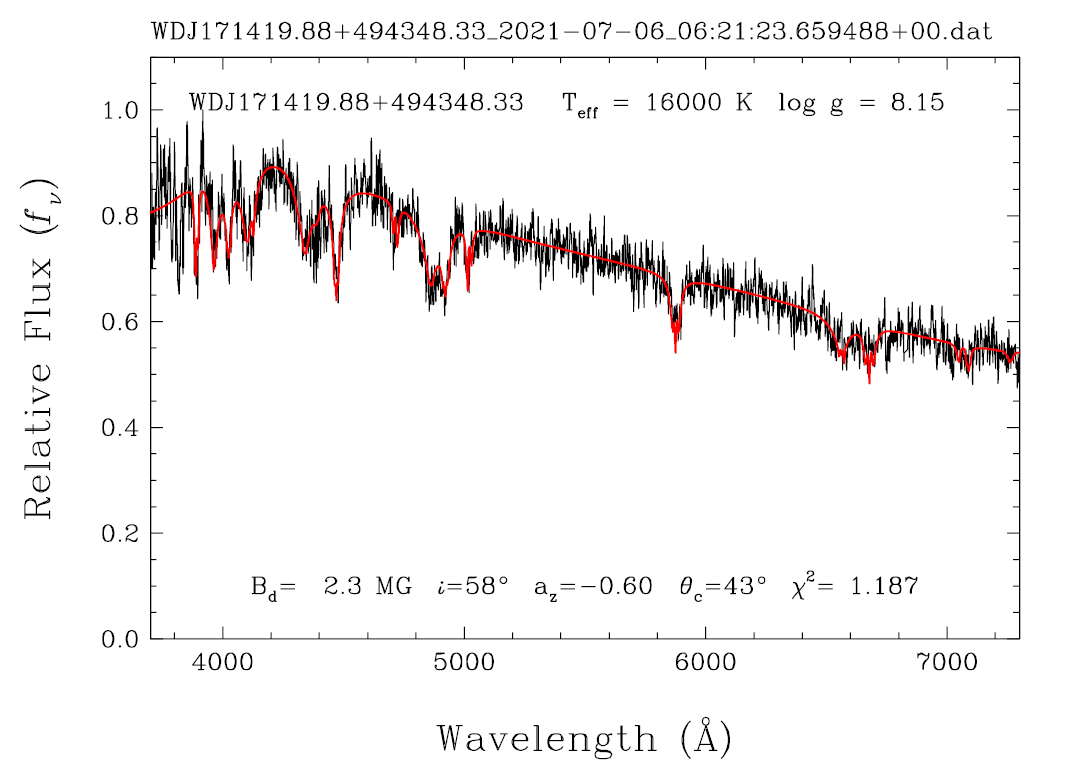}
\caption{Model fits to another magnetic and patchy atmosphere white dwarf identified in this work.}
\label{figpatch2} 
\end{figure}

We show another newly identified patchy atmosphere white dwarf in Figure \ref{figpatch2}. WDJ171419.88+494348.33 shows only modest Zeeman splitting that requires a field strength of 2.3 MG. Given its hotter temperature compared to WDJ233111.92+170812.81, the He lines are stronger. Our model fits under the assumption of homogeneous, mixed H/He atmosphere fits fail. However, a patchy atmosphere with a H-cap size of $43^\circ$ provides a good fit to the observed spectrum. These two objects are meant to highlight the patchy atmosphere white dwarfs in our sample. A comprehensive analysis of these systems and other magnetic white dwarfs with problematic fits will be presented elsewhere. 

\subsection{Unusual Objects}

\begin{figure*}
\center
\includegraphics[width=2.8in, clip=true, trim=0.4in 0.8in 0.1in 1.1in]{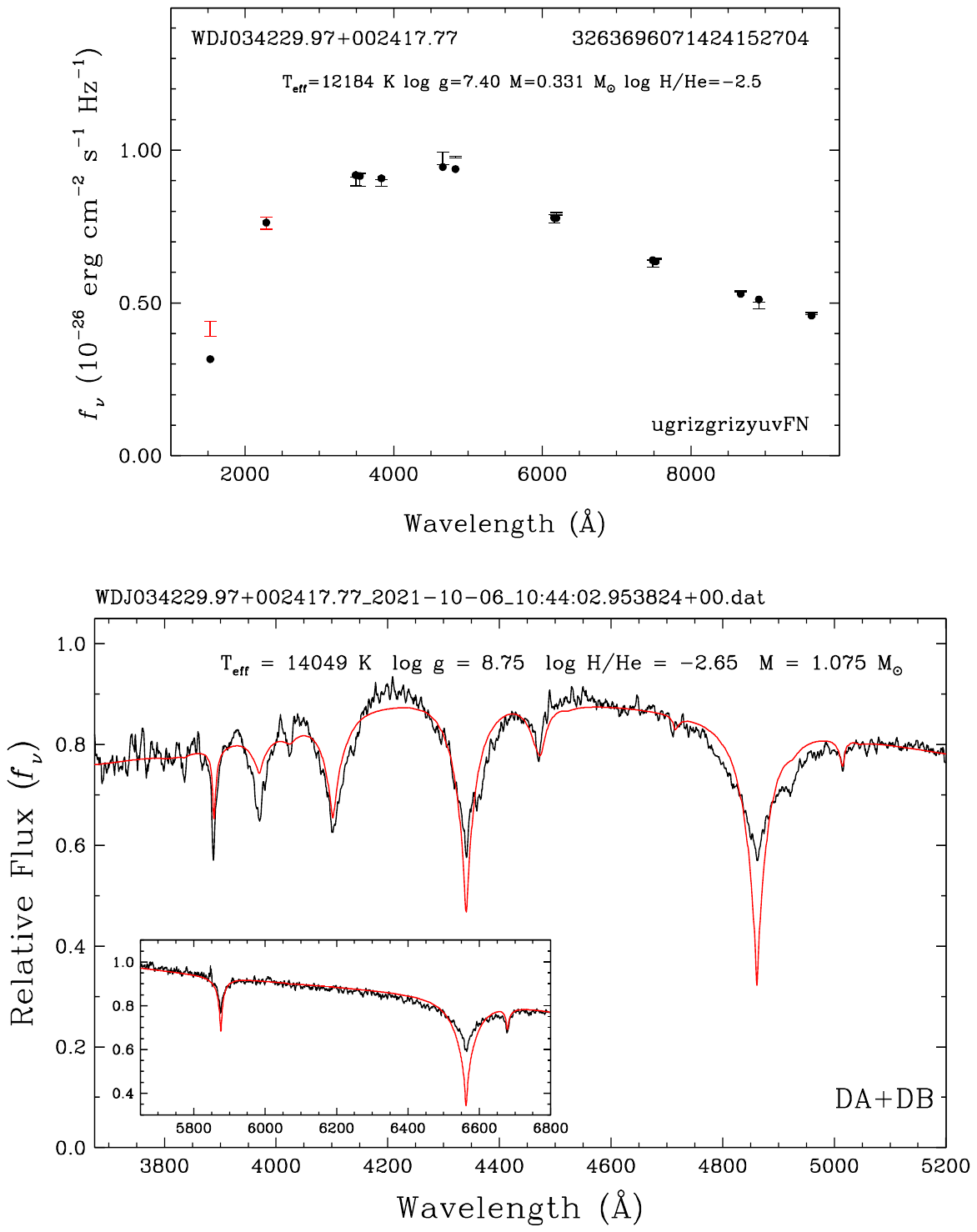}
\includegraphics[width=2.8in, clip=true, trim=0.4in 0.8in 0.1in 1.1in]{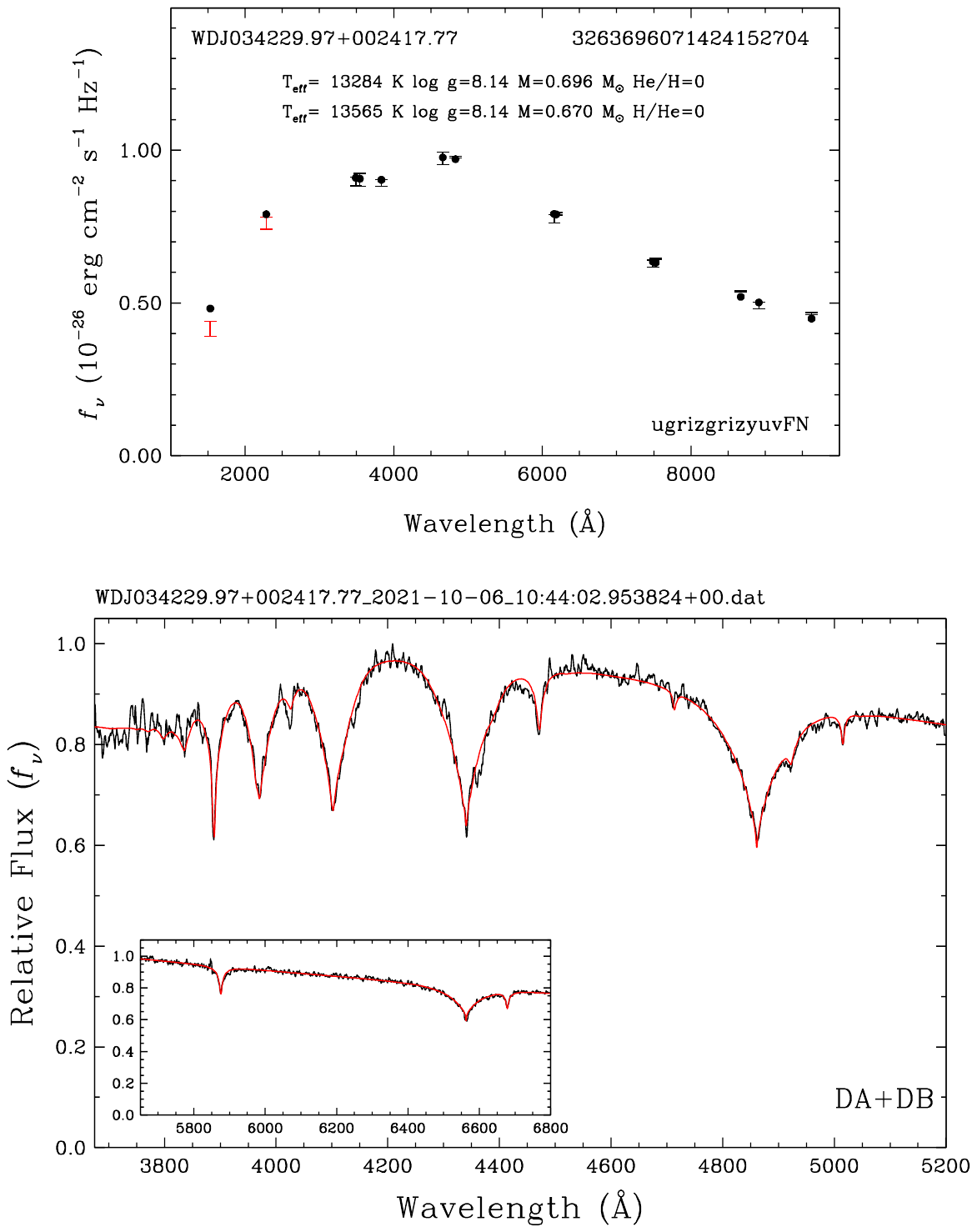}
\caption{Model fits to WDJ034229.97+002417.77 under the assumption of a single star (left panels) and a double degenerate binary containing a DA+DB system (right panels, the adopted photometric and spectroscopic parameters are given in the upper panel). The top and bottom panels show the photometric and spectroscopic fits, respectively.}
\label{figdadb} 
\end{figure*}

\begin{figure*}
\center
\includegraphics[width=2.8in, clip=true, trim=0.4in 0.8in 0.1in 1.1in]{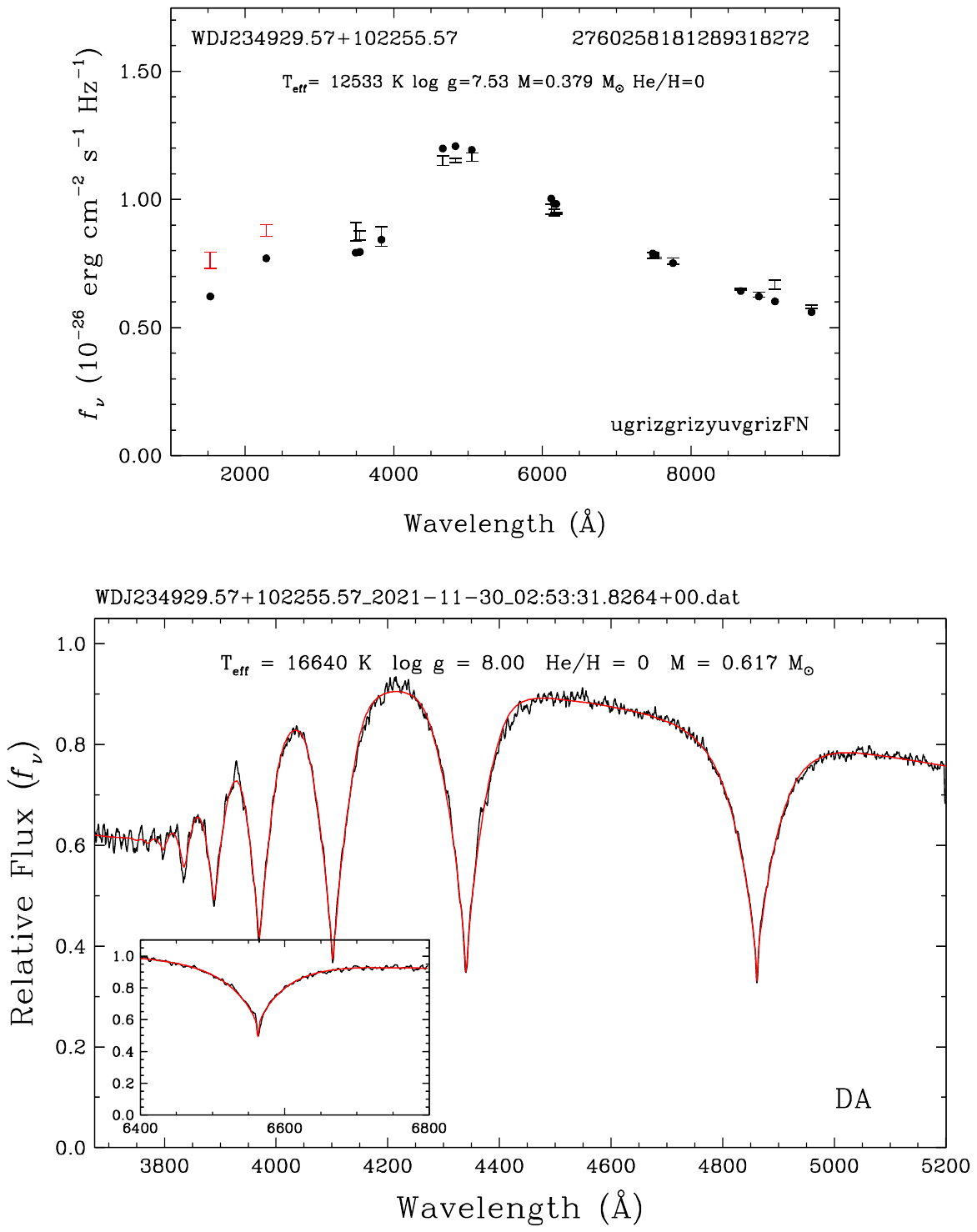}
\includegraphics[width=2.8in, clip=true, trim=0.4in 0.8in 0.1in 1.1in]{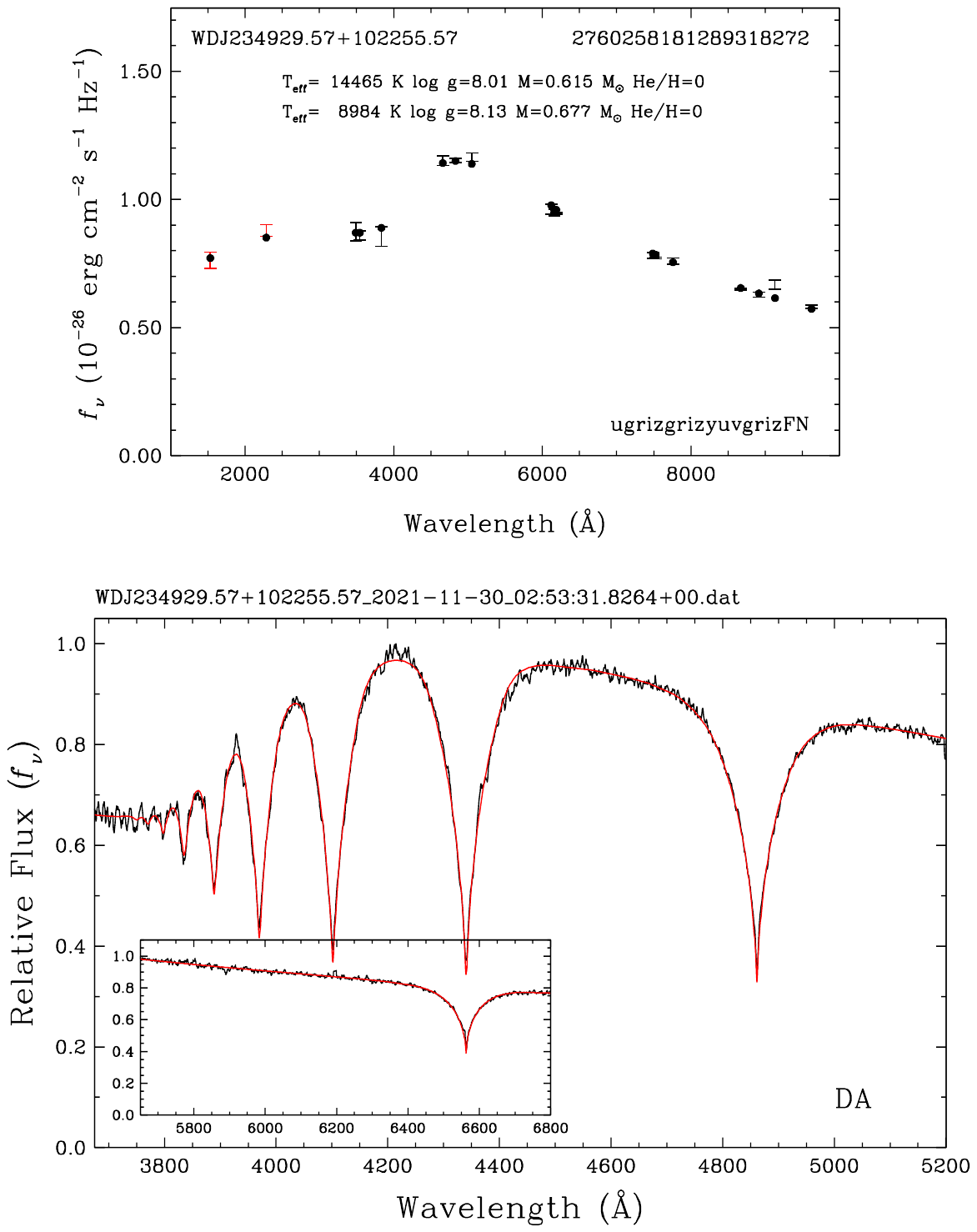}
\caption{Model fits to WDJ234929.57+102255.57 under the assumption of a single star (left panels) and a double degenerate binary containing a DA+DA system (right panels, the adopted photometric and spectroscopic parameters are given in the upper panel). The top and bottom panels show the photometric and spectroscopic fits, respectively.}
\label{figdouble} 
\end{figure*}

We identify 145 DC white dwarfs with featureless spectra. We show both pure H and pure He photometric fits for those stars, but display only the predicted pure He spectroscopic fits in the online materials. 
In the temperature range of this sample, a comparison with the pure He atmosphere model helps to see if the star is simply too cool to show He lines. Otherwise, if the predicted He lines are too strong (H lines would be even stronger at the same temperature), this implies that the DC white dwarf is probably magnetic. Some of the hotter DCs in our sample are likely magnetic, but the relatively noisy DESI spectra for those targets prohibit us from confirming that. In addition, our hot white dwarf sample in DESI DR1 contains 7 previously known AM CVn, 2 CVs, the oxygen atmosphere DS type white dwarf WDJ124043.46+671035.93 \citep{kepler16b}, and the LP40-365 type star WDJ110907.96+000132.92, which is suspected to be a partially burned runaway accretor \citep{elbadry23}. A detailed analysis of these unusual targets is beyond the scope of this paper. We simply provide the photometric fits for these objects for both pure H and pure He solutions with corresponding spectroscopic predictions (pure H in red; pure He in blue) in the online figures.

After a detailed model atmosphere analysis of the various spectral types discussed in previous sections, we identify two subsets of stars where our spectroscopic model fits fail to provide acceptable fits to the DESI spectra. These involve white dwarfs displaying both hydrogen and helium spectral features.
The first set of objects with spectroscopic model fits that require special attention is DA+DB white dwarf binaries. Our initial fits with DBA atmosphere models, with He dominated atmospheres and trace amounts of H, produced unacceptable fits to the spectra for some targets. 

The left panel in Figure \ref{figdadb} shows our model fits for one of these targets, WDJ034229.97+002417.77. Fitting the spectrum of this object under the assumption of a single white dwarf, the photometric fit indicates a mass of only $0.33~M_\odot$, whereas the spectroscopic fit gives $M=1.075~M_\odot$ and $\log$ H/He = $-2.65$. Such discrepancies between the photometric and spectroscopic parameters usually indicate a double degenerate system \citep[see for example][]{bedard17}. The best-fitting spectroscopic model clearly fails to match the DESI spectrum of this object. The right panel in Figure \ref{figdadb} shows our model fit under the assumption of a binary. We use the PIKAIA algorithm with $T_{\rm eff}$ and $\log{g}$ for the DA and DB components as free parameters, where the $\chi^2$ used is a combination of the photometric fit $\chi^2$ and the spectroscopic fit $\chi^2$, with more weight given to the spectroscopic fit. The composite
fit indicates a DA + DB binary with nearly equal temperatures and masses (0.70 and $0.67~M_\odot$). This composite fit matches the depths of both H and He lines. 

\begin{deluxetable*}{lrrccccc}
\tablecolumns{8} \tablewidth{0pt}
\tablefontsize{\tiny}
\tablecaption{DA+DB binary white dwarfs.\label{tabdadb}}
\tablehead{\colhead{Object} & \colhead{Gaia SourceID} & \colhead{$T_{\rm eff, DA}$} & \colhead{Mass, DA} & \colhead{$\log{g}$, DA} & \colhead{$T_{\rm eff,DB}$} & \colhead{Mass, DB} & \colhead{$\log{g}$, DB}\\
 & & (K) & ($M_\odot$) & (cm s$^{-2}$) & (K) & ($M_\odot$) & (cm s$^{-2}$)}
\startdata
WDJ002839.41+065529.97 & 2747880768440560640 & 20597 & 0.579 & 7.915 & 39684 & 0.538 & 7.807 \\
WDJ011356.38+301515.10 & 310054341933521664  &  8899 & 0.416 & 7.654 & 19856 & 0.712 & 8.190 \\
WDJ014921.98+314926.38 & 304407834329392768  & 12786 & 0.472 & 7.745 & 17439 & 0.431 & 7.679 \\
WDJ034229.97+002417.77 & 3263696071424152704 & 13284 & 0.696 & 8.144 & 13565 & 0.670 & 8.136 \\
WDJ043907.82+002728.74 & 3230857747910285696 &  8955 & 0.541 & 7.901 & 19371 & 0.632 & 8.060 \\
WDJ073525.02+320420.31 & 892357662805211776  &  8880 & 0.822 & 8.354 & 17813 & 1.030 & 8.693 \\
WDJ082022.88+033523.04 & 3091346038926485760 & 13698 & 0.657 & 8.079 & 23769 & 0.414 & 7.593 \\
WDJ083353.72+385218.90 & 911093409660826624  & 19315 & 0.518 & 7.802 & 14643 & 0.326 & 7.424 \\
WDJ090141.47+435952.90 & 1009267218061764736 & 14934 & 0.241 & 7.000 & 14388 & 0.218 & 7.000 \\
WDJ091227.58-053158.63 & 5758314781766103296 & 13907 & 0.624 & 8.023 & 20116 & 0.598 & 7.999 \\
WDJ093559.80+000828.89 & 3840274467176110848 & 15933 & 0.705 & 8.151 & 12024 & 0.242 & 7.151 \\
WDJ101316.02+075915.16 & 3874193610618742400 & 12383 & 0.563 & 7.924 & 25983 & 0.487 & 7.757 \\
WDJ103822.36+372112.03 & 751930335511863040  & 11730 & 1.027 & 8.669 & 13729 & 1.149 & 8.916 \\
WDJ112752.97+553522.03 & 843291028702104448  & 10413 & 0.622 & 8.035 & 23959 & 0.557 & 7.910 \\
WDJ131536.31+162415.38 & 3936757368427174912 & 16876 & 0.737 & 8.199 & 17068 & 0.295 & 7.285 \\
WDJ131928.37+522334.65 & 1562874316241555584 &  9310 & 0.869 & 8.424 & 13758 & 1.188 & 8.999 \\
WDJ133828.43+415943.79 & 1501073585142103424 & 11716 & 0.660 & 8.092 & 11693 & 0.425 & 7.696 \\
WDJ144055.25+025719.73 & 3655746248185163136 &  8965 & 0.673 & 8.124 & 16698 & 0.596 & 8.006 \\
WDJ150506.27+383017.58 & 1295485694092963968 &  9560 & 0.585 & 7.976 & 15314 & 0.590 & 7.999 \\
WDJ151134.14+274612.65 & 1274547767183148672 & 12011 & 0.645 & 8.066 & 20689 & 0.703 & 8.173 \\
WDJ173152.40+291142.97 & 4598771567165528448 & 11929 & 0.629 & 8.039 & 18795 & 0.802 & 8.332 \\
WDJ185432.43+584000.20 & 2155326017764931072 &  8834 & 0.519 & 7.862 & 16228 & 1.107 & 8.832 \\
WDJ211014.06+015830.65 & 2690860855634249856 & 10296 & 0.491 & 7.800 & 17183 & 0.559 & 7.939 \\
\enddata
\end{deluxetable*}

Table \ref{tabdadb} provides the list of 23 DA+DB binaries identified in this work. We provide the fits to all 23 systems in the online version of Figure \ref{figdadb}, but note that the set of regular fits provided on the \href{https://doi.org/10.5281/zenodo.18332706}{Zenodo archive} assume a single star. 
These are basically SB2 spectroscopic binaries. Follow-up radial velocity observations of these systems would be essential for constraining their orbital periods, and improving their binary parameters.

Our sample contains a large number of low-mass white dwarfs with photometric masses below $0.45~M_\odot$. A large fraction of these systems are likely in binary systems \citep{marsh95} with two DA white dwarfs. Figure \ref{figdouble} shows our model fits for one of these objects, where the photometric fit indicates $T_{\rm eff}=12,533$ K and $M=0.379~M_\odot$. Fitting the DESI spectrum under the assumption of a single star, the best-fitting model is significantly hotter and more massive with $T_{\rm eff}=16,640$ K and $M=0.617~M_\odot$. Such large discrepancies between the photometric and spectroscopic parameters can be explained by a double degenerate system \citep{bedard17}. The right panels show our model fits under the assumption of a DA+DA binary for the same system. Here, a binary system of DA white dwarfs with similar masses (0.62-$0.68~M_\odot$) but significantly different temperatures (8984 vs 14,465 K) provides a remarkable fit to both the photometry and spectroscopy. There are many other low-mass, and even normal-mass, white dwarfs with significantly discrepant photometric and spectroscopic parameters in the DESI sample.  
However, without radial velocity constraints on the mass ratios, there are degeneracies in the DA+DA atmosphere model fits, where an increase in surface gravity for one star can be compensated by a decrease in surface gravity of the companion to match the observed spectral energy distribution \citep[see for example Figure 6 in][]{kilic21}. A detailed model atmosphere analysis of the DA+DA candidates is beyond the scope of this paper.

\begin{figure}
\includegraphics[width=3.3in, clip=true, trim=0.4in 2.8in 0.2in 3.7in]{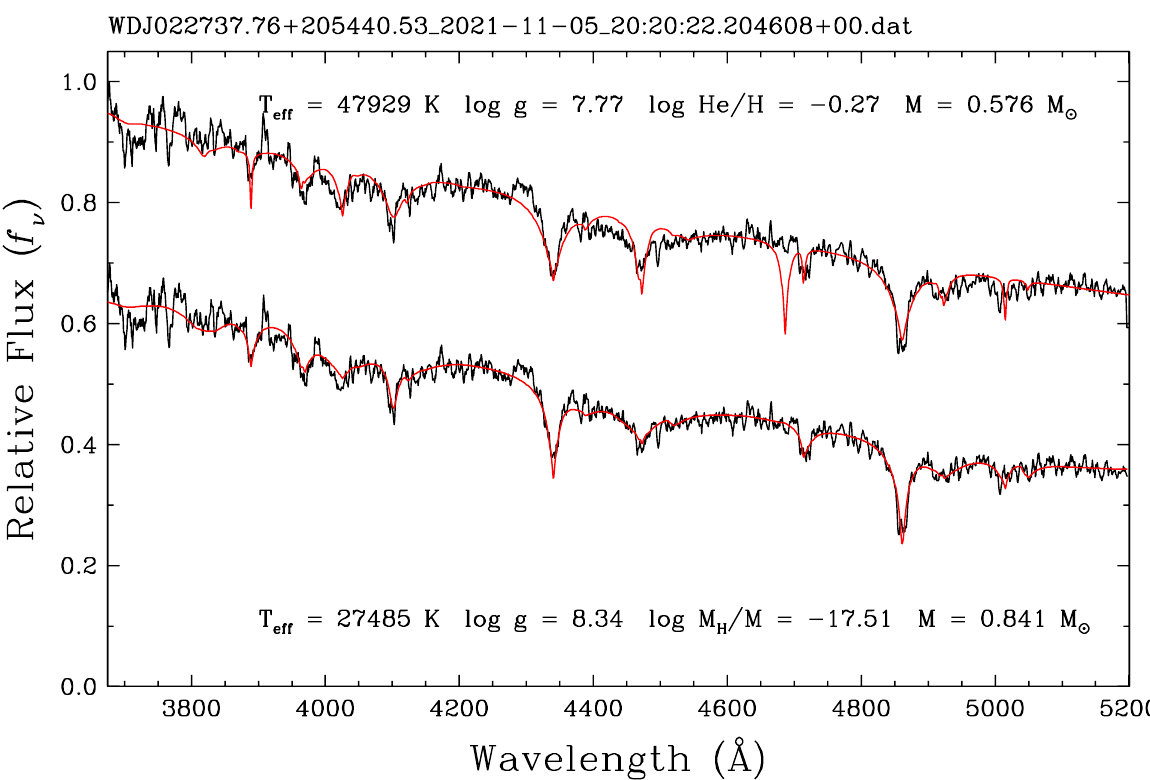}
\caption{Model fits to the hot white dwarf WDJ022737.76+205440.53 under the assumption of a mixed atmosphere (top) and a stratified atmosphere (bottom).}
\label{figstrat} 
\end{figure}

The second set of objects that require special attention includes hot white dwarfs, where the predicted strength of the \ion{He}{2} feature at 4686 \AA\ does not match the observations. Figure \ref{figstrat} shows our model fits to WDJ022737.76+205440.53 under the assumption of a homogeneous atmosphere (top), where hydrogen and helium are uniformly distributed throughout the atmosphere, and a stratified atmosphere (bottom), where a thin layer of hydrogen is floating on top of helium in diffusive equilibrium. We rely
on a model grid calculated using the framework described in \citet{manseau16} for stratified atmospheres. 
Clearly, the stratified atmosphere model with $\log$ $M_{\rm H}/M=-17.5$ provides a much better fit to the observed spectrum of this target. 

Stratified atmosphere white dwarfs have relatively diverse spectral types, including DAB, DAO, DBA, or DOA \citep{manseau16}. \citet{bedard20} found 31 stratified white dwarfs in the SDSS DR12 with $T_{\rm eff}<55,000$ K and with $\log$ $M_{\rm H}/M=-18.25$ to $-15.25$. We find 22 stratified atmosphere candidates in the hot DESI DR1 sample. These can be recognized in the general table of results because they have a $\log{q_H}$ measurement. However, these are approximate fits as they assume LTE.  

\section{Results}
\label{secres}

\subsection{DA White Dwarfs: DESI, we have a problem}

\begin{figure}
\center
\includegraphics[width=3.3in]{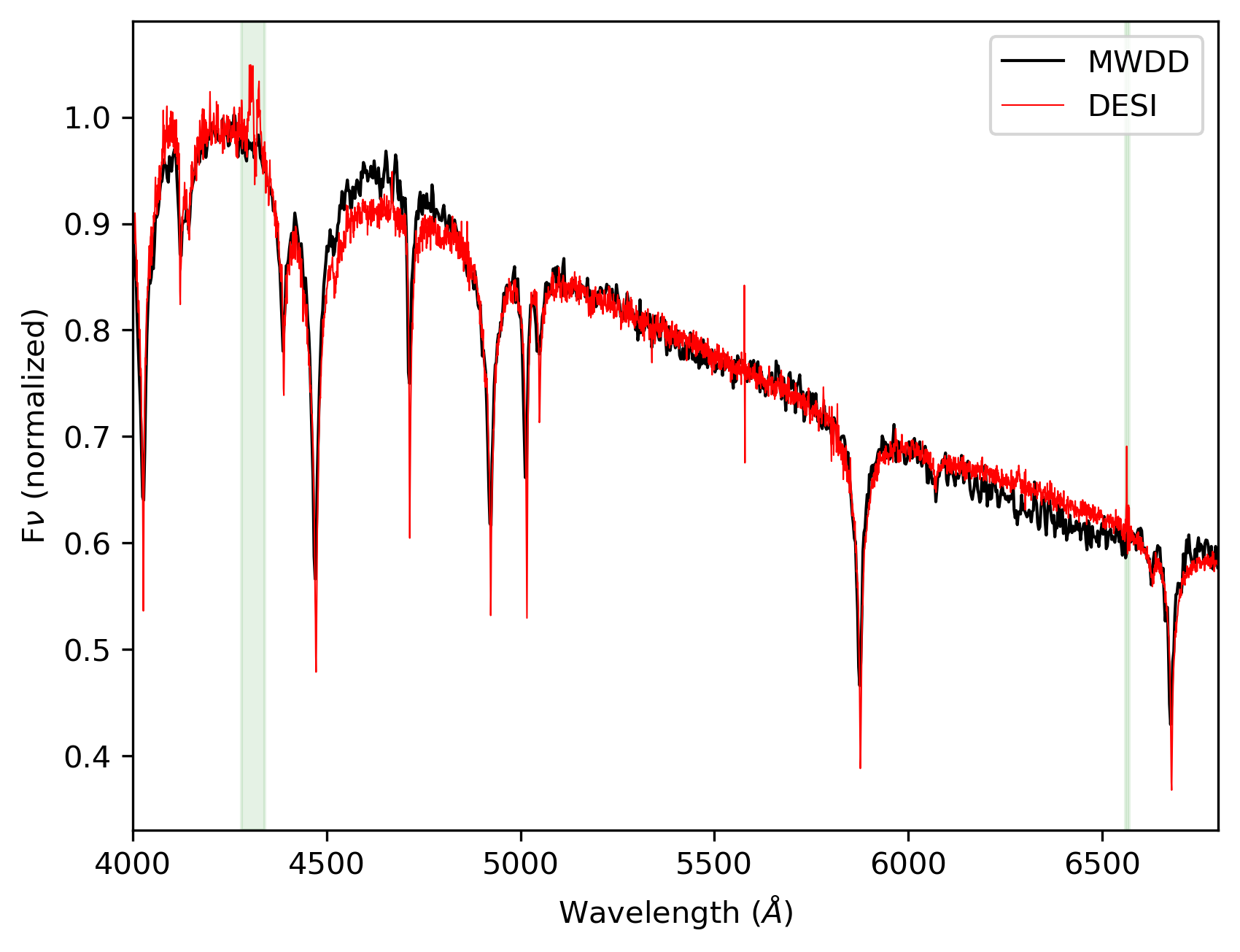}
\caption{DESI spectrum of the DB white dwarf GD 85 (red) compared to a lower resolution MWDD spectrum of the same object (black). Green shaded regions highlight the problem regions in DESI data.}
\label{figspike} 
\end{figure}

DA white dwarfs are commonly used as flux standards, thanks to their relatively simple, pure H atmospheres. DESI does not rely on DA white dwarfs for flux calibration. 
DESI spectroscopic calibration relies on main-sequence F stars, where lower metallicity halo stars are preferentially selected \citep[see][for a detailed description of DESI flux calibration]{guy23}. Theoretical stellar models appropriate for these stars are used to fit the temperature, surface gravity, metallicity, and radial velocity of each star to determine the flux calibration for each of the three DESI cameras. 

\citet{guy23} tested their flux calibration procedures using high S/N DA spectra obtained as part of the DESI survey. Figure 41 in their paper shows the average residuals between the DESI DA spectra and the best-fitting models: the flux calibration of the DESI DA white dwarfs appears to be good to within 2\% above 3700 \AA, but there are significant spikes in the hydrogen line cores. \citet{guy23} concluded that further work is needed to understand the origin of these features, and attributed them to the white dwarf fitting pipeline rather than the flux calibration itself. There is also a known issue related to a drop in sensitivity near 4300 \AA, which impacts the flux calibration in the 4200-4400 \AA\ range, and creates spurious features in some spectra. 

Figure \ref{figspike} shows an example DESI spectrum where these problems are clearly seen. Here we show a DB white dwarf, GD 85. Since it does not show any H lines, it is easier to see the problems in flux calibration in the H$\alpha$ line core. This figure shows a comparison between GD 85's DESI spectrum and a lower resolution spectrum available in the Montreal White Dwarf Database \citep[MWDD,][]{dufour17}. Green shaded regions highlight the problem regions in DESI data.
There are spurious features near 4300 \AA\ and the H$\alpha$ region, which complicate the detection of H lines in DBA white dwarfs, and impact the H line profiles in DA white dwarfs. Issues with both blue and red wings of H$\gamma$ are also common in DA white dwarfs (see below). 

\begin{figure}
\center
\includegraphics[width=3in, clip=true, trim=1.4in 0.6in 1.6in 1in]{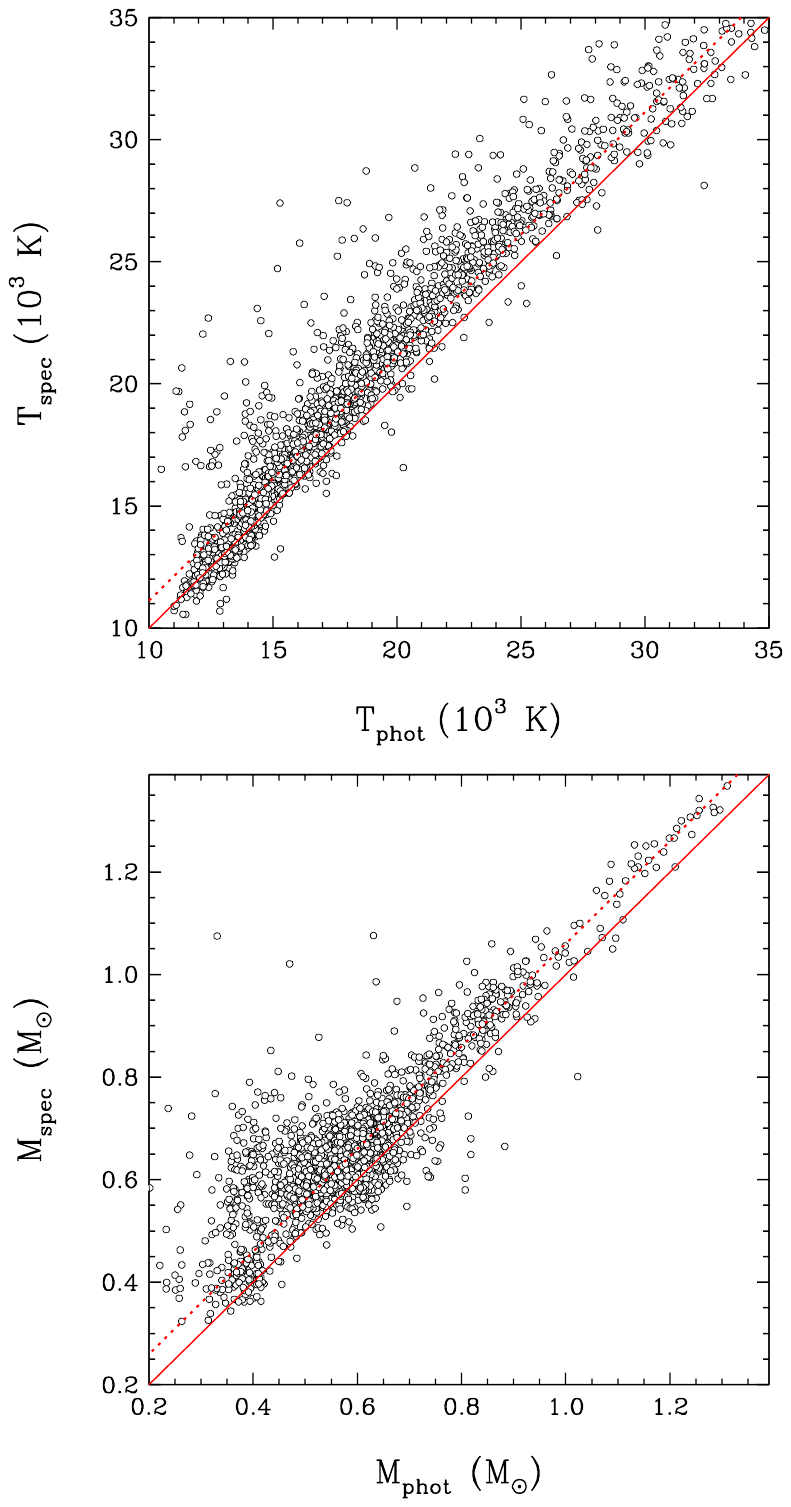}
\caption{Photometric and spectroscopic temperatures and masses for DA white dwarfs in DESI DR1. This figure is restricted to objects with distance accuracy better than 10\% and S/N $>20$ spectra in DESI. There are systematic offsets in temperatures and masses (shown as dotted lines) between the two measurement techniques.}
\label{figcorreltgDA} 
\end{figure}

Figure \ref{figcorreltgDA} shows the photometric and spectroscopic temperature and mass values for DA white dwarfs with S/N $>20$ DESI spectra in our sample. Surprisingly, we detect significant and systematic shifts in both temperatures and masses measured through DESI spectroscopy. We find the same problem even if we use the normalized Balmer line profiles for fitting the DESI spectra. The spectroscopic temperature and mass measurements are systematically higher than the photometric values. Since DESI is a magnitude limited survey, lower mass objects are over represented in this diagram. The relatively large concentration of objects near $M_{\rm phot}=0.4~M_\odot$ corresponds to double degenerates. Restricting our sample to objects with $M_{\rm phot}\geq0.5~M_\odot$, the spectroscopic temperatures and masses are higher on average by 1128 K and $0.06~M_\odot$, respectively. 

\begin{figure}
\center
\includegraphics[width=3in, clip=true, trim=0.4in 0.8in 0.1in 1.1in]{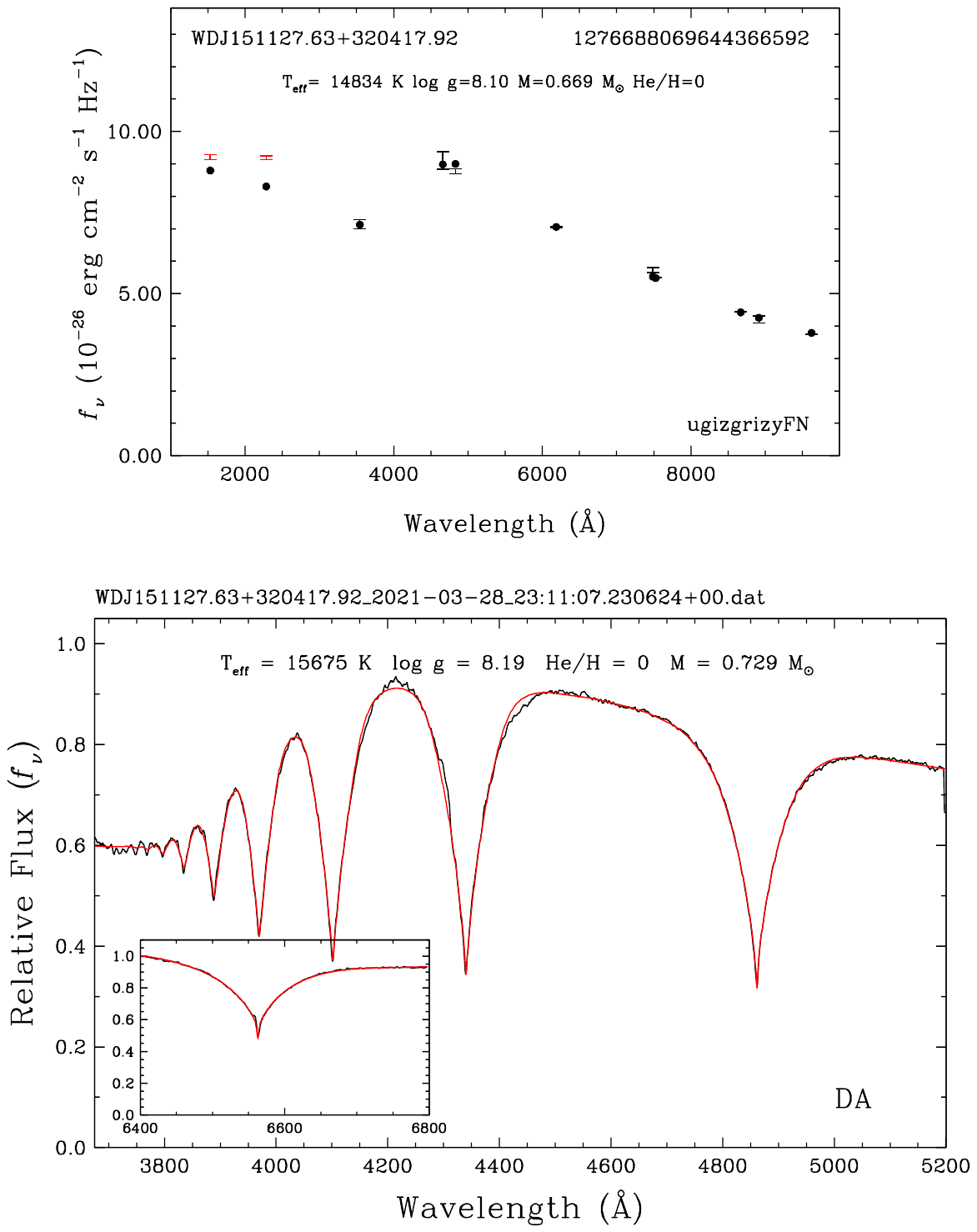}
\caption{Model fits to a DA white dwarf with a ${\rm S/N}=180$ DESI spectrum, highlighting the systematic differences between the photometric and spectroscopic values.}
\label{figdaprob} 
\end{figure}

The differences shown in the top panel of Figure \ref{figcorreltgDA} are due to a well known problem between the photometric and spectroscopic temperature scales for DA white dwarfs, and do not necessarily indicate a problem with the DESI spectra. \citet{genest19}, \citet{bergeron19}, and \citet{tremblay19b} also observed temperature offsets between the photometric and spectroscopic solutions for DA white dwarfs. For example, \citet[][see their Figure 13]{genest19} report spectroscopic temperatures that are 5 to 10\% hotter than the photometric values, a problem that they attribute to the uncertainties in the Stark broadening profiles. However, they do not find any systematic offsets between the photometric and spectroscopic masses (see their Figure 17), and they conclude that the spectroscopic mass scale is more reliable, and less affected by the problems with the physics of line broadening theory. Hence, the systematic mass offset seen in Figure \ref{figcorreltgDA} is significant, and this is the first time such a trend is reported in the literature. 

Figure \ref{figdaprob} shows the model fits to a ${\rm S/N}=180$ DESI spectrum of the DA white dwarf WDJ151127.63+320417.92 to illustrate the mass problem. The observed photometry is best-explained by a pure H atmosphere model with $T_{\rm eff}=14,834$ K and $M=0.669~M_\odot$, whereas the DESI spectrum indicates a significantly higher temperature and mass; $T_{\rm eff}=15,675$ K and $M=0.729~M_\odot$. These offsets are similar to the offsets seen for the DA white dwarf WDJ000139.70$-$083631.83 shown in the right panels of Figure \ref{fitda}, where the spectroscopic values are higher by 1074 K and $0.082~M_\odot$, respectively. Note that the higher mass and $\log{g}$ values for these two stars are not due to the 3D model atmosphere corrections, which are negligible at these temperatures and $\log{g}$ values \citep{tremblay13}. Even though the spectroscopic fit in Figure \ref{figdaprob} looks excellent overall, a close inspection of the H$\gamma$ region shows that the DESI spectrum shows discrepancies with the model in the 4200-4500 \AA\ region. This is likely due to the flux calibration problems discussed above. However, excluding the H$\gamma$ region from our model fits does not remedy the problem. This suggests that flux calibration based on F stars is not ideal for reproducing the shapes of the relatively broad Balmer lines in DA white dwarfs.

\begin{figure}
\center
\includegraphics[width=3in]{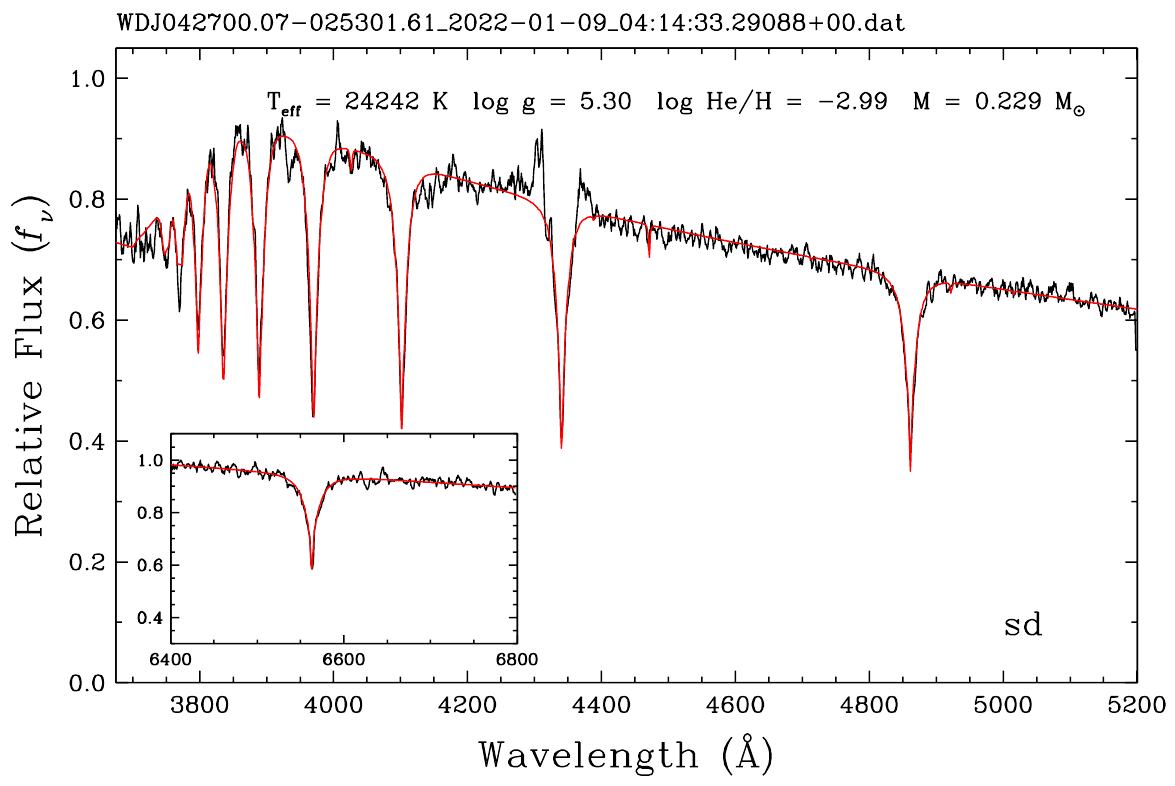}
\caption{Model fits to a subdwarf star where the DESI spectrum shows significant flux calibration issues near H$\gamma$.}
\label{figsdprob} 
\end{figure}

\begin{figure}
\center
\includegraphics[width=2.8in, clip=true, trim=1.2in 0.8in 1.5in 1in]{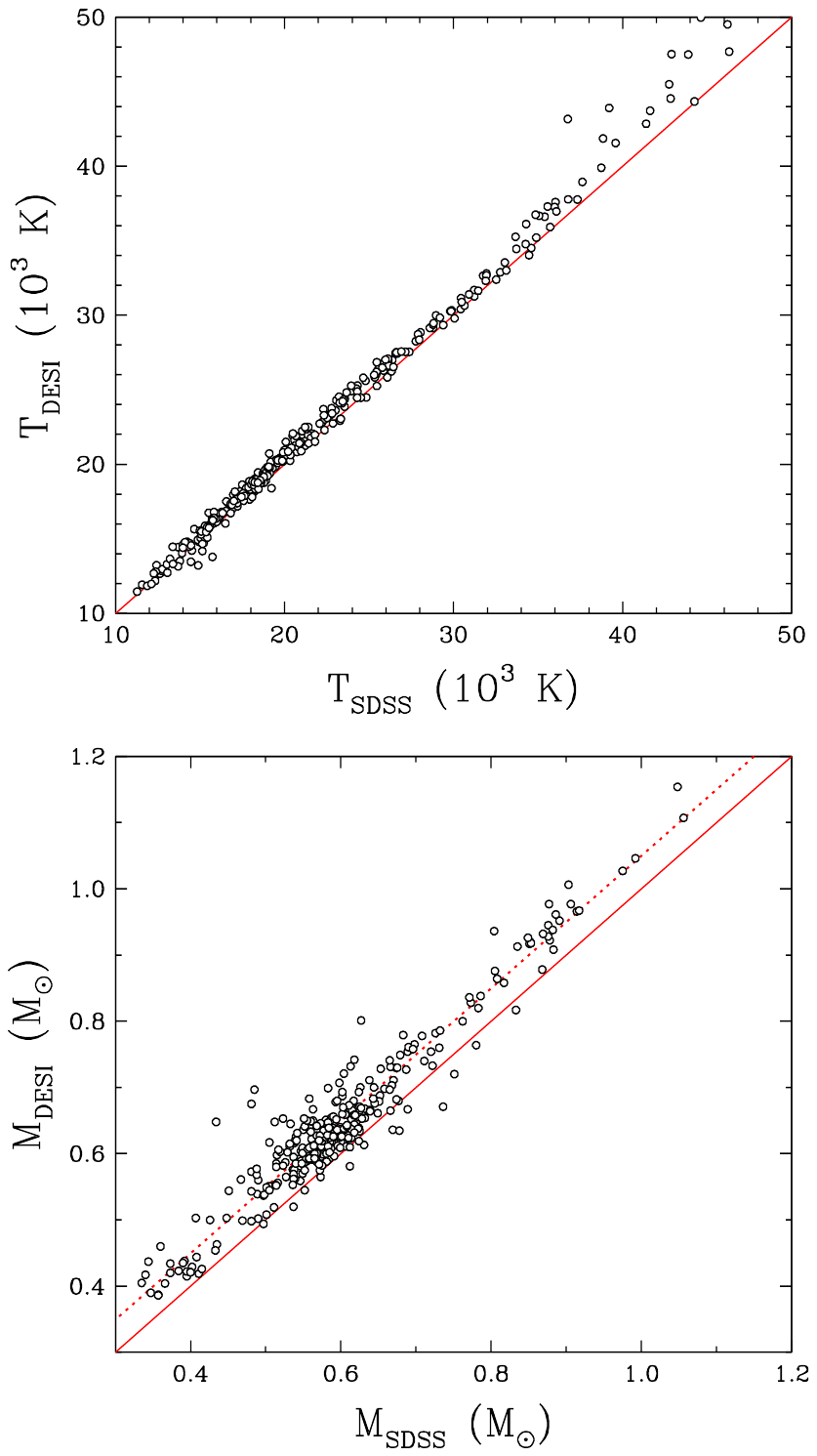}
\caption{Spectroscopic temperatures (top) and masses (bottom panel) obtained from DESI and SDSS spectra of 346 DA white dwarfs with high S/N spectroscopy available in both surveys. The spectroscopic fits to DESI data indicate systematic offsets of 570 K in temperature and 
$0.05~M_{\odot}$ in mass (shown as a dotted line) compared to the fits using SDSS spectroscopy data.}
\label{figcompsdss} 
\end{figure}

The most obvious flux calibration issues in DESI DR1 white dwarf sample are around H$\gamma$. These issues are most obvious in low $\log{g}$ objects,
because of the narrower lines. Figure \ref{figsdprob} shows our spectroscopic model fit to a subdwarf, where these issues are easily visible. The DESI DR1
spectrum over-estimates the flux in the H$\gamma$ wings for this star. Of course, this problem would be less apparent in normal DA stars because these defects would be lost within the broad lines, but they are likely impacting the spectroscopic model fits like the one shown in Figure \ref{figdaprob}.

To further demonstrate that this problem is unique to the DESI DR1 data, we compared the spectroscopic parameters of 346 DA white dwarfs with high signal-to-noise ratio DESI and SDSS spectra. For this particular experiment, we rely on normalized Balmer line profiles to fit both the DESI and SDSS spectra. Figure \ref{figcompsdss} shows the results from this experiment. The top and bottom panels show a comparison between the spectroscopic temperatures and masses obtained from DESI versus SDSS spectra, along with the 1:1 line shown in red in both panels.

Compared to the fits using the SDSS spectra, temperatures and masses obtained from the DESI data are on average higher by 570 K and $0.05~M_\odot$, respectively. This difference ($0.05~M_\odot$) is essentially identical to the systematic mass difference of $0.06~M_\odot$ against the photometric parameters discussed above. Hence, what we are seeing here with the DESI data is different; masses for DA white dwarfs are 0.05-$0.06~M_\odot$ higher than expected (from photometry and the SDSS spectra) based on DESI DR1 spectroscopy.

\citet{manser24} presented an analysis of 1958 DA white dwarfs within the DESI Early Data Release in their Figure 10. They discuss several issues with the $\log{g}$ distribution of their sample, including higher than expected surface gravities for DAs below $15,000$ K. They 
concluded that these artifacts are related to the spectroscopic analysis. For example, they attributed the higher than expected surface gravities to the so-called high $\log{g}$ problem due to the incorrect treatment of convection in 1D model atmospheres. Interestingly, they also noted that a small deviation from $\log{g}=8.0$ still persists even after the 3D model corrections from \citet{tremblay13} are included.

\begin{figure*}
\center
\includegraphics[width=4.5in, clip=true, trim=1in 0.3in 1.3in 0.4in]{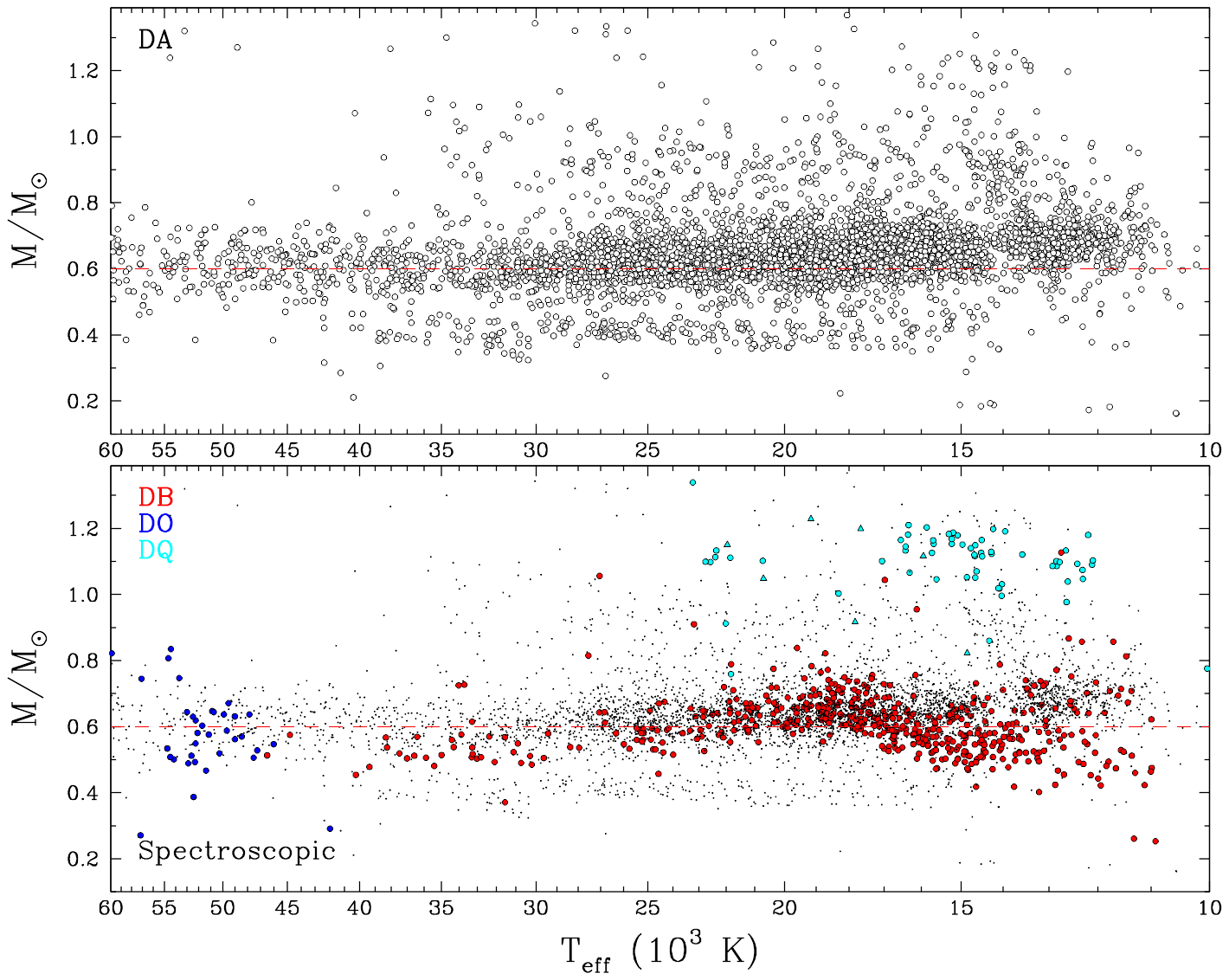}
\caption{Spectroscopic mass and temperature distributions of the DA (top panel) and non-DA (bottom panel) white dwarfs with S/N $>15$ spectra in DESI DR1. Different colors mark the various spectral types in the bottom panel, which also includes the DA white dwarfs as small dots for comparison. The dashed lines mark the canonical white dwarf mass ($0.6~M_\odot$) in the solar neighborhood.}
\label{figspec} 
\end{figure*}

Figure B2 in the Appendix of \citet{manser24} shows their results from a comparison between the spectroscopic and photometric parameters of their DESI Early Data Release DA white dwarf sample. They find higher temperatures from DESI spectra compared to the photometric temperatures, with a weighted mean difference of $15\pm22$\% in $T_{\rm eff}$ and $4\pm5$\% in $\log{g}$ between the spectroscopic and photometric parameters. Given the smaller sample size and the relatively large scatter, \citet{manser24} argue that there is overall a good agreement between the spectroscopic and photometric results. However, a systematic offset between the surface gravities derived from DESI spectra versus the photometric method is clearly visible in their Figure B2; the majority of the objects shown in that figure have spectroscopic surface gravities that are higher than the photometric values. For example, using the results from \citet{manser24}, the average $\log{g}$ increases from 7.95 for DAs between 18,000 - 20,000 K to 8.01 at 16,000 - 18,000 K, and to 8.17 for 14,000 - 16,000 K, and remains at that level down to 10,000 K. Hence, DA white dwarfs in the Early Data Release sample also show evidence of problems in spectroscopic surface gravities, but the smaller sample size likely made the interpretation more difficult in \citet{manser24}.

\subsection{DESI DR1 White Dwarf Sample Properties}
\label{secsample}

\begin{figure*}
\center
\includegraphics[width=4.5in, clip=true, trim=1in 0.3in 1.3in 0.4in]{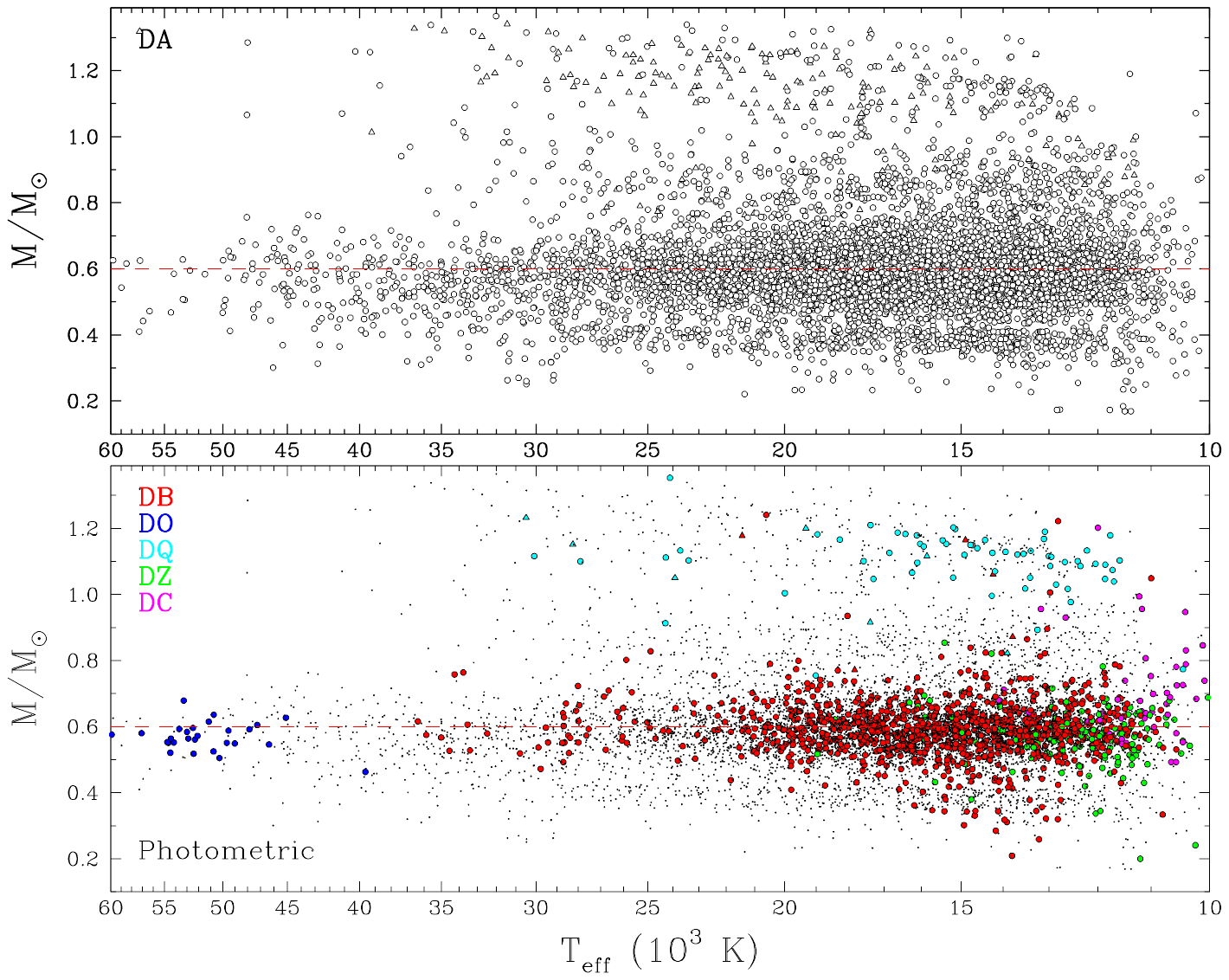}
\caption{Photometric mass and temperature distributions of the DA (top panel) and non-DA (bottom panel) white dwarfs in DESI DR1 with a distance uncertainty $<10$\%. Different colors mark the various spectral types in the bottom panel, which also includes the DA white dwarfs as small dots for comparison. 
The DA sample shown here includes DAO white dwarfs, DO includes DOZ, DZ includes DBZ and DBAZ, and DQ includes DAQ and DQA white dwarfs. Triangles mark the magnetic white dwarfs for each spectral type. This figure includes all DQs and DZs in our sample, regardless of their distance uncertainties. The dashed lines mark the canonical white dwarf mass of $0.6~M_\odot$.}
\label{figphot} 
\end{figure*}

Figure \ref{figspec} summarizes the spectroscopic results below 60,000 K for the DA and non-DA white dwarfs with S/N $>15$ spectra in DESI DR1 (see Figure \ref{figtg} for hotter stars). We find a broad range of masses for DA white dwarfs, including a significant population of low-mass DAs with $M\sim0.4~M_\odot$ and a smaller number of ultramassive DAs above $1~M_\odot$. More importantly, the spectroscopic masses for DA white dwarfs are concentrated above the $0.6~M_\odot$ line due to systematic issues with DESI spectra, as discussed in the previous section. 
The spectroscopic values also show a gap (or deficit of objects) near $T_{\rm eff}=14,000$ K, where Balmer lines reach their maximum strength. This indicates that the predicted line profiles are stronger than observed, thus pushing the objects on each side of the maximum. This may be due to an improper calibration of the convective efficiency \citep[see Figure 3 in][]{bergeron95}, or more likely here, a problem with the spectra \citep[e.g., Figure 7 in][]{genest19}. 

Figure \ref{figphot} shows the mass and temperature distribution of the DA and non-DA white dwarfs in our sample based on the photometric method. To reduce the scatter in this diagram, here we limit the sample to objects with a distance uncertainty $<10$\%, except for the DQs and DZs (all of which are included here). Unlike the spectroscopic mass distributions shown in the previous diagram, the photometric masses do not reveal any systematic trends. The spectroscopic masses appear larger than the photometric masses, as can be appreciated by using the $0.6~M_\odot$ sequence in both figures. On the other hand, the photometric masses are consistent with the canonical white dwarf mass of $0.6~M_\odot$ in the solar neighborhood. 
This is true for both DA and non-DA white dwarfs, including DB, DC, DO, and DZ white dwarfs. The exception to this rule is the DQ white dwarfs, which are ultramassive and further discussed below. 

The spectroscopic masses for DB white dwarfs also show an unusual pattern. The bottom panel in Figure \ref{figspec} shows that DO and DB white dwarfs with $T_{\rm eff}$ between 60,000 and 40,000 have similar spectroscopic mass estimates (in comparison to the DA population in the background of the figure), but the spectroscopic masses appear too low between 40,000 K and 30,000 K, and then too large near 20,000 K, and then down again. This pattern is not unique to DESI spectra however. Studying the SDSS DB white dwarf sample, \citet{genest19} also found the same pattern in the DB mass distribution (see their Figure 10). This is exactly the parameter space where the 1D atmosphere models have problems constraining the parameters of DB white dwarfs. The 3D model corrections for DB white dwarfs (not applied here) become important below 30,000 K \citep{cuka18,cuka21}, but they do not resolve this problem \citep[see Figure 7 in][]{genest19}. In addition, the treatment of van der Waals broadening is one of the largest sources of uncertainty below 16,000 K. Hence, the unusual pattern seen in the spectroscopic mass estimates for the DB white dwarfs in DESI DR1 is not surprising. 

The photometric masses for DB white dwarfs are not impacted by these issues.
The bottom panel in Figure \ref{figphot} demonstrates the overall sequence of helium-rich white dwarfs transitioning from DO to DB white dwarfs. DC and DZ white dwarfs form a natural extension of the DB sequence below about $T_{\rm eff}=12,000$ K. The so-called `DB gap', a deficiency in the number of DB white dwarfs between 50,000 and 40,000 K is also apparent in this diagram. 

An interesting feature of the photometric mass distributions shown in Figure \ref{figphot} is the prevalence of magnetic white dwarfs with $M\geq1~M_\odot$ among the DA population \citep[see also][]{bagnulo21,bagnulo22,moss25b} and that of DQs
in the non-DA population. A significant fraction of massive white dwarfs is expected to form through binary mergers, including double white dwarf mergers \citep{temmink20} that likely result in strongly magnetic ultramassive white dwarfs \citep{garciaberro12}. Previous work based on the $M>0.9~M_\odot$ and $T_{\rm eff}\geq11,000$ K white dwarfs in the 100 pc MWDD sample and the Pan-STARRS footprint found a magnetic fraction of 32\% \citep{jewett24}. We find a similar fraction of 36\% for the $M\geq1.0~M_\odot$ magnetic white dwarfs in our sample. 

\citet{jewett24} found that 10\% of their massive white dwarf sample with $T_{\rm eff}\geq11,000$ K is composed of DQs. Here, we also find a similar fraction; DQs make up 12\% of the $M\geq1.0~M_\odot$ white dwarf sample with $T_{\rm eff}\geq10,000$ K. Depending on their effective temperatures, DQs above 18,000 K have been traditionally classified as hot DQs \citep{dufour08}, and between 18,000-10,000 K as warm DQs. 
Hot and warm DQs form a continuous sequence in both Figure \ref{figspec} and Figure \ref{figphot}, and they are likely a homogeneous population
\citep[see also][]{dufour13,fortier15,koester19,coutu19,kilic24}.
Our detailed spectroscopic analysis in this paper strongly supports that conclusion. We discuss this sample further below, and simply refer to this population as warm DQs.

\citet{jewett24} also noted the absence of normal DBs in their massive white dwarf sample, as all six DBs in their sample are either strongly magnetic and/or rapidly rotating. The mass distribution of the DBs in the 100 pc SDSS sample is also clearly missing the high-mass tail \citep{kilic25a}, which favors a single star evolutionary channel for their formation \citep{hallakoun24}. Our sample includes a total of 11 massive DB white dwarfs (in a sample of 505 white dwarfs with $M\geq1.0~M_\odot$ and $T_{\rm eff}>10,000$ K), five of which are confirmed or suspected to be magnetic, including the double-faced object WDJ084716.22+484220.32 \citep{moss25}. There are also six ultramassive DC white dwarfs in our sample. These could be magnetic, but given their relatively cool temperatures, we cannot rule out non-magnetic white dwarfs with pure He atmospheres. 

Another feature of the photometric DA mass distribution is the prevalence of low-mass white dwarfs with $M\leq0.4~M_\odot$ in the top panel of Figure \ref{figphot}. They make up 16\% of the DA population, which is much higher than observed in volume-limited samples \citep[e.g.,][]{kilic20,kilic25a}. The over-abundance of these systems in DESI, as well as the Palomar-Green Survey \citep{liebert05} and the SDSS spectroscopy samples \citep[e.g.,][]{kleinman13},
is because DESI, Palomar-Green, and the SDSS are magnitude-limited surveys, which favor over-luminous systems. Note that our mass estimates for these low-mass white dwarfs are based on single star model fits. Many of these stars are likely over-luminous because they are in unresolved binary systems, and thus the stellar radius will be overestimated with the photometric method, and therefore the masses are underestimated. A  de-convolution may yield normal masses, such as the DA+DA binary shown in Figure \ref{figdouble}.
Follow-up radial velocity observations would be needed to properly characterize these systems. \citep[e.g.,][]{marsh95,munday24}. We discuss further a subsample of these systems, extremely low-mass white dwarfs, below. 

\subsection{White Dwarf Spectral Evolution}

A significant fraction of white dwarfs go through spectral evolution due to gravitational settling, radiative levitation, winds, convection, and external accretion. In a series of papers, \citet{bedard20,bedard22b,bedard22a,bedard23} studied the spectral evolution of hot white dwarfs, considering various transformations between the different spectral types.

\citet{bedard24} provide a summary of our current understanding of the white dwarf spectral evolution.
Figure 2 in that work presents the fraction of He-atmosphere white dwarfs as a function of temperature based on several recent studies. Unfortunately, all of those individual studies are limited to certain temperature ranges, for example the \citet{bedard20} sample is restricted to $T_{\rm eff}\geq30,000$ K, whereas \citet{ourique19} sample covers only the $T_{\rm eff}\leq30,000$ K region. By combining the results from various surveys, \citet{bedard24} show that 20\%–30\% of hot white dwarfs have He-rich atmospheres, but this fraction decreases to 5-15\% between 40,000 - 20,000 K, and increases back to 20-35\% at 10,000 K. These numbers suggest that about 75\% and 10\% of white dwarfs retain their H and He atmospheres, respectively, and 15\% transition from He to H atmospheres, and back to He atmospheres at cooler temperatures. However, there are significant discrepancies between various studies, and \citet{bedard24} highlight the need for better constraints on the spectral evolution covering a large temperature range.

\begin{figure}
\center
\includegraphics[width=3in, clip=true, trim=0.9in 2in 1in 2.2in]{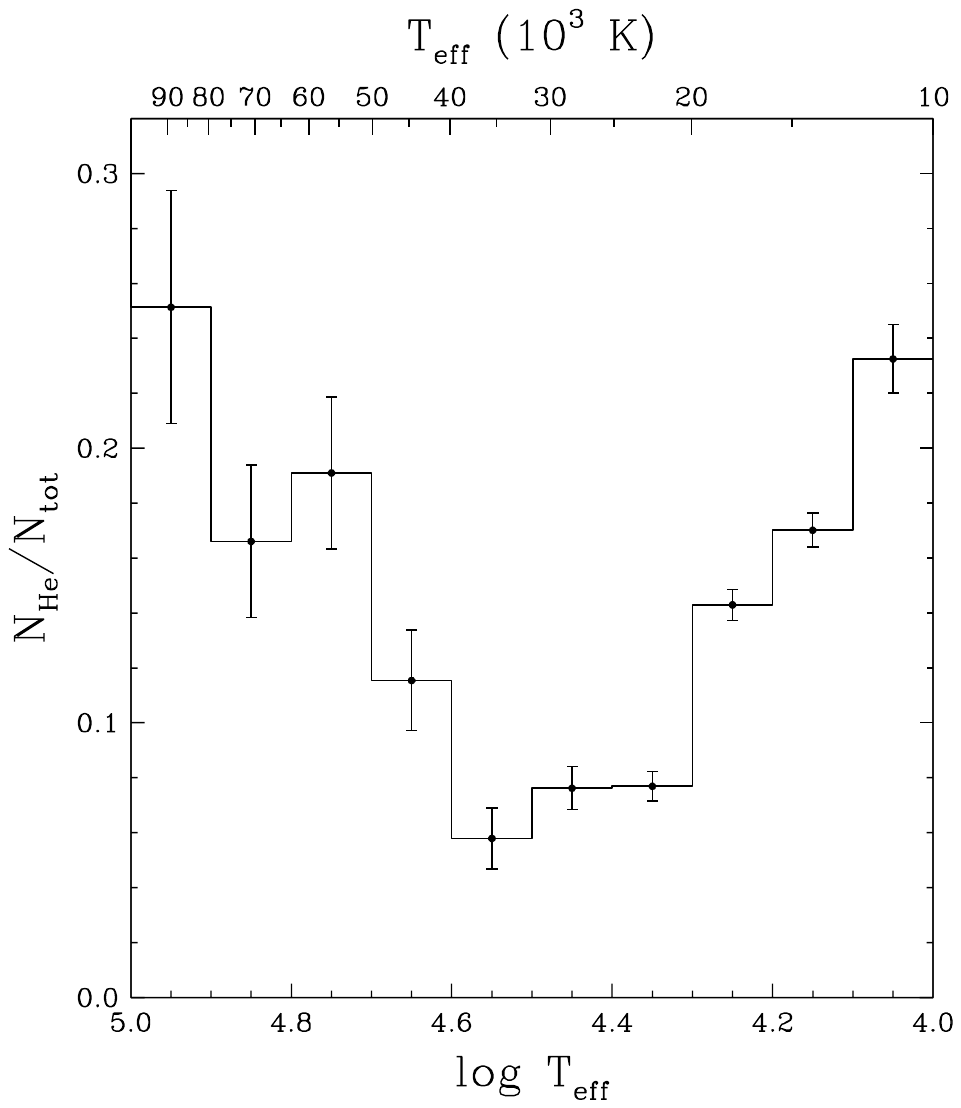}
\caption{Fraction of He-atmosphere white dwarfs as a function of temperature from the DESI DR1 sample.}
\label{fighetot} 
\end{figure}

DESI DR1 provides an excellent dataset to revisit white dwarf spectral evolution. Prior studies that relied on SDSS spectroscopy suffered from complicated selection biases \citep[see the discussion in][]{manser24}, while DESI's Gaia-based selection provides a more realistic representation of the local white dwarf sample. DESI still suffers from biases due to being a magnitude-limited survey, which favors hotter and brighter objects, and thus larger radii and low-mass white dwarfs. However, it is much easier to quantify and correct such a bias.

Figure \ref{fighetot} shows the fraction of He-atmosphere white dwarfs as a function of temperature from the DESI DR1 sample. Here, we use the complete DESI DR1 hot white dwarf sample without restrictions. In addition, we use the photometric temperatures for stars cooler than 30,000 K, and the spectroscopic temperatures for stars above that limit. The He-fraction starts at about 25\%, and decreases to about 7\% between 40,000 and 20,000 K, and then gradually rises to $\approx23$\% near 10,000 K. H- and He-atmosphere white dwarfs have different emergent fluxes due to their different atmospheric opacities. Their surface hydrogen layer masses are also different (thick versus thin layers). Hence, their overall luminosities are different even if the stars have identical atmospheric parameters. \citet{bedard20} present corrections for these effects for a magnitude limited survey, which tend to increase the He-fractions for the hottest bins. Since our results are based on LTE atmosphere models, the parameters for the hottest white dwarfs are relatively uncertain.
However, NLTE effects become negligible below 40,000 K. Hence, our results are more robust below that limit. The non-DA fractions found in this sample are comparable to the \citet{bedard20} estimates shown in their Figure 19 for $\log~T_{\rm eff}=$ 4.5-4.9. 

The non-DA fraction has been studied extensively in the literature. For example, \citet{genest19}, \citet{ourique19}, \citet{cunningham20}, \citet{bedard20},
\citet{lopez22}, \citet{jimenez23}, \citet{torres23}, and \citet{obrien24} present non-DA fractions in their samples covering parts of the temperature range
shown in Figure \ref{fighetot}. Collectively, these samples span the full temperature range shown in this figure. \citet{bedard24} present a summary 
of these results in their Figure 2; the non-DA fraction in these studies ranges from a few per cent to $\sim10$\% between 20,000 and 30,000 K, but there
is significant scatter, especially at lower temperatures. For example, near 10,000 K, the non-DA fraction varies from $\approx20$\% to 36\% between these different studies.
The majority of these studies either rely on SDSS spectroscopy (with a complicated selection bias), optical/UV photometry, or Gaia low-resolution XP spectra.
The non-DA fraction found in the DESI DR1 white dwarf sample presented here agrees with the previous studies, but provides more robust constraints, indicating
that the non-DA fraction is $\approx23$\% near 10,000 K.

Volume-limited surveys provide an independent check on these results. For example, using the 100 pc white dwarf sample in the SDSS footprint, which is limited to temperatures below 25,000 K, \citet{kilic25a} found He fraction increasing from about 9\% at 20,000 K to 24\% between 9000-10,000 K and
$\approx32$\% at 6000 K. Combining these numbers, the number of He-atmosphere white dwarfs decreases from about 25\% for the hottest white dwarfs down to 8\% for $T_{\rm eff}\sim20,000$ K white dwarfs, and increases back to $\sim23\%$ at 10,000 K and $\sim32$\% at 6000 K. The decrease from 25\% to 8\% for hot white dwarfs is explained by the H float-up process where the H diluted in the envelope floats to the surface due to gravitational settling.
DA white dwarfs may turn into non-DA white dwarfs below 20,000 K through the convective dilution process (mixing of the thin radiative H layer with the deeper He convection zone) and at cooler temperatures ($T_{\rm eff}\sim15,000$ K or less) due to convective mixing (when the bottom of the superficial H convection zone reaches the underlying He envelope) \citep[see][and references therein]{bedard24}. For the convective mixing process, depending on the thickness of the surface H envelope, stars with thicker layers mix at cooler temperatures, gradually increasing the number of non-DAs at cooler temperatures.   

\subsection{Warm DQs}
\label{secdq}

DQ white dwarfs above and below 10,000 K show wildly different properties. Cool DQs represent a natural extension of the average DB white dwarf sequence, where C is dredged up from the interior \citep{pelletier86}, changing the spectral type to a DQ white dwarf. On the other hand, warm DQs are different \citep{dufour08}. Their unusual kinematics, atmospheric abundances (dominated by carbon), and high masses indicate a merger origin \citep{dunlap15,coutu19,kawka23}.
Warm DQs are rare in the SDSS spectroscopy sample. \citet{dufour08} found nine hot DQs in the SDSS, whereas \citet{koester19} identified 26 warm DQs in a sample of 20,088 white dwarfs with SDSS spectra. As discussed in Section \ref{secsample}, hot and warm DQs seem to form a continuous sequence in mass vs temperature diagrams. We simply refer to the entire warm/hot DQ population as warm DQs going forward.

Recent discoveries of DAQ white dwarfs with C+H atmospheres \citep{hollands20,kilic24,jewett24} have brought new attention to these objects. \citet{kilic24} showed that warm DQ, DQA, and DAQ white dwarfs likely have similar C+H atmospheres, and that DAQs simply represent the most H-rich atmospheres in the warm DQ sample. \citet{kilic25b} took advantage of GALEX FUV photometry, which is significantly impacted by the presence of C in the atmosphere, to perform a targeted survey for warm DQs in the GALEX All Sky Imaging footprint, and discovered 75 warm DQs in the process, including 10 new DAQ white dwarfs. 

\begin{figure}
\center
\includegraphics[width=3.3in, clip=true, trim=0.3in 4.2in 0.4in 2.6in]{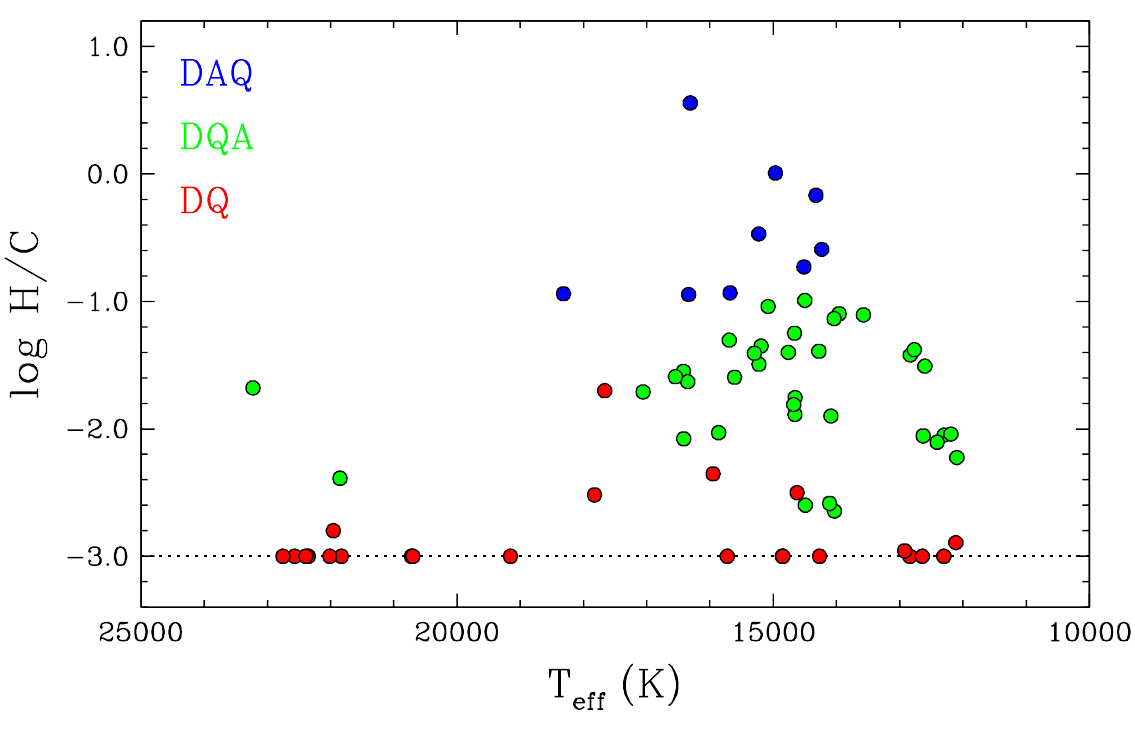}
\caption{H/C ratio vs.~spectroscopic effective temperature for the DQ, DQA, and DAQ white dwarfs in our sample. 
This figure excludes the two cool DQ stars with molecular carbon lines. The dotted line marks the lower limit of our model grid.}
\label{figwarmdq} 
\end{figure}

With a selection based on Gaia, DESI DR1 did much better than SDSS in finding warm DQ white dwarfs. We identify 68 warm DQs, including 9 DAQs and 36 DQAs in DESI DR1. Figure \ref{figwarmdq} shows the H/C ratios vs. spectroscopic temperatures for the DESI warm DQ sample. This figure presents $\log$ H/C, instead of $\log$ C/H as shown in the model fits, since hydrogen is a trace element for most of the DQs in our sample. Two thirds of the warm DQs in DESI are H-rich. H lines appear stronger than C lines in optical spectra for $\log$ H/C $\geq-1$; those objects are classified as DAQ. H is detected in warm DQs down to about $\log$ H/C = $-2.6$ in DQAs, demonstrating the hydrogen detection limit in these stars. It is possible that all warm DQs may have some H in their atmospheres. 

The lowest H/C ratio in our model grid is $\log$ H/C = $-3$. At such low H/C ratios, hydrogen is invisible in the spectra, and the model fits are identical to spectroscopic fits using pure carbon models. There are a number of DQs in our sample, where we had to force a solution at $\log$ H/C = $-3$. The H/C ratio in these stars are likely lower. 

WDJ005407.15+212141.23 is the most H-rich DQ in the DESI sample with H/C $>1$. Along with J0551+4135 \citep{hollands20,kilic24}, it seems to be an outlier among the DAQ population. However, given its relatively hot effective temperature near 16,000 K, the C lines are relatively weak compared to the Balmer lines (see Figure \ref{figmmt}). Hence, the most H-rich DAQs may escape detection in noisy optical spectra \citep[see for example][for the detection of a DAQ using UV spectroscopy]{sahu25}. 

The mass distribution of the warm DQ sample in DESI DR1 is included in Figure \ref{figphot}. The sample mean and standard deviation are $M=1.10 \pm 0.09~M_\odot$, which is very similar to the mass distribution of the warm DQ sample from \citet{kilic25b}. Hence, the DESI DR1 sample confirms the previous results that warm DQs are ultramassive. The most massive warm DQ in our sample is WDJ115305.54+005646.15 with $M=1.353 \pm 0.024~M_\odot$. The model fits to this DQ are shown in the right panels of Figure \ref{fitdq}. The best-fitting model provides an excellent fit to the photometry and spectroscopy of this target, giving us confidence in its unusually high mass. Along with WDJ205700.44$-$342556.37 presented in \citet{kilic25b}, WDJ115305.54+005646.15 is the only other warm DQ currently known with a mass above $1.3~M_\odot$. This is an important point, as it can help us understand the progenitors of these merger remnants.

\begin{figure}
\center
\includegraphics[width=3in]{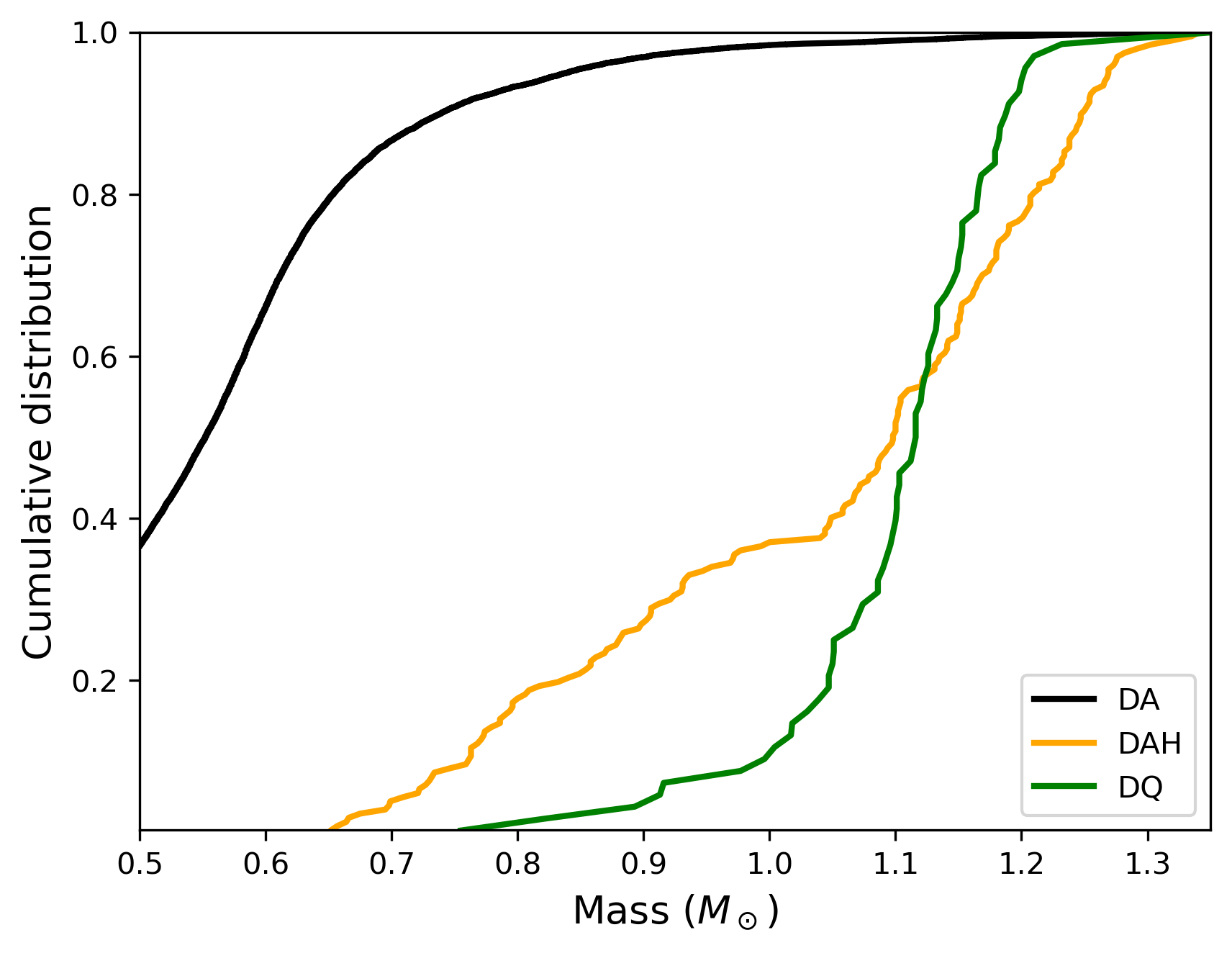}
\caption{Cumulative mass distributions of DA, DAH, and warm DQ white dwarfs in our sample.}
\label{figcumulative} 
\end{figure}

Figure \ref{figcumulative} shows the cumulative mass distributions of normal DA, magnetic DAH (excluding DAH? stars), and warm DQ white dwarfs in the 0.5-1.35 $M_\odot$ range. All three distributions are clearly very different; among the hot white dwarf sample, both magnetic and warm DQ white dwarfs are significantly more massive than normal white dwarfs. In addition, the distributions for magnetic and warm DQs are also different: a two sample Kolmogorov-Smirnov test rules out the null hypothesis at high significance with a $p-$value of 0.0005. Warm DQs are confined mostly to the 1.0-1.2 $M_\odot$ mass range, while magnetic DAs show a relatively broad mass distribution. For example, all but one of the warm DQs in the DESI sample have $M<1.25~M_\odot$, whereas 10\% of the magnetic DAs are more massive than that. A significant fraction of magnetic white dwarfs above $0.9~M_\odot$ are likely double white dwarf merger remnants \citep{garciaberro12,temmink20}. Hence, the differences in the mass distributions of DAH and warm DQ white dwarfs suggest an alternate formation scenario. 

\citet{shen23} proposed mergers of white dwarfs + subgiants as a potential solution to the cooling anomaly observed for the Q-branch stars \citep{cheng19}. In this scenario, the merger of a $1~M_\odot$ white dwarf with the helium-rich ($M\sim0.2~M_\odot$) core of a subgiant leads to the creation of large amounts of $^{26}$Mg. This, along with $^{22}$Ne causes a distillation process that can lead to multi-Gyr cooling delays. 

Based on a study of the Q-branch white dwarf population in the 100 pc sample, \citet{ould26} find that the delayed population on the Q-branch is made up of DA and warm DQ white dwarfs that are rarely magnetic \citep[see also][]{coutu19}. Double white dwarf mergers are expected to produce strongly magnetic remnants \citep{garciaberro12}. 
Hence, \citet{ould26} favor the white dwarf + subgiant merger origin for the delayed Q-branch population. However, it is not necessarily clear that a white dwarf + subgiant merger would not create a magnetic white dwarf. The lack of magnetic stars among the delayed population on the Q-branch \citep{ould26} may not necessarily be an evidence of white dwarf + subgiant mergers; more theoretical work is required to understand the end product of such mergers. 

An alternative idea is that the distillation process might kill the magnetic field. During distillation, there is turbulence in the carbon/oxygen core. Turbulence enhances magnetic diffusivity, which should lead to the field decaying faster. Since these stars get stuck on the Q branch for nearly a Hubble time, this leaves plenty of time for the field to decay and for the remnant to appear non-magnetic.

If this distillation-enhanced field decay idea is correct, then the expectation is that magnetic warm DQs should mostly appear before the Q-branch over-density (where the field decays). Interestingly, this is consistent with the fraction of magnetism among the warm DQ population shown in Figure \ref{figphot}; magnetic DQs are relatively common above $\sim15,000$ K, but not below that temperature \citep[see also][]{dufour08,coutu19}. In fact, if the hot DQs are the progenitors of warm DQs, then the argument that warm DQs are not magnetic because they are WD + subgiant merger products does not hold, because if that were true then the hot DQs should not be magnetic either.

\subsection{Metal-Rich White Dwarfs and the Curious Case of a Dimming DBZ}

\begin{figure}
\includegraphics[width=3.3in]{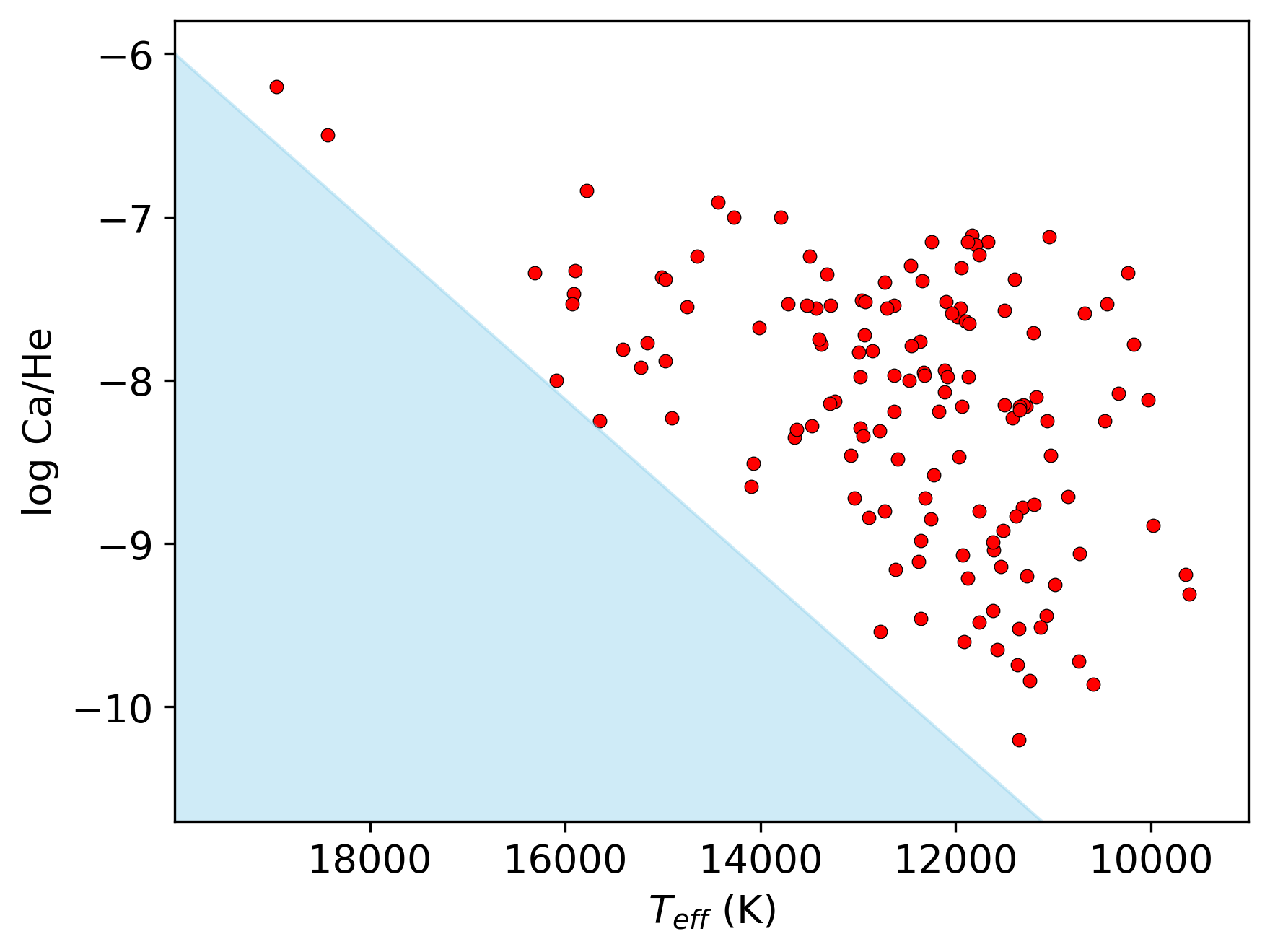}
\caption{Ca/He ratio vs. photometric effective temperature for the metal-rich white dwarfs in our sample. The shaded blue region is empty due to the detection limit of \ion{Ca}{2} H and K lines in DESI spectra.}
\label{figdz} 
\end{figure}

Figure \ref{figdz} shows the abundance of calcium as a function of effective temperature for the metal-rich white dwarfs in our sample. The bottom left portion of this diagram is empty (the shaded blue region) due to the detection limit of the \ion{Ca}{2} H and K lines in DESI spectra. Excluding the two hottest stars in the sample with $T_{\rm eff}\sim18,000$ K, the Ca abundances range from $\log$ Ca/He = $-10$ to $-7$. This is very similar to the range of Ca abundances seen in the DBZ/DZ white dwarfs analyzed in \citet[][see their Figure 8]{coutu19}. The bottom panel in Figure \ref{figphot} shows the masses of the metal-rich population in our sample, with an average of $0.565 \pm 0.097~M_\odot$. This is consistent with the mass distribution of the DB white dwarfs in the DESI sample. 

The DESI DR1 metal-rich white dwarf sample includes known dusty white dwarfs like GD 40 (WDJ030253.10$-$010833.80), as well as WD 1145+017 (WDJ114833.63+012859.42), which is eclipsed by a disintegrating minor body \citep{vanderburg15}. Our model fits under the assumption of chondritic abundance ratios provide excellent fits to the majority of targets in the DESI sample, indicating that these objects are accreting remnants of rocky bodies. However, high resolution spectroscopy of these targets can provide a more detailed picture of the types of objects accreted \citep[e.g.,][]{zuckerman07,xu19}. 

\begin{figure}
\includegraphics[width=3.3in, clip=true, trim=0.4in 0.8in 0.1in 1.1in]{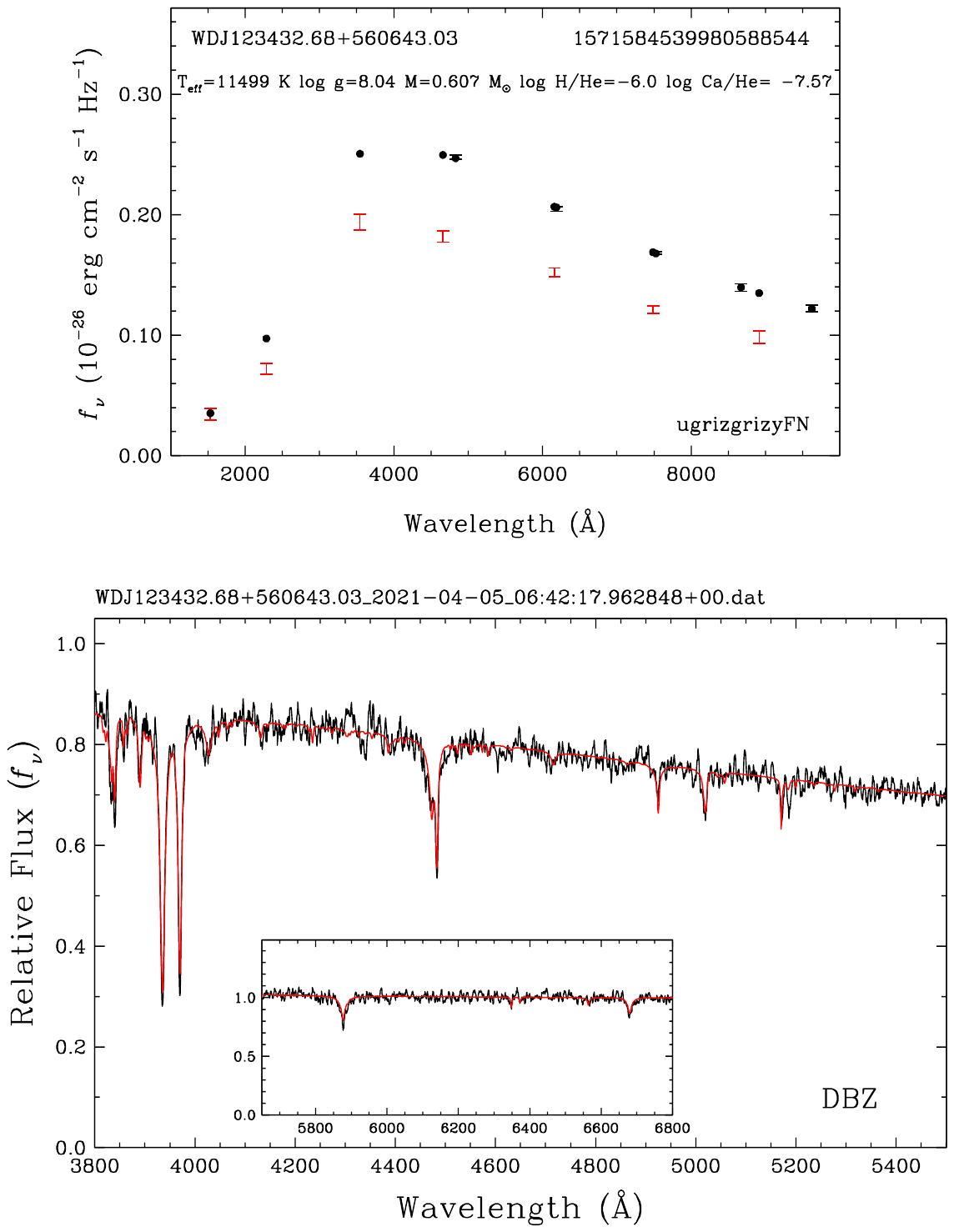}
\caption{Model fits to the DBZ white dwarf SBSS 1232+563, which was 0.3 mag dimmer in all filters during the SDSS observations.}
\label{fitdim} 
\end{figure}

An intriguing object included in our sample is SBSS 1232+563 (WDJ123432.68+560643.03), which was recently found to display sporadic dips from extended debris transits. \citet{hermes25} demonstrate that this star got fainter by $>40$\% in an 8 month long event in 2023, but these dimming events are not periodic. Figure \ref{fitdim} shows our model fits to this object. SBSS 1232+563 has optical photometry available in both SDSS and Pan-STARRS. However, the SDSS magnitudes are fainter by 0.34, 0.32, 0.34, and 0.35 mag in the $griz$ filters, respectively. \citet{coutu19} noted this discrepancy, but relied on the SDSS photometry for their model fits since the SDSS colors led to a much better spectroscopic fit of the helium lines indicating better constraints on the effective temperature of this star. \citet{coutu19} obtained $T_{\rm eff}=11,787$ K, $M=0.77~M_\odot$, and $\log$ Ca/He = $-7.41$, which were also adopted by \citet{hermes25}. 

Now that we know this object is going through sporadic dimming events, which happened to impact the SDSS photometry, we instead rely on the Pan-STARRS photometry along with the DESI spectrum of this object to constrain its parameters. The best-fitting model shown in Figure \ref{fitdim} provides an excellent fit to the DESI spectrum and the Pan-STARRS photometry. The SDSS (and GALEX UV) photometry is excluded from our fits and are shown as red error bars. The best-fitting model has $T_{\rm eff}=11,499$ K, $M=0.607~M_\odot$, and $\log$ Ca/He = $-7.57$. Hence, SBSS 1232+563 is not massive, and its parameters are consistent with a typical $0.6~M_\odot$ white dwarf that happens to be accreting from a debris disk that is causing sporadic dimming of this object.

We searched for similar dimming events among the metal-rich white dwarfs in our sample using Pan-STARRS and SDSS photometry with $\geq0.1$ mag differences in $gri$ filters. We found only one other potential candidate, WDJ214450.96+041847.23, that is dimmer in the SDSS by 0.17, 0.30, 0.64 mag in $gri$ filters, respectively. However, unlike SBSS 1232+563, the dimming is not consistent with eclipses from a dark body, because of the inconsistent dimming across the $gri$ bands. In addition, WDJ214450.96+041847.23 is near a bright star, hence its photometry is likely contaminated. 

\subsection{ELM White Dwarfs}
\label{secelm}

\begin{figure}
\includegraphics[width=3.3in]{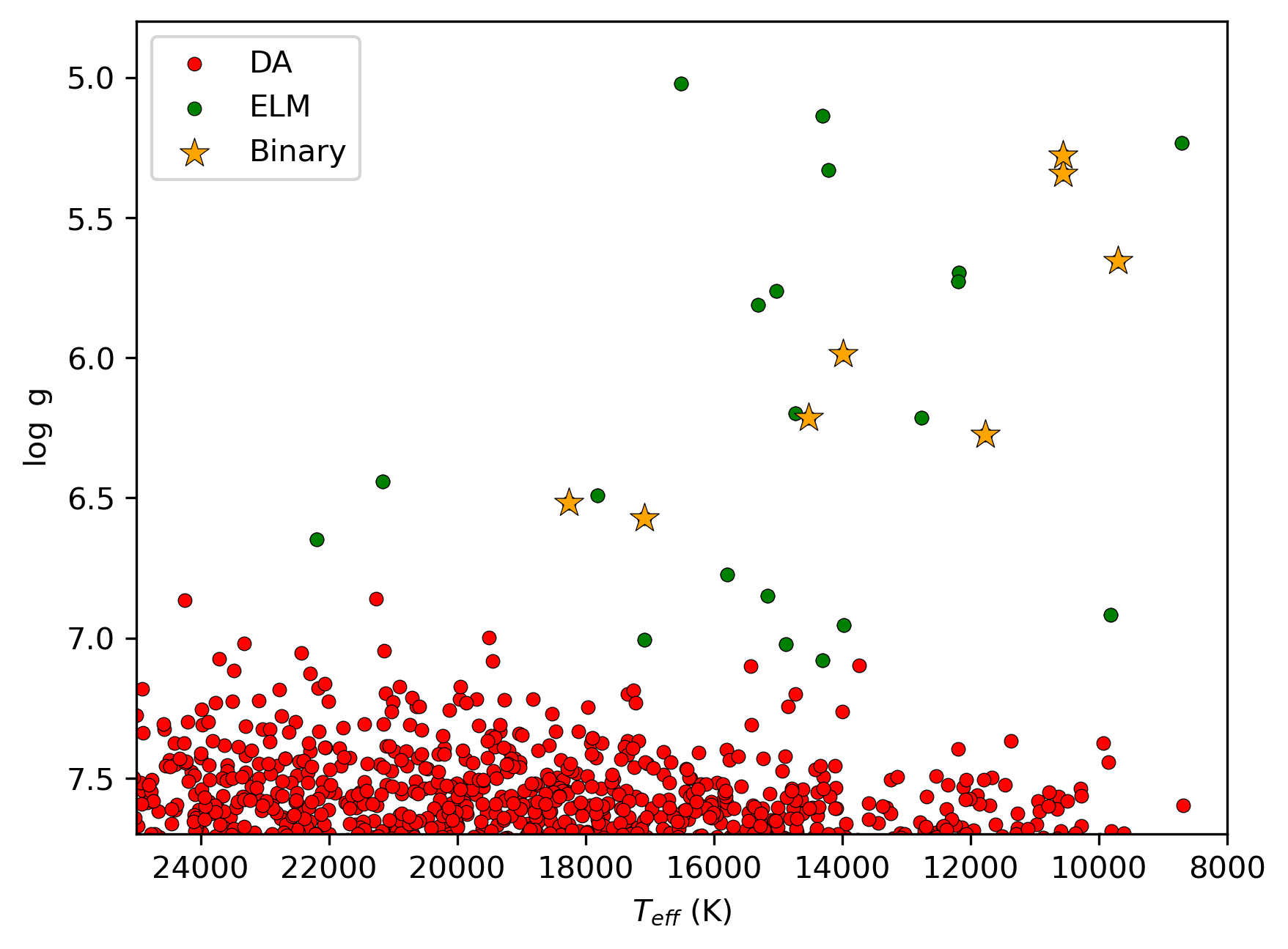}
\caption{Spectroscopic parameters of our DA white dwarf sample in the ELM region. Stars mark the previously known ELM white dwarf binaries, and green dots mark the remaining 20 ELM white dwarfs that are prime targets for follow-up radial velocity observations.}
\label{figelm} 
\end{figure}
  
\begin{deluxetable*}{lrlcrccc}
\tabletypesize{\tiny}
\tablecolumns{8} \tablewidth{0pt}
\tablecaption{ELM white dwarfs in our sample.\label{tabelm}}
\tablehead{\colhead{Name} & \colhead{SourceID} & \colhead{$T_{\rm eff,phot}$} & \colhead{$\log{g,phot}$} & \colhead{$T_{\rm eff,spec}$} & \colhead{$\log{g,spec}$} & \colhead{Mass,spec} & \colhead{References}\\
 &  & (K) & (cm s$^{-2}$) & (K) & (cm s$^{-2}$) & ($M_\odot$) & }
\startdata
WDJ004449.47$-$182625.85 & 2369026857221068160 & 12233 $\pm$  182 & 6.58 $\pm$ 0.11 & 14877 & 7.02 & 0.288 &  \\
WDJ081353.06+752111.68   & 1135693806667747840 & 12138 $\pm$  312 & 5.76 $\pm$ 0.09 & 12179 & 5.70 & 0.173 &  \\
WDJ083107.92+001331.38   & 3076962575704962176 & 14197 $\pm$  183 & 5.93 $\pm$ 0.16 & 15028 & 5.76 & 0.188 &  \\
WDJ102153.13+054322.46   & 3861429723729285376 & 17644 $\pm$  839 & 7.40 $\pm$ 0.35 & 17811 & 6.49 & 0.215 &  \\
WDJ112157.14+605210.42   & 861011995046220544  & 11484 $\pm$   73 & 5.37 $\pm$ 0.03 & 10557 & 5.34 & 0.162 & \citet{kosakowski23} \\
WDJ112914.17+471501.77   & 785814333240812544  & 11371 $\pm$   93 & 5.29 $\pm$ 0.03 & 10565 & 5.28 & 0.163 & \citet{kosakowski23} \\
WDJ115935.70$-$063347.03 & 3595644434349398272 & 10629 $\pm$  159 & 6.11 $\pm$ 0.57 &  8708 & 5.23 & 0.149 &  \\
WDJ123410.37$-$022802.81 & 3683189405578704640 & 17020 $\pm$  253 & 6.48 $\pm$ 0.12 & 18265 & 6.52 & 0.223 & \citet{kilic11} \\
WDJ123619.70$-$044437.32 & 3680368505418792320 & 11516 $\pm$   90 & 6.26 $\pm$ 0.04 & 11770 & 6.27 & 0.182 & \citet{kosakowski20} \\
WDJ131349.97+582801.50   & 1566990986557895040 & 16088 $\pm$  369 & 6.72 $\pm$ 0.09 & 17084 & 7.01 & 0.296 &  \\
WDJ134221.63+082949.71   & 3724718990552618624 & 12515 $\pm$  427 & 6.55 $\pm$ 0.19 & 13971 & 6.96 & 0.269 &  \\
WDJ134602.58+443151.67   & 1503230380279211776 & 15018 $\pm$  496 & 7.19 $\pm$ 0.26 & 15169 & 6.85 & 0.254 &  \\
WDJ141934.99+045736.13   & 3669654795398003456 & 11211 $\pm$  365 & 7.25 $\pm$ 0.49 &  9815 & 6.92 & 0.241 &  \\
WDJ143633.29+501026.93   & 1603554764703627520 & 16919 $\pm$  407 & 6.44 $\pm$ 0.12 & 17088 & 6.57 & 0.221 & \citet{mullally09,kilic10a} \\
WDJ143948.40+100221.72   & 1174732619846880896 & 14470 $\pm$  285 & 6.39 $\pm$ 0.10 & 14519 & 6.21 & 0.184 & \citet{brown10} \\
WDJ152854.00+410755.29   & 1390546686291717120 & 13347 $\pm$  493 & 6.07 $\pm$ 0.34 & 12195 & 5.73 & 0.173 &  \\
WDJ153844.22+025209.58   & 4424162321742140928 & 10977 $\pm$  180 & 5.70 $\pm$ 0.32 &  9699 & 5.65 & 0.157 & \citet{brown13} \\
WDJ165349.31+321257.13   & 1313285000641259264 & 13284 $\pm$  314 & 6.70 $\pm$ 0.16 & 14309 & 7.08 & 0.299 &  \\
WDJ170816.36+222551.02   & 4568269229124267904 & 18555 $\pm$  711 & 6.92 $\pm$ 0.28 & 21157 & 6.44 & 0.243 &  \\
WDJ174253.42+164127.73   & 4549325743288467968 & 17533 $\pm$  913 & 6.72 $\pm$ 0.26 & 22188 & 6.65 & 0.272 &  \\
WDJ181643.27+330254.85   & 4592975732137952640 & 14244 $\pm$ 1054 & 6.29 $\pm$ 0.39 & 16517 & 5.02 & 0.215 &  \\
WDJ185702.62+621655.20   & 2252937090963935360 & 10683 $\pm$  255 & 6.04 $\pm$ 0.34 & 12762 & 6.22 & 0.173 &  \\
WDJ190600.88+623923.82   & 2252265701675503616 & 12737 $\pm$  238 & 5.32 $\pm$ 0.19 & 14220 & 5.33 & 0.188 &  \\
WDJ211752.68+161829.37   & 1784606808964103168 & 13208 $\pm$  304 & 5.25 $\pm$ 0.29 & 14302 & 5.14 & 0.193 &  \\
WDJ212157.77+084729.33   & 1740525772898328832 & 15740 $\pm$  433 & 6.91 $\pm$ 0.24 & 15789 & 6.77 & 0.243 &  \\
WDJ213228.36+075428.27   & 1740741380258586624 & 13615 $\pm$  205 & 5.90 $\pm$ 0.18 & 13993 & 5.99 & 0.181 & \citet{brown13} \\
WDJ233938.46$-$034734.45 & 2639275992010565376 & 15269 $\pm$  432 & 6.09 $\pm$ 0.36 & 15311 & 5.81 & 0.188 &  \\
WDJ234852.30+280438.36   & 2866648537004311808 & 14392 $\pm$  357 & 6.16 $\pm$ 0.21 & 14728 & 6.20 & 0.185 &  \\
\enddata
\end{deluxetable*}

Extremely low-mass (ELM) white dwarfs with $M\leq0.3~M_\odot$ must form through close binary evolution, as the Universe is not old enough to produce them through single star evolution. The shortest period ELM binaries are some of the strongest sources of gravitational waves in the mHz frequency range of the Laser Interferometer Space Antenna \citep[LISA,][]{lisa12,kupfer24,kosakowski25,chickles25}. Electromagnetic constraints on the positions and inclinations of these systems can help significantly reduce the uncertainties of the gravitational wave parameters \citep{shah12,shah13,barrientos25}. Hence, the discovery of additional ELM white dwarfs before LISA's launch can aid their detection in gravitational waves.

\citet{liebert04} identified the first ELM white dwarf in the field population based on the SDSS DR1 spectroscopy. \citet{eisenstein06} expanded this sample to 13 objects among a sample of 9316 white dwarfs from the SDSS DR4. 
Due to their low surface gravities, ELM white dwarfs display a stronger Balmer jump, and occupy a unique parameter space in the $u-g$ vs. $g-r$ color-color diagram; at a given temperature (or $g-r$ color), they have redder $u-g$ colors. \citet{brown10} took advantage of this color-selection (see their Figure 1) to perform a spectroscopic survey in the SDSS footprint. Along with its recent extension to the southern hemisphere, the ELM Survey has discovered $>150$ systems with orbital periods ranging from 12 min \citep{brown11} to $\sim1.5$ days \citep{brown10,brown20b,kilic10a,kosakowski23}. \citet{burdge19a,burdge20a} used a photometric approach to search for short period binary systems using the Zwicky Transient Facility (ZTF) data, and identified 15 photometrically variable systems with periods ranging from 7 to 56 min, including seven detached systems. Given the cadence of the ZTF observations, these photometric searches are biased towards the shortest period systems. 

With a Gaia-based target selection for white dwarfs, DESI DR1 provides an excellent opportunity to identify ELM white dwarfs. Figure \ref{figelm} shows the spectroscopic surface gravities and temperatures for the DA white dwarfs in the ELM region. We identify 28 targets with $M\leq0.3~M_\odot$. Stars mark the eight previously confirmed ELM binaries, whereas green dots mark the remaining 20 ELM white dwarfs identified in this work. Note that some of these systems were previously identified as ELM or hot subdwarf candidates in the literature based on Gaia photometry and astrometry \citep{pelisoli19,geier20}. However, DESI spectra clearly indicate that these objects are ELM white dwarfs. We present the physical parameters, both from the photometric and spectroscopic methods, for the ELM sample in Table \ref{tabelm}. We plan on obtaining follow-up radial velocity observations of the 20 systems without prior orbital constraints.

\section{Conclusions}
\label{seccon}

We present a detailed model atmosphere analysis of hot ($T_{\rm eff}\gtrsim 10,000$ K), blue white dwarfs with $G_{\rm BP}-G_{\rm RP}\leq0$ in DESI DR1. Our sample is dominated by DA white dwarfs, which we use to verify DESI flux calibration. Unfortunately, our analysis reveals a significant discrepancy between the photometric and spectroscopic masses for DA white dwarfs of order 0.05-$0.06~M_\odot$, indicating problems with the broad hydrogen line profiles in DESI DR1 spectroscopy. Regardless of these issues, the photometric mass distribution of the DA sample is consistent with a peak at the canonical mass of $0.6~M_\odot$.

A remarkable feature of the photometric mass distribution of the DESI DR1 sample is the prevalence
of magnetic white dwarfs among the ultramassive DA population \citep{bagnulo21,bagnulo22,moss25b} and that of warm DQs in the non-DA population. The fraction of magnetic white dwarfs is 36\% for $M\geq1.0~M_\odot$, white dwarfs. A large fraction of these objects are likely double white dwarf merger remnants \citep{garciaberro12,temmink20}. Warm DQ white dwarfs are also believed to be merger remnants due to their unusual atmospheric composition, unusually high masses, and kinematics. Yet, magnetism is relatively rare in the warm DQ population \citep{coutu19}. We find significant differences in the mass distributions between DQs and magnetic DAs, favoring a different origin than the magnetic white dwarfs. As suggested by \citet{shen23} and \citet{ould26}, white dwarf + subgiant mergers are the likely progenitors of warm DQ white dwarfs.

We use the DESI data to constrain the non-DA to DA ratio over a broad temperature range ($10^5$ to $10^4$ K), and demonstrate that the number of He-atmosphere white dwarfs decreases from about 25\% for the hottest white dwarfs down to 8\% for $T_{\rm eff}\sim20,000$ K white dwarfs, and increases back to $\sim23\%$ at 10,000 K. The decrease from 25\% to 8\% for hot white dwarfs is due to the H float-up process where the H diluted in the envelope floats to the surface due to gravitational settling, whereas the gradual increase from 8\% to 23\% at cooler temperatures is best explained by convective mixing \citep[see][and references therein]{bedard24}. 

We highlight unusual objects in our sample, including DA+DB binaries, warm DQs, ELMs, and a relatively large number of metal-rich DBZ/DZ white dwarfs. High-resolution follow-up spectroscopy of the latter sample can provide interesting and additional constraints on the bulk composition of the accreted material in these systems \citep[e.g.,][]{zuckerman07,klein11,xu14,xu19}.

In this paper, we limited our selection to blue/hot white dwarfs to keep the sample size manageable for visual inspection. A detailed model atmosphere analysis of the significantly larger sample of cool white dwarfs in DESI DR1 could provide a relatively complete picture of white dwarf spectral evolution based on a single, well-defined Gaia-selected sample. Such a study would also be important for improving our understanding of the cool DQ and DZ white dwarf populations in the solar neighborhood, and may reveal additional rare white dwarfs. However, accomplishing that task requires pushing the limits of human capacity for visual inspection, as noted in \citet{caron23}.

\begin{acknowledgements}

This work is supported in part by the NSF under grant  AST-2508429, the NASA under grants 80NSSC22K0479, 80NSSC24K0380, and 80NSSC24K0436, the NSERC Canada, the Fund FRQ-NT (Qu\'ebec), and the Smithsonian Institution.

This research used data obtained with the Dark Energy Spectroscopic Instrument (DESI). DESI construction and operations is managed by the Lawrence Berkeley National Laboratory. This material is based upon work supported by the U.S. Department of Energy, Office of Science, Office of High-Energy Physics, under Contract No. DE–AC02–05CH11231, and by the National Energy Research Scientific Computing Center, a DOE Office of Science User Facility under the same contract. Additional support for DESI was provided by the U.S. National Science Foundation (NSF), Division of Astronomical Sciences under Contract No. AST-0950945 to the NSF’s National Optical-Infrared Astronomy Research Laboratory; the Science and Technology Facilities Council of the United Kingdom; the Gordon and Betty Moore Foundation; the Heising-Simons Foundation; the French Alternative Energies and Atomic Energy Commission (CEA); the National Council of Humanities, Science and Technology of Mexico (CONAHCYT); the Ministry of Science and Innovation of Spain (MICINN), and by the DESI Member Institutions: www.desi.lbl.gov/collaborating-institutions. The DESI collaboration is honored to be permitted to conduct scientific research on I’oligam Du’ag (Kitt Peak), a mountain with particular significance to the Tohono O’odham Nation. Any opinions, findings, and conclusions or recommendations expressed in this material are those of the author(s) and do not necessarily reflect the views of the U.S. National Science Foundation, the U.S. Department of Energy, or any of the listed funding agencies.

\end{acknowledgements}

\facilities{Mayall (DESI), MMT (Blue Channel spectrograph)}

\bibliographystyle{aasjournal}

\end{document}